%% file: main_lncs.tex
\documentclass[twocolumn,runningheads]{llncs}
\usepackage[T1]{fontenc}
\usepackage{graphicx}
\usepackage{lineno,hyperref}
\usepackage{subcaption}
\usepackage{tikz}
\usepackage{pgfplotstable}
\usepackage{pgfplots}
\usepackage{multirow}
\usepackage{float}
\usepackage{textcomp}
\usepackage{amsmath,amsfonts,amssymb}
\usepackage{multicol}
\usepackage{cuted}
\usepackage{makecell}
\usepackage{colortbl}
\usepackage{hyperref}
\usepackage{booktabs}
\usepackage{longtable}
\usepackage{array,tabularx}

\usepackage{adjustbox}
\usepackage{rotating}
\usepackage{dblfloatfix}

\usepackage{xcolor,soul}
\hypersetup{hidelinks}

\pgfplotsset{compat=1.18}

\input{Definitions/methods}

\usepackage{blindtext}

\usepackage[margin=0.75in]{geometry}
\begin{document}
\title{SMILE-UHURA Challenge - Small Vessel Segmentation at Mesoscopic Scale from Ultra-High Resolution 7T Magnetic Resonance Angiograms}
\titlerunning{SMILE-UHURA Challenge}
%
\input{Definitions/authors_lncs}
\authorrunning{S. Chatterjee et al.}
%
%
\maketitle              
\begin{abstract}
The human brain receives nutrients and oxygen through an intricate network of blood vessels. Pathology affecting small vessels, at the mesoscopic scale, represents a critical vulnerability within the cerebral blood supply and can lead to severe conditions, such as Cerebral Small Vessel Diseases. The advent of 7 Tesla MRI systems has enabled the acquisition of higher spatial resolution images, making it possible to visualise such vessels in the brain. However, the lack of publicly available annotated datasets has impeded the development of robust, machine learning-driven segmentation algorithms. To address the complexities of mesoscopic vessel segmentation and to highlight the need for advanced techniques to manage the high noise levels and poor vessel-to-background contrast inherent in "ultra-high-resolution" data, the SMILE-UHURA challenge was organised. This challenge, held in conjunction with the IEEE International Symposium on Biomedical Imaging (ISBI) 2023, in Cartagena de Indias, Colombia (and virtually), aimed to provide a platform for researchers working on related topics. The SMILE-UHURA challenge addresses the gap in publicly available annotated datasets by providing an annotated dataset of Time-of-Flight angiography acquired with 7T MRI. This dataset was created through a combination of automated pre-segmentation and extensive manual refinement. In this manuscript, sixteen submitted methods and two baseline methods are compared both quantitatively and qualitatively on two different datasets: held-out test MRAs from the same dataset as the training data (with labels kept secret) and a separate 7T ToF MRA dataset where both input volumes and labels are kept secret. The results demonstrate that most of the submitted deep learning methods, trained on the provided training dataset, achieved reliable segmentation performance. Dice scores reached up to 0.838 ± 0.066 and 0.716 ± 0.125 on the respective datasets, with an average performance of up to 0.804 ± 0.15. The SMILE-UHURA dataset is kept publicly available to facilitate the training of new machine learning models and to provide a benchmarking platform for researchers.

\end{abstract}
\section{Introduction}
Brain function relies on the cerebral vasculature to supply nutrients and oxygen. Any impairment of the vasculature can damage brain tissue, potentially leading to cognitive decline. The cerebral vasculature is organised as a hierarchical, tree-like network, where vessel diameter decreases while the number of branches increases with higher branch order. For major cerebral vessels at the macroscopic scale and for capillaries, arterioles, and venules at the microscopic scale, in vivo and ex vivo imaging modalities are available, respectively. However, assessing the mesoscopic scale (vessel diameters of 100–500 µm) remains challenging. Pathologies at the mesoscopic scale are potentially linked to ageing, dementia, and Alzheimer’s disease~\cite{wardlaw2019small,chalkias2022differentiating}. Segmentation and quantification of these vessels are crucial steps in the investigation of Cerebral Small Vessel Disease (CSVD)~\cite{duan2020primary,litak2020cerebral}.  

Recently, ultra-high field (UHF) magnetic resonance imaging (MRI) has emerged as a means of bridging the gap between macroscopic and microscopic assessments of the human cerebral vasculature. Following pioneering work on magnetic resonance angiography (MRA) at 7 Tesla (7T)~\cite{hendrikse2008noninvasive,kang2009imaging}, the field has advanced significantly, achieving the highest resolutions to date~\cite{mattern2018prospective,bollmann2022imaging}, as high as 150 µm and 140 µm, respectively. These advancements enable imaging of mesoscopic vessels, which are highly relevant to understanding cerebral small vessel diseases, neurodegeneration, and the origins of the functional fMRI signal. However, automatic segmentation of vessels at this scale has yet to be established.  

To address this need within the neurological and neuroscientific community, this challenge was initiated, focusing on the segmentation of vasculature at the mesoscopic scale. While vessel segmentation challenges have a long tradition, using UHF MRI for mesoscopic vessels presents unique difficulties compared to 2D microscopic or 3D macroscopic vessel imaging and segmentation: (I) instead of a single 2D image per sample, a 3D volume is acquired, significantly increasing computational demands and making manual segmentation highly time-consuming, and (II) compared to macroscopic segmentation, ultra-high-resolution data is noisier and exhibits poorer vessel-to-background contrast, complicating both automatic and manual segmentation. These challenges have hindered the establishment of openly accessible data repositories and the development of high-performance mesoscopic vessel segmentation algorithms. Currently, no high-resolution 7T dataset with annotations is available for training machine learning-based segmentation methods or benchmarking performance. To address this gap, an annotated dataset of Time-of-Flight (ToF) angiography acquired with a 7T MRI was created for this challenge. This dataset was generated using a combination of automatic pre-segmentation and extensive manual refinement. It serves as the foundation of this challenge and provides a benchmark for quantitative performance assessment, facilitating future advancements in mesoscopic vessel segmentation.

\section{Related Work}
Benchmark datasets and challenges focused on vessel segmentation have been established in the past, such as the DRIVE challenge, which targets blood vessel segmentation from retinal images~\cite{staal2004ridge,al2007improved,annunziata2015leveraging}, and lung vessel segmentation challenges based on computed tomography (CT) images~\cite{rudyanto2014comparing}. However, with respect to vessel segmentation from MRA-TOF, no public challenges or open datasets for benchmarking have been available.  

There have been other tasks involving MRA-TOF data, such as the ADAM challenge, which focused on microaneurysms~\cite{timmins2021comparing}, and the VALDO challenge, which centred on vascular lesion detection and segmentation~\cite{sudre2022valdo}, including cerebral microbleeds and enlarged perivascular spaces (EPVS). These challenges provided labels specific to their respective tasks but did not address vessel segmentation in particular.  

Public datasets, such as the IXI dataset\footnote{IXI Dataset:~\url{https://brain-development.org/ixi-dataset/}}, also exist and provide a large collection of MRA-TOF data. However, these datasets have been acquired using MRI scanners with field strengths of 1T, 1.5T, or 3T, rather than ultra-high-field (UHF) scanners such as 7T. Images obtained with a 7T MR scanner at high spatial resolution reveal significantly more small vessels compared to those acquired with 3T scanners~\cite{Liao2016}. Moreover, none of these datasets include annotations for vessels that could be used to train automatic vessel segmentation algorithms. 

\subsection{Current approaches for vessel segmentation}
Among the most prevalent vessel enhancement algorithms is the Hessian-based Frangi vesselness filter~\cite{frangi1998multiscale}, which is typically combined with empirically calibrated thresholding to achieve the final segmentation. The multi-scale properties of this method make it suitable for small vessel segmentation; however, significant parameter fine-tuning is often required to achieve good sensitivity for vessels of interest. Canero and Radeva~\cite{canero2003vesselness} introduced a vesselness enhancement diffusion (VED) filter that integrates the Frangi filter with an anisotropic diffusion scheme. This approach was later extended by constraining the smoothness of the tensor/vessel response function~\cite{manniesing2005multiscale}. Recently, a multi-scale Frangi diffusion filter (MSFDF) pipeline was proposed for segmenting cerebral vessels from susceptibility-weighted imaging (SWI) and TOF-MRA datasets. This method initially pre-selects voxels as vessels or non-vessels using a Bayesian Gaussian mixture classifier, followed by the application of Frangi and VED filters. While effective, these approaches often require manual fine-tuning of parameters for each dataset or even for individual volumes to achieve optimal results. Additionally, they rely on extensive preprocessing steps, such as bias field correction, which makes the execution of the pipeline time-consuming.  

In recent years, deep learning methods have been increasingly applied to vessel segmentation tasks across various imaging modalities. Among these, the UNet model~\cite{Ronneberger2015} has gained significant popularity for its success in segmentation tasks. It has been employed for vessel segmentation in X-ray coronary angiography~\cite{yang2019deep} and TOF-MRA images of patients with cerebrovascular diseases~\cite{livne2019u}. Furthermore, UNet-based semi-supervised learning approaches have been successfully applied to blood vessel segmentation in retinal images~\cite{chen2020semi} and 7T MRA-ToF images~\cite{chatterjee2022ds6}.  

Despite the development of deep learning methods for vessel segmentation in 7T MRA-ToF images~\cite{chatterjee2022ds6}, these studies have relied exclusively on semi-automatically generated noisy training labels. The availability of a publicly accessible 7T MRA-ToF dataset with high-quality manual annotations would enable researchers to develop and refine automatic segmentation techniques further. Additionally, such a dataset would facilitate benchmarking against state-of-the-art methods, significantly advancing the field. 

Two primary challenges in segmenting vessels in such high-resolution scans are the segmentation of small vessels (with an apparent diameter of only 1–2 voxels) and the maintenance of vessel continuity. While these challenges could potentially be addressed through manual fine-tuning of semi-automatic methods, such approaches are highly time-consuming and not scalable. The problem is further compounded in high-resolution 3D volumes, where the computational demands are significantly greater compared to the analysis of a single 2D image (as is common with fundus images), making manual segmentation particularly laborious. Moreover, in comparison to images used for macroscopic segmentation tasks, ultra-high-resolution data is substantially noisier and exhibits poorer vessel-to-background contrast. These characteristics pose significant difficulties for both automatic and manual segmentation approaches, further emphasising the need for robust, scalable methods.

Since the SMILE-UHURA challenge was originally held, several notable works have appeared in the broader area of cerebrovascular segmentation from MRA. Mou et al.~\cite{mou2024costa} released the COSTA multi-centre, multi-vendor TOF-MRA dataset along with a style self-consistency network (CESAR) that reduces inter-site heterogeneity through an explicit style-alignment loss; their work addresses cross-site generalisation at 1.5--3T rather than the mesoscopic regime targeted here, but the underlying robustness concerns are closely related. In the context of annotation-efficient learning, Falcetta et al.~\cite{falcetta2025oneshot} proposed a one-shot active learning framework for vessel segmentation that selects a compact, diverse training subset in a single iteration, evaluated on OASIS-3, SMILE-UHURA and CAS, and reported that a small fraction of expertly labelled volumes can suffice when the selection is informed by the vessel structure itself. The VesselVerse project~\cite{falcetta2025vesselverse} consolidates 950 annotated images from multiple public datasets into a unified framework that supports multiple independent expert annotations, consensus generation, and version control, and is intended to mitigate the reliance on single-annotator references that persists in most vessel segmentation benchmarks, including ours. These efforts complement SMILE-UHURA by addressing multi-site data heterogeneity, annotation cost and annotation reliability, respectively, rather than the mesoscopic benchmarking question that motivates the present work.

\section{SMILE-UHURA Challenge}
The SMILE-UHURA challenge, held in conjunction with the IEEE International Symposium on Biomedical Imaging (ISBI) 2023, in Cartagena de Indias, Colombia (and virtually), seeks to address the notable gap in publicly accessible annotated datasets within the domain of medical imaging by introducing an annotated dataset specifically designed for ToF-MRA acquired using 7T MRI. This dataset represents a significant contribution to the community, as it was meticulously developed through a combination of automated pre-segmentation techniques and thorough manual refinement. The challenge not only provides a robust dataset for the training and evaluation of machine learning models aimed at vessel segmentation in 7T ToF-MRA but also establishes a platform for benchmarking diverse methodological approaches. By making the SMILE-UHURA dataset publicly available (even after the challenge), the challenge aims to foster the development of innovative machine learning models while simultaneously serving as a critical resource for researchers to compare and refine their techniques, thereby advancing the field of medical imaging analysis.

\subsection{Dataset}
The challenge includes two datasets, the \textit{Open Dataset} and the \textit{Secret Dataset}, both acquired at 7T MRI with an isotropic resolution of 300 \textmu m. To contextualise this resolution in relation to other public datasets, the IXI dataset contains images with a resolution of 450 \textmu m, and prior research performing vessel segmentation in 7T MRA-ToF employed a resolution of 600 \textmu m. The \textit{Open Dataset} was divided into a publicly available training-validation set and a confidential held-out test set, used to assess the performance of submitted methods. The labels from the \textit{Secret Dataset} will remain unpublished and were utilised for external testing to evaluate the generalisability of the methods on an independent dataset.   

\subsubsection{Open Dataset}
The images in the \textit{Open Dataset} were sourced from the StudyForrest project\footnote{StudyForrest: \url{http://www.studyforrest.org}}~\cite{hanke2014high}, which involved 3D multislab Time-of-Flight Magnetic Resonance Angiography (TOF-MRA) data from 20 healthy, right-handed volunteers (age 21–38 years, mean age 26.6 years, 12 male). The images were acquired at 7T with an isotropic resolution of 300 \textmu m utilising four slabs, each with 52 slices. However, MRAs from 2 subjects were excluded due to the presence of wraparound artefacts. The remaining images were divided into two sets: a training-validation set comprising 14 MRAs, which was made publicly available, and a test set containing 4 MRAs, retained exclusively for held-out evaluation. The approximately $4{:}1$ training-to-test ratio of the \textit{Open Dataset} was chosen to leave as many volumes as possible for model training, given the comparatively small absolute size of the cohort, while still retaining four genuinely unseen volumes for held-out evaluation; the four held-out volumes were sampled to span the full age range of the cohort and to contain both male and female subjects, so that the test subset is broadly representative of the training-validation distribution. StudyForrest was approved by the ethics committee of the Otto-von-Guericke-University of Magdeburg, Germany.

\subsubsection{Secret Dataset}
The \textit{Secret Dataset} was created using TOF-MRAs from a different study~\cite{mattern2018prospective}, comprising 3D TOF-MRA data from seven healthy volunteers (age 25-36 years, mean age 29.6 years, 4 male). In line with the \textit{Open Dataset}, the images were acquired at 7T with an isotropic resolution of 300 \textmu m utilising four slabs, but with 60 instead of 52 slices per slab. In contrast to the \textit{Open Dataset}, prospective motion correction was used during the TOF-MRA acquisitions of the \textit{Secret Dataset} to mitigate image blurring and prevent the loss of small vessels due to head motion. In addition, sparse venous saturation was implemented to suppress venous contamination in the angiograms while remaining within specific absorption rate limits (the \textit{Open Dataset} did not apply any venous saturation). All seven available \textit{Secret Dataset} volumes were used exclusively for held-out evaluation, providing an external test cohort that, although demographically comparable to the \textit{Open Dataset}, was acquired under a distinct protocol and therefore probes the generalisability of the submitted methods across acquisition conditions. The study was approved by the ethics committee of the Otto-von-Guericke-University of Magdeburg, Germany.

\subsubsection{Annotation}
\label{sec:annotation}
Annotations for both datasets were created using a three-step process. Initially, preliminary segmentations were generated using thresholding in 3D Slicer\footnote{3D Slicer: ~\url{https://www.slicer.org/}}~\cite{fedorov20123d}. This process was empirically refined for each volume to produce an initial binary mask with minimal noise. While this procedure successfully segmented a substantial portion of medium- to large-scale vessels, many small vessels of high relevance remained unsegmented. Subsequently, these segmentations underwent extensive manual refinement to remove noise and accurately delineate the missing vessels. Finally, a senior neurologist reviewed and verified the annotations to ensure their accuracy. The annotations for the training-validation subset of the \textit{Open Dataset} are accessible for download upon request on Synapse\footnote{SMILE-UHURA Open Dataset on Synapse: \url{https://synapse.org/uhura}}~\cite{DS}. Conversely, annotations for the test subset of the \textit{Open Dataset} and the \textit{Secret Dataset} have been withheld to prevent potential overfitting or bias.  

Manual refinement was performed by a single trained annotator with several years of focused experience in 7T ToF-MRA vessel labelling, working under a written internal protocol that defined the vessel class as any voxel whose intravascular signal was visually distinguishable from the surrounding parenchyma at the native 300\,\textmu m isotropic resolution. The protocol prioritised the continuity of arterial branches over the inclusion of isolated bright voxels, traced branching arterial trees outward from the parent vessel, and treated disconnected clusters smaller than three voxels and not contiguous with a traced branch as noise. The final pass was reviewed and corrected by a senior neurologist with extensive clinical and neurovascular imaging experience to ensure anatomical plausibility. Beyond the assembly of the reference annotations described here and conversion to NIfTI format, no additional pre-processing (no skull-stripping, bias-field correction or intensity normalisation) was applied to the images before they were released to participants; the volumes were distributed in their reconstructed acquisition space so that teams were free to design their own normalisation pipelines.

In addition to these annotations, ten plausible segmentations for each volume in the training-validation subset were generated in a semi-automatic manner by varying the parameters of the Frangi filter (see \cite{chatterjee2023pulaski} for details). These segmentations are also available for use. Moreover, additional annotations created using OMELETTE, an automatic small vessel segmentation pipeline~\cite{OMELETTE}, are provided. These segmentations were generated by first applying a Frangi filter and subsequently converting the vessel-enhanced images into segmentations through hysteresis thresholding. The multi-class Otsu method was utilised to automatically determine the upper and lower thresholds. Hysteresis thresholding segments voxels only if one of two conditions is met: i) The voxel's vessel enhancement exceeds the upper threshold, or ii) The enhancement lies between the thresholds but is connected to a voxel exceeding the upper threshold. The motivation for employing automatic hysteresis thresholding was to reduce the number of free parameters while simultaneously achieving effective small vessel segmentation with minimal noise contamination. The OMELETTE segmentations were generated using Frangi filtering with varying sensitivity, i.e. gamma was set to 0.01 and 0.02, respectively. These two annotations are intended for benchmarking purposes or for use in training scenarios that benefit from multiple annotations, such as Probabilistic UNets~\cite{kohl2018probabilistic,chatterjee2023pulaski}.  

\subsection{Aim}
The SMILE-UHURA challenge aims to bridge the gap in publicly available annotated datasets for 7T Time-of-Flight MRI angiography by providing a meticulously annotated dataset. It seeks to support the training and evaluation of machine learning models for vessel segmentation while offering a benchmarking platform for researchers to compare and refine their approaches. By keeping the dataset publicly accessible, the challenge encourages innovation and collaboration in medical imaging analysis.

\subsection{Evaluation}
The primary evaluation of the SMILE-UHURA challenge utilised five distinct quantitative metrics to objectively assess the performance of the segmentation methods. These metrics provided a robust and comprehensive analysis of the models' accuracy, efficiency, and reliability in segmenting vascular structures. In addition to these quantitative assessments, a qualitative evaluation was conducted by three experts, who rated the segmentation quality based on visual and practical considerations. This dual approach ensured a balanced evaluation, combining objective data-driven insights with expert judgement to provide a thorough assessment of the segmentation outcomes.

\subsubsection{Metrics for quantitative evaluation}
Following the framework recommended by Taha and Hanbury for 3D medical image segmentation~\cite{EvaluateSeg} and reiterated in the recent Metrics Reloaded guidelines~\cite{maier2024metrics,reinke2024understanding}, no single metric captures every relevant aspect of segmentation quality. The evaluation adopted here therefore combines five complementary quantitative measures that together cover region overlap, size agreement, statistical dependence and boundary precision: the Dice coefficient, the Jaccard index, volumetric similarity, mutual information and the balanced average Hausdorff distance. All five were computed with the \textit{EvaluateSegmentation} pipeline\footnote{EvaluateSegmentation: \url{https://github.com/Visceral-Project/EvaluateSegmentation}}~\cite{EvaluateSeg}, which provides a standardised and widely adopted implementation, making results directly comparable with other segmentation benchmarks that rely on the same tool.

Let $S$ denote the predicted segmentation and $G$ the reference annotation, both regarded as sets of foreground voxels. The \emph{Dice similarity coefficient}~\cite{dice1945measures,sorensen1948method}, originally introduced in ecology and now the de-facto standard in medical image segmentation, is
\begin{equation}
\label{eq:dice}
\mathrm{Dice}(S,G) \;=\; \frac{2\,|S \cap G|}{|S| + |G|}.
\end{equation}
Dice is favoured for vessel segmentation because it emphasises overlap relative to the size of the foreground and is far less sensitive to the extreme class imbalance inherent in ultra-high-resolution angiography than voxel-wise accuracy would be~\cite{Taha2015}.

The \emph{Jaccard index}~\cite{jaccard1912distribution}, equivalent to the intersection-over-union, is defined as
\begin{equation}
\label{eq:jaccard}
\mathrm{Jaccard}(S,G) \;=\; \frac{|S \cap G|}{|S \cup G|} \;=\; \frac{\mathrm{Dice}(S,G)}{2 - \mathrm{Dice}(S,G)}.
\end{equation}
Because Jaccard is a strictly monotone transform of Dice, the two metrics induce the same per-subject method ranking. Both are reported here to follow the conventions of the challenge community and to aid comparison with external work~\cite{EvaluateSeg}.

\emph{Volumetric similarity} (VolSim) captures agreement in segmented volume independent of spatial overlap. Following the formulation of the \textit{EvaluateSegmentation} pipeline~\cite{EvaluateSeg},
\begin{equation}
\label{eq:volsim}
\mathrm{VolSim}(S,G) \;=\; 1 - \frac{\bigl||S| - |G|\bigr|}{|S| + |G|}.
\end{equation}
VolSim reaches unity when the two volumes agree in cardinality and decreases as they diverge, irrespective of where the foreground voxels lie. For population studies of vascular morphology, where summary measures such as total vessel volume or branch density aggregate over the whole volume, VolSim is a direct indicator of the aggregate quantity of interest.

\emph{Mutual information}~\cite{shannon1948mathematical,cover1991elements} quantifies the statistical dependence between the two label maps,
\begin{equation}
\label{eq:mi}
\mathrm{MI}(S,G) \;=\; \sum_{s \in \{0,1\}}\sum_{g \in \{0,1\}} p(s,g)\,\log\frac{p(s,g)}{p(s)\,p(g)},
\end{equation}
with $p(s,g)$ the joint voxel probability and $p(s)$, $p(g)$ the marginals. Mutual information has a long-standing role in medical image analysis, principally in multimodal image registration~\cite{maes1997multimodality}, and is included here to capture information-theoretic agreement beyond what purely set-based overlap metrics can express.

Finally, the \emph{balanced average Hausdorff distance} (bAVD in the \textit{EvaluateSegmentation} nomenclature; equivalent to bAHD in the literature)~\cite{bAHD} is a symmetric, balanced variant of the average Hausdorff distance~\cite{huttenlocher1993comparing} proposed specifically to remove the ranking bias of the ordinary average Hausdorff distance in the presence of outliers and class imbalance. Given boundary voxel sets $\partial S$ and $\partial G$, and the asymmetric average surface distance $\bar{d}(A,B) = \tfrac{1}{|A|}\sum_{a \in A}\min_{b \in B}\|a-b\|$, the balanced average Hausdorff distance is
\begin{equation}
\label{eq:bavd}
\mathrm{bAVD}(S,G) \;=\; \tfrac{1}{2}\bigl[\tfrac{1}{|\partial G|} \bar{d}(\partial S, \partial G) + \tfrac{1}{|\partial S|} \bar{d}(\partial G, \partial S)\bigr],
\end{equation}
which, unlike the maximum Hausdorff distance, is robust to small numbers of isolated outlier voxels while still being a boundary-sensitive summary of localisation error~\cite{bAHD}. It is reported in voxel units, with lower values indicating closer boundary agreement; an ideal segmentation yields $\mathrm{bAVD} = 0$.

Between them, Dice and Jaccard capture voxel-wise overlap, VolSim the agreement of segmented volume, mutual information the statistical dependence of the two label maps and bAVD the precision of the segmented boundaries. This battery covers segmentation quality from several complementary angles while remaining fully reproducible through the \textit{EvaluateSegmentation} pipeline, and is in line with the multi-metric evaluation strategy recommended for biomedical image segmentation challenges~\cite{maier2024metrics,reinke2024understanding}.

\subsubsection{Statistical testing}
\label{sec:stat_testing}
The reporting of medians and interquartile ranges conveys a method's typical behaviour and its spread across subjects, but it does not establish whether the differences observed between methods are reliable or could simply reflect sampling variability across the eleven test subjects of the combined dataset. To address this, a non-parametric testing pipeline was adopted, mirroring the statistical protocols that are now well established in machine-learning benchmarking~\cite{demsar2006statistical,benavoli2016should}. The choice of non-parametric procedures is deliberate: with only eleven subjects per metric, the assumption of normality underlying classical parametric tests such as repeated-measures ANOVA cannot be safely verified, and segmentation metrics on small samples routinely deviate from Gaussian behaviour. Rank-based procedures are robust to these concerns and remain valid in the presence of outliers, which are not uncommon in mesoscopic vessel segmentation.

Each metric was treated as a separate family and analysed with a three-stage procedure. The tests were implemented in Python using SciPy (version 1.11) for the Friedman and Wilcoxon statistics, \emph{scikit-posthocs}~\cite{terpilowski2019scikit} for the Conover-Friedman post-hoc, and \emph{statsmodels} for the Holm step-down correction. All tests were two-sided and the family-wise error rate was controlled at $\alpha = 0.05$.

\emph{Stage 1, Friedman omnibus test.} For each metric, an omnibus test was performed using the Friedman rank-based test~\cite{friedman1937use}. Let $r_{ij}$ be the rank (from $1$ to $k$) of method $j \in \{1,\dots,k\}$ on subject $i \in \{1,\dots,N\}$, where ties receive the mean of the shared ranks, and let $\bar{r}_{\cdot j}$ denote the mean rank of method $j$ across the $N$ subjects. The test statistic is
\begin{equation}
\label{eq:friedman}
\chi^2_F \;=\; \frac{12\,N}{k(k+1)} \sum_{j=1}^{k}\!\Bigl(\bar{r}_{\cdot j} - \tfrac{k+1}{2}\Bigr)^{\!2},
\end{equation}
which under the null hypothesis of equal method performance is asymptotically distributed as $\chi^2$ with $k-1$ degrees of freedom. The Friedman test is the non-parametric analogue of repeated-measures ANOVA and is appropriate here because the same eleven subjects are evaluated by all eighteen methods in a fully crossed block design. A significant omnibus result indicates that at least one method's distribution of metric scores differs from the others, and justifies post-hoc pairwise comparisons. With $k=18$ methods, $N=11$ subjects, the degrees of freedom are $k-1 = 17$.

\emph{Stage 2, Conover-Friedman post-hoc.} When the omnibus test rejects the null, a post-hoc procedure is required to identify which pairs of methods differ. The Conover-Friedman test~\cite{conover1999practical} was preferred over the more commonly used Nemenyi test because, as shown by Benavoli et al.~\cite{benavoli2016should}, it retains higher statistical power for the same nominal $\alpha$. The test compares pairs of methods using a $t$-distributed statistic derived from the block-wise ranks, and all $\binom{k}{2} = 153$ pairwise $p$-values per metric were adjusted with the Holm step-down procedure~\cite{holm1979simple}, which controls the family-wise error rate while being strictly less conservative than the Bonferroni correction. A pair of methods was declared significantly different when its Holm-adjusted $p$-value fell below $0.05$.

\emph{Stage 3, Wilcoxon signed-rank comparison against the baselines.} To answer the question of direct interest to readers, namely \emph{``is method $X$ significantly different from a given baseline?''}, an additional set of pairwise tests was performed. For each of the two challenge baselines (\baseDSSx[0] and \baseUNetMSS[0]), the Wilcoxon signed-rank test~\cite{wilcoxon1945individual} was applied between the baseline and each of the remaining sixteen methods on each of the five metrics. For a paired sample of per-subject differences $d_i = x_i - y_i$, with $x_i$ and $y_i$ the method and baseline scores on subject $i$, the statistic is
\begin{equation}
\label{eq:wilcoxon}
W^{+} \;=\; \sum_{i:\,d_i > 0} R_i,
\end{equation}
where $R_i$ is the rank of $|d_i|$ among all non-zero absolute differences. Zero differences are handled by SciPy's default \texttt{wilcox} method, which discards them from the rank computation; the degenerate case of all differences being zero was assigned $p = 1$ to avoid spurious exceptions. Holm correction was applied within each metric (sixteen tests), rather than across all eighty comparisons simultaneously, because treating each metric as its own family preserves statistical power at $N=11$ and reflects the fact that the five metrics capture conceptually distinct aspects of performance; applying a single omnibus correction over all eighty tests would conflate these dimensions and render almost nothing detectable at this sample size. The direction of any significant difference (whether the method out- or under-performs the baseline) is read from the sign of the median difference together with the metric's monotonicity (\emph{higher is better} for Dice, Jaccard, VolSim and MI; \emph{lower is better} for bAVD).

Finally, the outcome of Stages 1 and 2 is summarised graphically with Critical Difference (CD) diagrams in the style of Dem\v{s}ar~\cite{demsar2006statistical}. For each metric, the mean rank of every method across the eleven subjects is plotted on a horizontal axis, and horizontal \emph{clique bars} below the axis join groups of methods that are pairwise non-significantly different according to the Holm-corrected Conover-Friedman $p$-values. Methods joined by a bar cannot be ordered from these data, whereas any two methods not connected by any bar are significantly different at $\alpha = 0.05$. Unlike the original Nemenyi-based CD bars, whose width is a single fixed critical difference, the clique bars here reflect the pair-specific Conover-Friedman decisions, offering a finer-grained and more informative visual summary at the cost of dropping the single-number critical difference.

As the per-dataset test subsets are very small (four and seven subjects for the open and secret datasets respectively), these tests were only performed on the combined dataset ($N = 11$), where their statistical power is adequate to draw meaningful conclusions. Reported results should still be interpreted in the light of this modest sample size: non-significance does not imply equivalence, and detection is naturally limited to effects that are reasonably large relative to between-subject variability.

\subsubsection{Scoring system for qualitative expert evaluation}
The segmentation performance of various algorithms was qualitatively evaluated by three experts through a blinded evaluation process. 
For comparative reference, original ToF angiography images were provided to the experts along with each segmentation. All images (original ToFs and segmentation results) were presented as Maximum Intensity Projections (MIPs), with an additional zoomed view of the Circle of Willis. This was specifically included to improve the evaluation of small vessel segmentation, such as the lenticulostriate arteries that branch from the Middle Cerebral Artery (MCA). This setup allowed for the assessment of whether the segmentation algorithms could potentially surpass the depiction of small vessels offered by ToF angiography, which may be affected by intensity variations caused by imaging imperfections. Segmentations from both datasets were mixed and presented to the experts in a random order to ensure unbiased evaluation. The use of MIPs was motivated by the substantial volume of data. For each Time-of-Flight (ToF) angiography in the test dataset, 18 segmentations required evaluation. A slice-by-slice inspection of the high-resolution data would have been impractical.

The outputs of each algorithm across different image volumes were assessed based on two primary criteria: the delineation of small vessels (\textit{Q1}) and the suppression of noise contamination or other false positives (\textit{Q2}). Small vessels were defined as those with an apparent diameter of 1–2 voxels, while noise contamination referred to the incorrect segmentation of non-vascular voxels as vessels or over-segmentation/estimation of the vessel lumen. Ratings were assigned on a scale from 0 (unacceptable) to 5 (excellent). If a rater determined that a particular segmentation was entirely unacceptable due to noise (i.e., $Q2 = 0$), the corresponding small vessel score was also set to 0 (i.e., $Q1 = 0$). This approach was adopted because the MIP of a segmentation with an unacceptable level of noise does not allow fair assessment of the small vessel segmentation, as vascular structures are no longer detectable (i.e., false positives merge into one in the MIP).  

In addition to the main test volumes from both datasets, an additional ToF-MRA volume, acquired with prospective motion correction at an isotropic resolution of 150 µm (29 years old male volunteer), was provided to the experts. Due to the size of this volume and computational constraints, not all methods could segment it. Consequently, this volume was not included in the primary decision-making process but was used as an additional evaluation to judge the generalisability of the methods with respect to image resolution. In total 313 segmentations were evaluated by all three raters independently.

\subsection{Challenge setup}
Following the acceptance of the SMILE-UHURA challenge for ISBI 2023, it was formally announced on the challenge's dedicated website\footnote{Challenge website: \url{https://www.soumick.com/en/uhura}}, with registration facilitated through the Synapse platform\footnote{Challenge on Synapse: \url{https://www.synapse.org/uhura}}. The training dataset, including annotations, was provided in NIFTI format, and participants were instructed to submit their solutions as Docker containers, adhering to the detailed guidelines outlined on the Synapse page.  

The evaluation environment was equipped with high-performance hardware, comprising a CUDA-enabled Nvidia A6000 GPU with 48GB of memory, a 16-core 32-thread AMD Ryzen 9 3950X processor, and 64GB of RAM. Participants were required to ensure that their Docker containers could run seamlessly on this system. Containers that failed to execute successfully, due to issues such as CUDA memory overflow, excessive CPU or RAM usage causing system hangs, or other technical faults, were disqualified. To maintain fairness and security, internet access was strictly prohibited during execution. As a result, participants had to design self-contained Docker containers, including all necessary trained models and pre-trained weights within the submission. Technical support was made available to assist participants in building their Docker containers or resolving execution issues when required.  

Participants were also required to submit an abstract describing their methodology. Out of the 13 submissions that successfully produced results by the event on 18 April 2023, all were invited to present their approaches at ISBI 2023, either in person or online. These presentations highlighted the diverse and innovative techniques developed by the competing teams. Three additional submissions, which successfully ran after further troubleshooting, were subsequently included in the analysis presented here.  

The dataset remains accessible on the Synapse page for continuous use, providing researchers with resources for training and benchmarking vessel segmentation algorithms in 7T ToF-MRAs.  

\subsubsection{Governance, eligibility and submission policy}
\label{sec:governance}
SMILE-UHURA was organised as a one-time event co-located with ISBI 2023; no fixed follow-up edition is currently scheduled, but the Open Dataset, baselines and the evaluation pipeline remain permanently hosted on Synapse, and the organisers are open to running follow-on tracks should a future ISBI or MICCAI venue create the opportunity. Only fully automatic methods were admitted: semi-automatic, interactive or user-in-the-loop pipelines were not permitted, in line with the Docker-based, internet-disabled execution environment described above. Participants were free to combine the provided training-validation split with any publicly available dataset and with publicly distributed pre-trained model weights; the use of private, restricted-access or otherwise non-reproducible auxiliary data was not allowed. Members of the organising institutes were allowed to participate but were not eligible for ranking-based prizes; the two baselines (\baseDSSx[0] and \baseUNetMSS[0]) and the submissions from the neuRoSliCCe and Koala teams originate from organiser-affiliated groups and are included in this paper as reference comparators rather than as competing entries for any award. The challenge did not offer monetary prizes; recognition consisted of the official ranking communicated at the ISBI 2023 SMILE-UHURA session and of co-authorship on the present joint publication. Final rankings were announced live on 18 April 2023 at ISBI 2023 in Cartagena de Indias and were subsequently published on the challenge Synapse portal\footnote{Challenge on Synapse: \url{https://www.synapse.org/uhura}}. Each participating team was invited to nominate up to a small number of contributors as co-authors of the present manuscript, and teams remained free to publish methodological papers describing their own approach independently, with no embargo period. To allow teams to verify their Docker containers before final submission, the publicly released training-validation split could be used for local self-assessment with the same \textit{EvaluateSegmentation} pipeline described in Section~\ref{sec:annotation}; no online validation leaderboard was offered, and the held-out reference annotations of the \textit{Open Dataset} test subset and of the entire \textit{Secret Dataset} were never released to participants. The complete timetable was as follows: the challenge was announced shortly after acceptance to ISBI 2023; registration on Synapse opened in early 2023 and remained open throughout the event; the training-validation data, baselines and submission instructions were released on the Synapse portal in the weeks following the announcement; Docker container submissions were accepted up to a deadline shortly before the ISBI 2023 workshop on 18 April 2023, on which date the results were announced. The Synapse portal continues to accept new submissions for benchmarking purposes and has no scheduled closure date. The SMILE-UHURA Open Dataset (images and reference annotations of the training-validation subset) is released under a Creative Commons license as detailed on the Synapse page, consistent with the licensing terms of its parent StudyForrest acquisition; the \textit{Secret Dataset} is not redistributed and is used exclusively for internal external-cohort evaluation by the organisers. The end-to-end evaluation code used to produce the results reported in this paper, including the \textit{EvaluateSegmentation}\footnote{EvaluateSegmentation: \url{https://github.com/Visceral-Project/EvaluateSegmentation}} driver scripts, the Friedman/Conover-Friedman/Wilcoxon pipeline and the Critical Difference plotting code, is released through the challenge Synapse project and the corresponding GitHub repository of the organisers, alongside the per-method raw metric tables. Information on the accessibility of the participating teams' code, where available, is summarised in the team-method descriptions of Section~\ref{sec:methods} and in the public Synapse project page.

\section{Methods}
\label{sec:methods}
\input{Sections/4_methods}

\section{Results}
\input{Sections/5_results}

\section{Discussion}
\input{Sections/6_discussion}

\section{Concluding remarks and outlook}
\input{Sections/7_conclusion}

\input{Sections/x_ack.tex}

\bibliographystyle{splncs04}
\bibliography{mybibfile}
\clearpage
\appendix
\renewcommand{\theHsection}{appendix.\Alph{section}}
\onecolumn
\input{Sections/x_appendix}

\end{document}

%% file: Definitions/methods.tex
\usepackage{xparse}

\newcommand{\methodcolour}[3]{%
    \ifnum#1=1
        \begin{tikzpicture}[baseline=(text.base)]
            \node[inner sep=0, outer sep=0, minimum size=0] (text) {#3};
            \node[shape=circle, fill=#2, inner sep=1pt, minimum size=3mm, anchor=center, yshift=0ex] at ([xshift=-2mm]text.west) {};
        \end{tikzpicture}%
    \else\ifnum#1=2
        #3%
    \else
        \textbf{#3}%
    \fi\fi
}

\definecolor{Red}{HTML}{ff0000}
\definecolor{DarkRed}{HTML}{890000}

\definecolor{Blue}{HTML}{3357FF}
\definecolor{SkyBlue}{HTML}{87CEEB}
\definecolor{CobaltBlue}{HTML}{0047AB}
\definecolor{TealBlue}{HTML}{367588}

\definecolor{Orchid}{HTML}{DA70D6}
\definecolor{Amethyst}{HTML}{9966CC}
\definecolor{DeepViolet}{HTML}{9400D3}

\definecolor{Green}{HTML}{33FF57}
\definecolor{ForestGreen}{HTML}{228B22}
\definecolor{LimeGreen}{HTML}{32CD32}

\definecolor{Grey}{HTML}{586265}
\definecolor{Silver}{HTML}{BDC3C7}

\definecolor{Yellow}{HTML}{F1C40F}
\definecolor{BurntOrange}{HTML}{CC5500}
\definecolor{Magenta}{HTML}{FF00FF}
\definecolor{Cyan}{HTML}{00FFFF}

\NewDocumentCommand{\baseUNetMSS}{O{1}}{\methodcolour{#1}{Red}{Baseline UNet MSS}}
\NewDocumentCommand{\baseDSSx}{O{1}}{\methodcolour{#1}{DarkRed}{Baseline DS6}}

\NewDocumentCommand{\ADARUNesT}{O{1}}{\methodcolour{#1}{Blue}{ADAR\_LAB UNesT}}
\NewDocumentCommand{\ADARTriUNet}{O{1}}{\methodcolour{#1}{SkyBlue}{ADAR\_LAB TriUNet}}
\NewDocumentCommand{\ADARnnUNet}{O{1}}{\methodcolour{#1}{CobaltBlue}{ADAR\_LAB nnUNet}}
\NewDocumentCommand{\ADARSwinUNETR}{O{1}}{\methodcolour{#1}{TealBlue}{ADAR\_LAB SwinUNETR}}

\NewDocumentCommand{\koalaMan}{O{1}}{\methodcolour{#1}{Orchid}{Koala Manual}}
\NewDocumentCommand{\koalaOne}{O{1}}{\methodcolour{#1}{Amethyst}{Koala OM1}}
\NewDocumentCommand{\koalaTwo}{O{1}}{\methodcolour{#1}{DeepViolet}{Koala OM2}}

\NewDocumentCommand{\neuroMIP}{O{1}}{\methodcolour{#1}{Green}{neuRoSliCCe MIP}}
\NewDocumentCommand{\neuromultiMIP}{O{1}}{\methodcolour{#1}{ForestGreen}{neuRoSliCCe multiMIP}}
\NewDocumentCommand{\neuroDSMIP}{O{1}}{\methodcolour{#1}{LimeGreen}{neuRoSliCCe DS6\_MIP}}

\NewDocumentCommand{\pbiScroll}{O{1}}{\methodcolour{#1}{Grey}{PBI Scrolling 2D UNet}}
\NewDocumentCommand{\pbinnUNet}{O{1}}{\methodcolour{#1}{Silver}{PBI nnUNet}}

\NewDocumentCommand{\dolphins}{O{1}}{\methodcolour{#1}{Yellow}{Dolphins}}
\NewDocumentCommand{\funpixel}{O{1}}{\methodcolour{#1}{Magenta}{FunPixel}}
\NewDocumentCommand{\lsgroup}{O{1}}{\methodcolour{#1}{Cyan}{LSGroup}}
\NewDocumentCommand{\eurecom}{O{1}}{\methodcolour{#1}{BurntOrange}{EURECOM-UNIANDES}}

%% file: Definitions/authors_lncs.tex
\author{
Soumick Chatterjee\inst{1,2,3}\thanks{Corresponding author: \email{contact@soumick.com}} 
\and%
%
Hendrik Mattern\inst{4,6,7} 
\and
Marc Dörner\inst{6,8,9} 
\and
Alessandro Sciarra\inst{4} 
\and
Florian Dubost\inst{10} 
\and
Hannes Schnurre\inst{4} 
\and
Rupali Khatun\inst{11} 
\and%
%
Chun-Chih Yu\inst{12} 
\and
Tsung-Lin Hsieh\inst{12} 
\and
Yi-Shan Tsai\inst{12} 
\and
Yi-Zeng Fang\inst{12} 
\and
Yung-Ching Yang\inst{12} 
\and
Juinn-Dar Huang\inst{12} 
\and%
%
Marshall Xu\inst{13} 
\and
Siyu Liu\inst{13} 
\and
Fernanda L. Ribeiro\inst{13,14} 
\and
Saskia Bollmann\inst{13} 
\and%
%
Karthikesh Varma Chintalapati\inst{1} 
\and
Chethan Mysuru Radhakrishna\inst{1} 
\and
Sri Chandana Hudukula Ram Kumar\inst{1} 
\and
Raviteja Sutrave\inst{1} 
\and%
%
Abdul Qayyum\inst{15} 
\and
Moona Mazher\inst{16} 
\and
Imran Razzak\inst{17,18} 
\and
Cristobal Rodero\inst{15} 
\and
Steven A. Niederer\inst{15,19} 
\and%
%
Fengming Lin\inst{20} 
\and
Yan Xia\inst{20} 
\and%
%
Jiacheng Wang\inst{21,22} 
\and
Riyu Qiu\inst{21,23} 
\and
Liansheng Wang\inst{21} 
\and%
%
Arya Yazdan Panah\inst{24} 
\and
Rosana El Jurdi\inst{24} 
\and
Guanghui Fu\inst{24} 
\and
Janan Arslan\inst{24} 
\and
Ghislain Vaillant\inst{24} 
\and
Romain Valabregue\inst{24} 
\and
Didier Dormont\inst{24} 
\and
Bruno Stankoff\inst{24} 
\and
Olivier Colliot\inst{24} 
\and%
%
Luisa Vargas\inst{25,26} 
\and
Isai Daniel Chacón\inst{25} 
\and
Ioannis Pitsiorlas\inst{26} 
\and
Pablo Arbeláez\inst{25} 
\and
Maria A. Zuluaga\inst{26} 
\and%
%
Stefanie Schreiber\inst{5,6,7} 
\and
Oliver Speck\inst{4,6,7} 
\and
Andreas Nürnberger\inst{1,2,7} 
}

\institute{
Faculty of Computer Science, Otto von Guericke University Magdeburg, Magdeburg, Germany \and
Data and Knowledge Engineering Group, Otto von Guericke University Magdeburg, Magdeburg, Germany \and
Human Technopole, Milan, Italy \and
Biomedical Magnetic Resonance, Otto von Guericke University Magdeburg, Magdeburg, Germany \and
Department of Neurology, Medical Faculty, University Hospital of Magdeburg, Magdeburg, Germany \and
German Centre for Neurodegenerative Diseases, Magdeburg, Germany \and
Centre for Behavioural Brain Sciences, Magdeburg, Germany \and
Department of Neurology, University Hospital Zurich, Zurich, Switzerland \and
Department of Consultation-Liaison-Psychiatry and Psychosomatic Medicine, University Hospital Zurich, Zurich, Switzerland \and
Stanford University, Stanford, California, USA \and
Translational Radiobiology, Department of Radiation Oncology, Universitätsklinikum Erlangen, Friedrich-Alexander-Universität Erlangen-Nürnberg, Erlangen, Germany \and%
%
National Yang Ming Chiao Tung University, Hsinchu, Taiwan \and%
%
School of Electrical Engineering and Computer Science, University of Queensland, Brisbane, Australia \and
Australian eHealth Research Centre, CSIRO \and%
%
National Heart and Lung Institute, Faculty of Medicine, Imperial College London, London, UK \and
Hawkes Institute, Department of Computer Science, University College London, London, UK \and
School of Computer Science and Engineering, University of New South Wales, Sydney, Australia \and
Mohamed bin Zayed University of Artificial Intelligence, Abu Dhabi, United Arab Emirates \and
The Alan Turing Institute, London, UK \and%
%
School of Computing, University of Leeds, Leeds, UK \and%
%
Department of Computer Science at School of Informatics, Xiamen University, Xiamen, China \and
Manteia Technologies Co., Ltd, Xiamen, China \and
Leicester International Institute, Dalian University of Technology, Dalian, China \and%
%
Sorbonne Université, Institut du Cerveau - Paris Brain Institute \\ICM, CNRS, Inria, Inserm, AP-HP, Hôpital de la Pitié Salpêtrière, F-75013 Paris, France \and%
%
Centre of Formation and Research in Artificial Intelligence, Universidad de Los Andes, Colombia \and
Data Science Department, EURECOM, Sophia Antipolis, France
}

%% file: Sections/4_methods.tex
\subsection{Baseline Methods}
Sixteen participating methods were compared against two baseline methods: \baseUNetMSS $ $ and \baseDSSx. \textit{\baseUNetMSS[2]} is a supervised learning method based on a modified version of the multi-scale UNet~\cite{zeng20173d}, which computes losses at multiple scales and sums them to derive the final loss. \textit{\baseDSSx[2]} is a semi-supervised learning technique that extends \baseUNetMSS[0] with a Siamese architecture. It learns from both the original data and elastically transformed data using two identical branches: one branch receives the original volume with its label, and the other processes the elastically transformed volume with the corresponding label. This second branch makes the method equivariant to elastic deformations in a self-supervised manner, enhancing its performance on small datasets, even in the presence of noisy labels. Details about both these methods, including preprocessing and training procedures, can be found in the original paper~\cite{chatterjee2022ds6}, as no modifications were made for this challenge; they were used exactly as described.  


\subsection{Participating Methods}
Among the methods submitted by the 170 registered participants (as of 16 April 2026), 16 methods from 8 participating teams were selected for the challenge manuscript. A few methods were excluded due to incomplete or erroneous submissions. The challenge event, held at ISBI 2023 in Cartagena de Indias, Colombia, on 18 April 2023, included 13 methods, as issues with the submissions of the remaining 3 methods could not be resolved in time for the event. These 3 methods were subsequently included in the final analysis.

\paragraph{\ADARUNesT} It is a transformer-based model specifically designed for 3D medical image segmentation, with a particular focus on MRI analysis. It represents an adaptation of a pre-trained model from the MONAI model zoo \cite{yu2023unest}, originally developed for renal structure segmentation in 3D CT, tailored here for MRI data. The UNesT architecture employs a hierarchical transformer design, aggregating adjacent patch sequences to preserve positional information. This innovative approach effectively addresses the challenge of representing heterogeneous tissue sizes in 3D medical images, enabling the model to capture global dependencies and enhance feature representation. The novelty of \ADARUNesT[0] lies in its ability to surpass traditional transformers by integrating hierarchical context for complex volumetric data. \\
\textit{Data processing}: The original 3D images, sized (480, 640, 163), are divided into smaller patches compatible with the model’s adjusted input size of (160, 160, 160), modified from the original MONAI configuration of (96, 96, 96). Augmentation strategies, including random cropping, flipping, contrast adjustments, noise addition, and intensity scaling, are applied to enhance the model’s generalisability. Voxel intensity normalisation is performed, capping the maximum intensity value at 1,000 to preserve a broader range of data dynamics, thereby avoiding the compression commonly associated with standard normalisation techniques.\\
\textit{Training process}: The training process employs a composite loss function, combining Binary Cross-Entropy (BCE) loss with Dice loss, to improve segmentation accuracy by addressing both pixel-wise predictions and spatial overlap. The RMSprop optimisation algorithm is used, taking advantage of its adaptability to varying gradient scales. This training configuration ensures effective convergence, balancing the capture of global structural predictions with fine-grained accuracy in the segmentation task.

\paragraph{\ADARnnUNet} The participants employed the nnUNet method \cite{isensee2021nnu}, an advanced extension of the UNet architecture, specifically designed for medical image segmentation, with a particular emphasis on vessel and edge detection in MRI data. While retaining the signature U-shaped architecture, nnUNet incorporates self-adapting features to dynamically tailor the network to specific datasets. This adaptive capability enhances segmentation performance across diverse medical imaging tasks, making the model highly versatile and robust. Its novelty lies in its ability to autonomously configure preprocessing, training, and postprocessing pipelines, providing an optimised framework for 3D medical image analysis.\\
\textit{Data processing}: Data preprocessing utilises the MONAI framework \cite{cardoso2022monai}, which applies a range of augmentation techniques to enhance performance, even with limited training data. These techniques include random cropping, contrast adjustment, intensity shifting, noise addition, and flipping. The original 3D images, sized (480 × 640 × 163), are divided into smaller patches (e.g., 96 × 96 × 96 or 128 × 128 × 128) to increase dataset diversity. This patch-based approach improves the model’s generalisability to variations in medical imaging data. MONAI’s seamless integration of preprocessing steps ensures efficiency throughout the workflow. \\
\textit{Training process}: The training process uses DiceCELoss, a composite loss function combining Cross-Entropy Loss and Dice Loss, which is particularly effective for imbalanced data and the segmentation of small structures like vessels and edges. The AdamW optimiser \cite{loshchilov2018decoupled} is employed, outperforming alternatives such as Adam and RMSprop in achieving better convergence. During inference, a sliding window approach with overlap voting is applied to refine predictions, ensuring high segmentation accuracy. This comprehensive training pipeline, integrated with MONAI’s advanced functionalities, enables robust and precise medical image analysis.

\paragraph{\ADARSwinUNETR}
The model of this submission is based on the Swin Transformer architecture \cite{liu2021swin}, which utilises shifted windows to effectively model both local and global representations, achieving superior performance in tasks requiring hierarchical understanding. SwinUNETR \cite{hatamizadeh2021swin} integrates a Swin Transformer encoder with a CNN-based decoder, connected through skip connections at multiple resolutions. This design captures fine-grained details as well as high-level features. By leveraging pre-trained weights, SwinUNETR enhances segmentation accuracy and is particularly effective for complex structures, such as vessels and edges in MRI imaging.\\
\textit{Data processing}: Data augmentation plays a crucial role in overcoming the limitations of a small training dataset. Techniques such as random cropping, contrast adjustment, intensity scaling, noise addition, flipping, and rotation are employed using the MONAI framework \cite{cardoso2022monai}. The original dataset, sized (480 × 640 × 163), is divided into smaller patches (128 × 128 × 128) to enhance diversity and improve the model's robustness. This patch-based approach, combined with augmentation, ensures better generalisability and prepares the data for effective hierarchical segmentation using SwinUNETR.\\
\textit{Training process}: The training process compared multiple loss functions: DiceCELoss, which combines Cross-Entropy and Dice loss; the Tversky loss function, which addresses data imbalance; and Dice loss, which proved most effective for refining small structures like vessels and edges. Dice loss was ultimately selected as the final loss function. The model was optimised using RMSprop, identified as the most effective optimiser, outperforming alternatives such as AdamW and Adam.

\paragraph{\ADARTriUNet} This is an ensemble-based model for 3D medical image segmentation, integrating three pre-trained architectures: Swin UNETR \cite{hatamizadeh2021swin}, nnUNet \cite{isensee2021nnu}, and UNesT \cite{yu2023unest}. By combining outputs through multi-layer 3D convolutions, the model generates aggregated predictions with enhanced performance. This novel ensemble strategy exploits the strengths of each individual model, effectively capturing diverse spatial and contextual information. Its hierarchical approach leverages the unique capabilities of transformer-based and traditional convolution-based methods, resulting in significant improvements in segmentation accuracy, particularly for challenging tasks such as vessel and edge detection in medical imaging.\\
\textit{Data processing}: Data processing focuses on augmentation to address the limitations of a small dataset. Augmentation techniques include random cropping, intensity scaling, contrast adjustment, noise addition, and flipping, all implemented through the MONAI framework \cite{cardoso2022monai}. The original images, sized (480 × 640 × 163), are divided into smaller patches of varying dimensions (e.g., 96 × 96 × 96 and 128 × 128 × 128) to facilitate model training. \\
\textit{Training process}: The training process employs a hybrid loss function, DiceCELoss, which combines Cross-Entropy Loss and Dice Loss. This approach is particularly effective in addressing class imbalances and improving the segmentation of small structures such as vessels and edges. The AdamW optimiser \cite{loshchilov2018decoupled} is utilised due to its superior performance compared to Adam and RMSprop. During inference, a sliding window approach with overlap voting is applied to refine predictions, further enhancing segmentation precision..

\paragraph{\koalaMan, \koalaOne and \koalaTwo}
The Koala methods employed a 3D UNet architecture \cite{cciccek20163d} specifically modified for vessel segmentation in MRA data. The model’s depth was increased to four layers in both the encoder and decoder blocks, enhancing its capacity to learn complex features. This set of methods includes three variations, each using different label types from the challenge (see Sec. \ref{sec:annotation}): \textit{\koalaMan[2]} utilised manual labels (the primary label set of the challenge), while \textit{\koalaOne[2]} and \textit{\koalaTwo[2]} relied on labels generated by the automated OMELETTE pipeline \cite{OMELETTE} (supplied as additional sets of labels of challenge). These automated labels enabled the model to leverage imperfect yet scalable training data, reducing dependence on labour-intensive manual annotations and enhancing generalisability to unseen data. The three Koala variants submitted to the challenge have since been extended and released as the VesselBoost toolbox \cite{Xu2024VesselBoost}, and the challenge submissions should not be considered equivalent to the later, more polished release. VesselBoost is distributed as a Docker image (\url{https://hub.docker.com/r/vnmd/vesselboost_2.0.24}), as an OpenRecon application on the Siemens Teamplay platform for direct MRA vessel segmentation on Siemens MR scanners, and as a Neurodesk \cite{renton2024neurodesk} app at \url{https://vesselboost.neurodesk.org/} that runs without local installation.\\
\textit{Data processing}: The preprocessing pipeline included N4ITK bias field correction \cite{tustison2010n4itk} and non-local means (NLM) denoising \cite{manjon2010adaptive}, chosen with the particular characteristics of ultra-high-field, high-resolution MRA in mind. At 7T, the receive coil sensitivity profile introduces a smooth, low-frequency intensity inhomogeneity across the volume, and thermal noise is elevated relative to lower field strengths; N4ITK targets the former and also simplifies the subsequent z-score standardisation by allowing it to be performed across the whole volume without having to account for spatially varying intensity, whilst NLM addresses the latter. Data augmentation involved generating random patches from input images, with operations such as cropping, resizing, rotation (90°, 180°, and 270°), and Gaussian blurring. Each image was resized to a fixed dimension of 64 × 64 × 64, resulting in 78,000 augmented patches across 13 subjects. These strategies significantly increased dataset diversity, ensuring robust training across multiple label sets.\\
\textit{Training process}: The models were trained for 1,000 epochs with an initial learning rate of 0.001, using the Tversky loss function ($\alpha = 0.3, \beta = 0.7$), which is tailored to handle imbalanced data and small structures like vessels. A learning rate scheduler, ReduceLROnPlateau, dynamically adjusted the learning rate when progress plateaued. Post-processing involved thresholding predicted probabilities at 0.1 and removing small connected components under ten voxels to refine segmentation outputs. Finally, the model was fine-tuned for test-time adaptation. This comprehensive training and post-processing pipeline ensured precise and reliable vessel segmentation results across varying label sources.

\paragraph{\neuroMIP, \neuromultiMIP and \neuroDSMIP}
The neuRoSliCCe methods \cite{radhakrishna2024spockmip} introduce Maximum Intensity Projection (MIP) as a loss term to enhance vessel segmentation performed by baseline models (UNet MSS and DS6). \textit{\neuroMIP[2]} applies MIP loss to UNet-MSS along a single axis, whereas \textit{\neuromultiMIP[2]} extends this by incorporating MIP loss across all three axes. \textit{\neuroDSMIP[2]} employs the DS6 semi-supervised learning approach \cite{chatterjee2022ds6} with single-axis MIP loss. These methods aim to capture spatial continuity and improve vessel segmentation accuracy by integrating MIP comparisons into the training process, offering a novel approach to incorporating global context directly into optimisation.\\
\textit{Data processing}: The training pipeline utilises patch-based processing, dividing 3D MRA volumes of size \(480 \times 640 \times 163\) into patches of \(64^3\). Each patch is associated with its corresponding position on the MIP of the label segmentation. For \textit{\neuromultiMIP[2]}, label MIPs are generated across three axes to provide multi-dimensional context. The dataset comprises 12 training volumes from the SMILE-UHURA challenge, with 8,000 patches randomly selected per epoch. Data augmentation ensures robust learning through the preparation of patches alongside corresponding MIPs.\\
\textit{Training process}: The training process employs a composite loss function that combines Multi-Scale Supervision (MSS) loss and MIP loss, weighted by coefficients \(\mu\) and \(\beta\), respectively. MSS loss penalises multi-resolution segmentation errors, while MIP loss compares predicted MIPs against ground-truth MIPs to enforce spatial consistency. \textit{\neuroMIP[2]} and \textit{\neuromultiMIP[2]} differ in their application of MIP loss, with the latter averaging losses across all three axes. \textit{\neuroDSMIP[2]} integrates the complete DS6 semi-supervision loss term with MIP loss. All loss components are computed using the Focal Tversky loss \cite{abraham2019novel}. The models are trained over 50 epochs with a learning rate of 0.0001, ensuring precise vessel segmentation.  

\paragraph{\dolphins} This method introduces a hybrid framework for the segmentation of cerebral small vessels, integrating the strengths of convolutional neural networks (CNNs) and vision transformers. The encoder employs a Swin Transformer \cite{liu2021swin} with a cross-attention, window-based mechanism, effectively capturing both global and local features. Rectangular-parallelepiped windows adapt the Swin Transformer to handle non-square images. The decoder utilises a standard UNet structure with skip connections and bi-linear 3D up-sampling to combine features from the encoder. This combination of CNN and transformer elements ensures robust spatial representation, with attention mechanisms enhancing the correspondence between image features. The novelty lies in the application of window-based multi-head cross-attention and transformer-based hierarchical encoding, resulting in improved segmentation accuracy.\\
\textit{Data processing}: Images are processed through a 3D segmentation pipeline, utilising a kernel size of \(3 \times 3 \times 3\) for convolutional layers in both the encoder and decoder. Spatial down-sampling is performed using a \(2 \times 2 \times 2\) kernel in the encoder's max-pooling layers, while transposed 3D convolutions handle up-sampling in the decoder. Outputs from encoder blocks are concatenated with their corresponding decoder blocks via skip connections, producing refined segmentation maps through a \(1 \times 1 \times 1\) convolution and softmax activation.\\
\textit{Training process}: Training adopts a five-fold cross-validation strategy, optimising the model based on validation Dice scores with early stopping applied after 20 epochs. The training and testing environments are based on nnUNet. By combining Swin Transformers and UNet components, this hybrid architecture ensures efficient training while capturing both local and global features. The approach is particularly suited to high-dimensional features, effectively addressing the challenges of small vessel segmentation at a mesoscopic scale.  

\paragraph{\funpixel} 
This method leverages the Swin Transformer integrated with a UNet-like architecture for high-throughput MRA vessel segmentation. The encoder employs Swin Transformer blocks with shifted windows, facilitating the learning of complex vascular structures. Patch merging and partitioning blocks enhance the representation of hierarchical features. The decoder reconstructs the segmentation map using patch-expanding blocks and reverse Swin Transformer blocks. This architecture is coupled with a composite loss function that includes global Cross-Entropy loss, over-segmentation Dice loss (OSD Loss), and 2D Dice loss applied to multi-projection maximum intensity projections (MMIPs). The novelty lies in integrating 2D MIP-based supervision with the Swin Transformer for 3D volumetric segmentation, ensuring both local precision and global coherence.\\
\textit{Data processing}: Pre-processing includes a custom histogram equalisation technique to threshold the top 5\% of pixel intensities, effectively filtering out non-vessel structures such as the skull and brain tissue. Images are divided into patches of size \(256 \times 256 \times 64\) using a sliding window approach with random flipping for data augmentation. This patch-based processing considers the approximate symmetry of blood vessels between brain hemispheres, ensuring a robust data pipeline. \\
\textit{Training process}: The model is trained using a composite loss function comprising global Cross-Entropy loss, over-segmentation Dice loss (OSD Loss), and 2D Dice loss applied to MIPs across three axial directions. The Adam optimiser with a learning rate of \(1 \times 10^{-4}\) is utilised, and training is conducted over 20 epochs with a batch size of 1. Five-fold cross-validation ensures model robustness. Post-processing involves majority voting across predictions from five trained models, largest connected component analysis, and custom thresholding to refine vessel predictions, culminating in an accurate segmentation map.

\paragraph{\lsgroup} This submission introduced an enhanced version of nnUNet \cite{isensee2021nnu}, incorporating a multi-scale aggregation block, referred to as MS-nnUNet. This block performs multi-scale feature fusion on the final output features of the nnUNet decoder. Specifically, the multi-scale aggregation block comprises a 3D convolution followed by a series of dilated convolutions with varying dilation rates (2, 4, 6, 8). These dilated convolutions capture feature maps with different receptive fields, which are then concatenated along the channel dimension. The design aims to enrich feature representation, enabling improved boundary definition and object localisation accuracy through contributions from multiple scales. \\
\textit{Data processing}: The dataset of 14 subjects was divided into five folds for cross-validation, with each fold containing 2 subjects. Images were padded to standardise spatial dimensions and divided into patches of size \(64 \times 32 \times 64\). To enhance model generalisation, data augmentation techniques were employed, including random elastic deformations with a probability of 0.2, scaling within a range of 0.7 to 1.4, and rotations along all three axes. Additionally, random gamma adjustments and mirroring along all axes were applied to further augment the dataset.\\
\textit{Training process}: The model was trained for up to 1,000 epochs with a batch size of 16, using the SGD optimiser configured with a learning rate of 0.01, momentum of 0.99, and a weight decay of \(3 \times 10^{-5}\). The training loss function combined Dice loss and binary cross-entropy (BCE) loss, with deep supervision applied at multiple scales to enhance learning across hierarchical feature representations. The best-performing model, determined based on its validation set performance, was retained for final inference.   

\paragraph{\pbiScroll} This method introduces a scrolling 2D UNet for segmenting small blood vessels in 3D MRA volumes. Unlike conventional 3D segmentation methods, this approach processes 3D volumes by sliding along six anatomical directions (AP, PA, LR, RL, IS, SI), enabling 2D segmentation while retaining 3D spatial information. The method employs a modified 2D UNet with group normalisation, Leaky ReLU activation, and average pooling for down-sampling, ensuring computational efficiency. Its novelty lies in combining directional scrolling with a lightweight 2D architecture, achieving segmentation performance comparable to state-of-the-art 3D models such as nnUNet.\\
\textit{Data processing}: The dataset consists of 7T MRI images with dimensions \([480, 640, 163]\), divided into training, validation, and testing sets comprising 8, 3, and 3 subjects, respectively. During pre-processing, the images are padded and sliced into stacks of 10 channels, representing anatomical slices. Data augmentation includes random flipping along all dimensions with a probability of 0.5. This approach integrates 2D slices into a pipeline that preserves 3D spatial continuity.\\
\textit{Training process}: The model is trained over 1,000 epochs with a batch size of 8, using the Adam optimiser with an initial learning rate of 0.001. The loss function combines Dice loss and binary cross-entropy (BCE) loss. During training, slices are reshaped and processed in batches, accumulating gradients for optimisation. For inference, outputs from all anatomical directions are summed and normalised, with a detection threshold of 0.3 applied to identify vessel voxels. This configuration delivers performance comparable to 3D methods while significantly reducing computational demands.

\paragraph{\pbinnUNet} This submission utilised nnUNet \cite{isensee2021nnu}, a self-configuring, state-of-the-art model designed for biomedical image segmentation, particularly excelling in high-resolution MRI vessel segmentation. Renowned for its robustness and flexibility, nnUNet automatically adapts to dataset-specific features, such as voxel spacing, intensity distributions, and class ratios. This adaptability makes it especially effective for tasks requiring precise boundary delineation, such as vessel segmentation. The model’s ability to self-configure without extensive manual tuning enables it to capture fine, complex structures within MRI data, which is critical for accurately segmenting small vessels.\\
\textit{Data processing}: The original data processing steps were adhered to in this submission. The input images were preprocessed by cropping non-zero volumes, and patches were created with a patch size of [224, 64, 160] to reduce computational load. Resampling based on voxel spacing was performed to maintain spatial semantics, and z-score normalisation was applied where necessary. This automated preprocessing pipeline ensures efficiency and consistency across diverse input datasets.\\
\textit{Training process}: The dataset was divided into training, validation, and testing subsets, comprising 8, 3, and 3 samples respectively. nnUNet was trained for 1,000 epochs using the Adam optimiser with an initial learning rate of 0.001 and a composite loss function combining Dice loss and binary cross-entropy (BCE) loss, balancing pixel-wise precision with spatial overlap. Following the methodology outlined in the nnUNet paper, the framework automatically determined the batch size as 2, based on GPU memory constraints. The training process incorporated dynamic adaptation of patch size, extensive data augmentation (including scaling, rotation, mirroring, and elastic deformations), and robust training configurations to optimise nnUNet's performance across diverse tasks. This configuration highlights nnUNet's adaptability and precision, making it particularly well-suited for vessel segmentation, where fine-grained anatomical structures demand a high level of accuracy and consistency.

\paragraph{\eurecom} This submitted method, based on the JoB-VS framework \cite{valderrama2023job}, is tailored for the segmentation of brain vasculature using ultra-high-resolution 7T Time-of-Flight (ToF) Magnetic Resonance Angiography (MRA) images. JoB-VS employs a triangular lattice structure to facilitate multi-scale processing, making it particularly effective for segmenting vessels of varying sizes. The framework has been adapted to focus exclusively on vessel segmentation by configuring the loss function to exclude brain segmentation (\(\alpha = 0, \beta = 1\)). This adaptation enhances precision in segmenting vessels, particularly small and intricate structures, while interpolated high-resolution data further improves sensitivity.\\
\textit{Data processing}: Pre-processing involves Z-score intensity normalisation to standardise input images. To optimise the segmentation of smaller vessels, the data is interpolated to twice its original resolution. The dataset is divided into two folds of seven subjects each for cross-validation, ensuring the robustness of the model through balanced training and evaluation datasets.\\
\textit{Training process}: The JoB-VS framework is trained using a combination of Dice and Cross-Entropy loss terms for vessel segmentation. The Adam optimiser, with a weight decay of \(1 \times 10^{-5}\), is employed alongside a learning rate scheduler to ensure convergence. The model is trained with a batch size of 1 and an initial learning rate of \(5 \times 10^{-4}\), continuing until optimal performance is achieved. 

Table \ref{tab:methods_all} presents a comprehensive summary of all the submitted methods. While some of these methods were based on previously published work by the participants, others were developed specifically for this challenge, some of which have since been published as complete studies. Table \ref{tab:methods_papers_codes} provides a detailed list of the published methods, including links to their corresponding codebases.

\begin{table*}[!tbp]
\centering
\caption{Brief comparison of the methods}
\label{tab:methods_all}
\footnotesize
\setlength{\tabcolsep}{3pt}
\renewcommand{\arraystretch}{1.14}
\begin{tabularx}{\textwidth}{@{}>{\raggedright\arraybackslash}p{4.15cm}>{\raggedright\arraybackslash}p{2.25cm}>{\raggedright\arraybackslash}X>{\raggedright\arraybackslash}p{2.35cm}>{\raggedright\arraybackslash}p{1.65cm}@{}}
\toprule
\textbf{Method} & \textbf{Base model} & \textbf{Method details} & \textbf{Loss function} & \textbf{Optimiser} \\ \midrule\midrule
\baseUNetMSS & UNet MSS & UNet with multi-scale supervision. & Focal Tversky & Adam \\ \midrule
\baseDSSx & UNet MSS & UNet with multi-scale supervision and deformation-aware semi-supervised learning. & Focal Tversky & Adam \\ \midrule\midrule
\ADARUNesT & UNesT & Hierarchical transformer adapted from a pre-trained MONAI model-zoo configuration. & Dice + binary cross-entropy & RMSprop \\ \midrule
\ADARnnUNet & nnUNet & Self-configuring nnUNet pipeline adapted to the challenge data. & Dice + cross-entropy & AdamW \\ \midrule
\ADARSwinUNETR & SwinUNETR & Swin Transformer encoder with a CNN-based decoder. & Dice + cross-entropy & RMSprop \\ \midrule
\ADARTriUNet & UNesT + nnUNet + SwinUNETR & Ensemble that aggregates the outputs of three pre-trained architectures with 3D convolutions. & Dice + cross-entropy & AdamW \\ \midrule
\makecell[l]{\koalaMan\\ \koalaOne\\ \koalaTwo} & UNet & 3D UNet trained with manual labels and two OMELETTE-derived label sets, followed by test-time adaptation. & Tversky & Adam \\ \midrule
\neuroMIP & UNet MSS & UNet MSS with single-axis MIP loss. & Focal Tversky & Adam \\ \midrule
\neuromultiMIP & UNet MSS & UNet MSS with MIP loss across all three axes. & Focal Tversky & Adam \\ \midrule
\neuroDSMIP & DS6 & DS6 semi-supervised learning with single-axis MIP loss. & Focal Tversky & Adam \\ \midrule
\dolphins & SwinUNETR & Hybrid Swin Transformer encoder and UNet-like decoder with cross-attention. & Dice & Not reported \\ \midrule
\funpixel & SwinUNETR & Swin Transformer with UNet-like decoding and multi-projection MIP supervision. & Cross-entropy + OSD + 2D Dice & Adam \\ \midrule
\lsgroup & MS-nnUNet & nnUNet enhanced with multi-scale feature fusion in the decoder. & Dice + binary cross-entropy & SGD \\ \midrule
\pbiScroll & UNet & 2D UNet applied to 3D volumes by scrolling along six anatomical directions. & Dice + binary cross-entropy & Adam \\ \midrule
\pbinnUNet & nnUNet & Self-configuring nnUNet pipeline with automated preprocessing and patch-size adaptation. & Dice + binary cross-entropy & Adam \\ \midrule
\eurecom & JoB-VS & Triangular lattice framework for multi-scale vessel segmentation. & Dice + cross-entropy & Adam \\ \bottomrule
\end{tabularx}
\end{table*}

\begin{table*}[!tbp]
\centering
\caption{Published manuscripts and codes of some of the submitted methods. Displayed repository and Docker Hub entries omit URL prefixes; GitHub links start with \textbf{https://github.com/} and Docker Hub links start with \textbf{https://hub.docker.com/r/}.}
\label{tab:methods_papers_codes}
\footnotesize
\setlength{\tabcolsep}{2pt}
\renewcommand{\arraystretch}{1.12}
\begin{tabularx}{\textwidth}{@{}>{\raggedright\arraybackslash}p{4.0cm}>{\centering\arraybackslash}p{1.5cm}>{\raggedright\arraybackslash}p{3.2cm}>{\raggedright\arraybackslash}X@{}}
\toprule
\multicolumn{1}{c}{\textbf{Method}} & \textbf{Paper} & \multicolumn{1}{c}{\textbf{GitHub}} & \multicolumn{1}{c}{\textbf{Docker Hub}} \\ \midrule
\baseUNetMSS & \multirow{2}{1.5cm}{\centering\cite{chatterjee2022ds6}} & \multirow{2}{3.2cm}{\href{https://github.com/soumickmj/DS6}{soumickmj/DS6}} & \href{https://hub.docker.com/r/smileuhura/baseline_unet_mss}{smileuhura/\allowbreak baseline\_unet\_mss} \\
\baseDSSx &  &  & \href{https://hub.docker.com/r/smileuhura/baseline_ds6}{smileuhura/baseline\_ds6} \\ \midrule\midrule
\ADARUNesT &  &  & \href{https://hub.docker.com/r/vanessa0716/adar_lab_unest}{vanessa0716/\allowbreak adar\_lab\_unest} \\ \midrule
\ADARnnUNet &  &  & \href{https://hub.docker.com/r/chunchih/adar_lab_nnunet}{chunchih/\allowbreak adar\_lab\_nnunet} \\ \midrule
\ADARSwinUNETR &  &  & \href{https://hub.docker.com/r/chunchih/adar_lab_swinunetr}{chunchih/\allowbreak adar\_lab\_\allowbreak swinunetr} \\ \midrule
\ADARTriUNet &  &  & \href{https://hub.docker.com/r/chunchih/adar_lab_triunet}{chunchih/\allowbreak adar\_lab\_triunet} \\ \midrule
\koalaMan & \multirow{3}{1.5cm}{\centering\cite{Xu2024VesselBoost}} & \multirow{3}{3.2cm}{\href{https://github.com/KMarshallX/VesselBoost}{KMarshallX/\allowbreak VesselBoost}} & \href{https://hub.docker.com/r/vnmd/koala_manual}{vnmd/koala\_manual} \\
\koalaOne &  &  & \href{https://hub.docker.com/r/vnmd/koala_om1}{vnmd/koala\_om1} \\
\koalaTwo &  &  & \href{https://hub.docker.com/r/vnmd/koala_om2}{vnmd/koala\_om2} \\ \midrule
\neuroMIP & \multirow{3}{1.5cm}{\centering\cite{radhakrishna2024spockmip}} & \multirow{3}{3.2cm}{\href{https://github.com/soumickmj/SPOCKMIP}{soumickmj/\allowbreak SPOCKMIP}} & \href{https://hub.docker.com/r/cradhakr/uhura_mip}{cradhakr/uhura\_mip} \\
\neuromultiMIP &  &  & \href{https://hub.docker.com/r/cradhakr/uhura_multi_mip}{cradhakr/\allowbreak uhura\_multi\_mip} \\
\neuroDSMIP &  &  & \href{https://hub.docker.com/r/cradhakr/uhura_ds6_mip}{cradhakr/uhura\_ds6\_mip} \\ \midrule
\dolphins &  &  & \href{https://hub.docker.com/r/abdulenib/dolphins}{abdulenib/dolphins} \\ \midrule
\funpixel &  &  & \href{https://hub.docker.com/r/lfm840731775/funpixel_swinunet}{lfm840731775/\allowbreak funpixel\_swinunet} \\ \midrule
\lsgroup &  & \href{https://github.com/SMILE-UHURA/LSGroup}{SMILE-UHURA/\allowbreak LSGroup} & \href{https://hub.docker.com/r/rq16/smile}{rq16/smile} \\ \midrule
\pbiScroll &  &  & \href{https://hub.docker.com/r/aryayazdanpanah/scrolling2dunetsmileuhura}{\makecell[l]{aryayazdanpanah/\\scrolling2dunet\\smileuhura}} \\ \midrule
\pbinnUNet &  &  & \href{https://hub.docker.com/r/ghisvail/paris_brain_institute_nnunet}{ghisvail/\allowbreak paris\_brain\_\allowbreak institute\_nnunet} \\ \midrule
\eurecom & \cite{valderrama2023job} & \href{https://github.com/BCV-Uniandes/JoB-VS}{BCV-Uniandes/\allowbreak JoB-VS} &  \\ \bottomrule
\end{tabularx}
\end{table*}

%% file: Sections/5_results.tex
\subsection{Quantitative Results}

\subsubsection{Open Dataset}

\begin{figure*}[!htbp]
    \centering
    \includegraphics[width=0.77\textwidth]{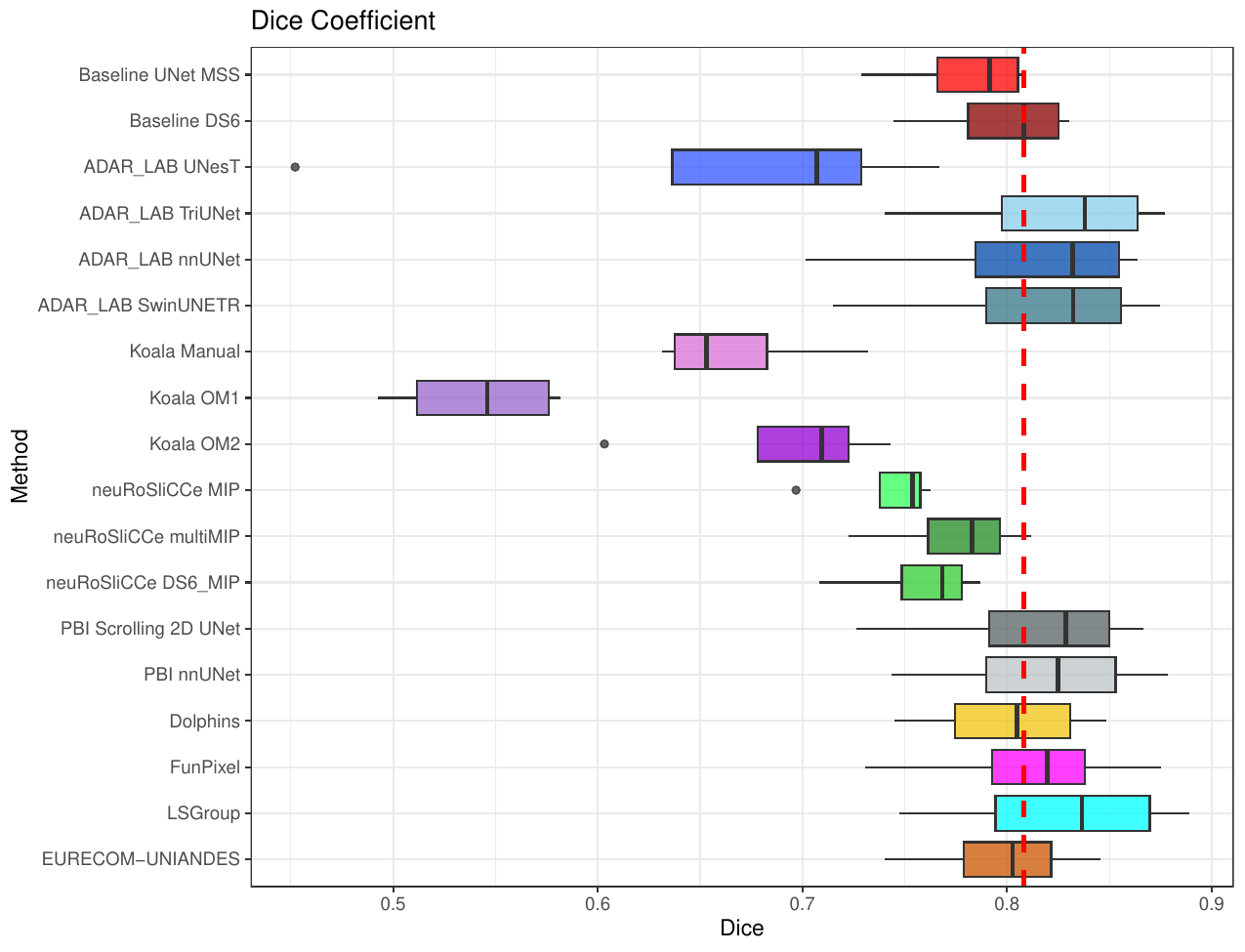}
    \caption{Dice scores on the test subset from the open dataset. The red dashed line denotes the median of the better-performing baseline method (i.e., DS6).}
    \label{fig:box_Forrest_Dice}
\end{figure*}

\begin{figure*}[!htbp]
    \centering
    \begin{subfigure}[t]{\textwidth}
        \centering
        \includegraphics[width=0.95\textwidth]{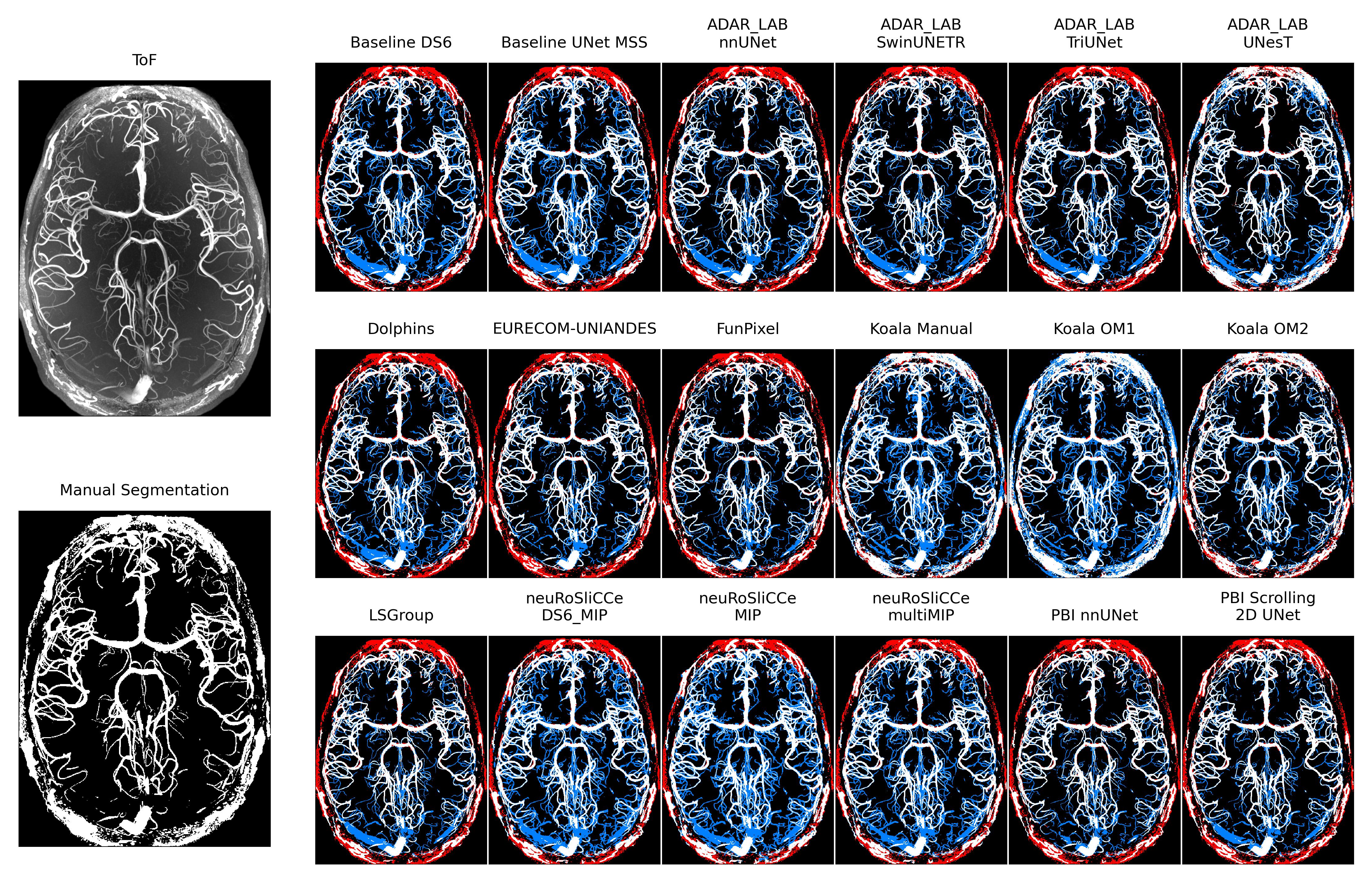}
        \caption{Full maximum intensity projection.}
        \label{fig:qual_open_full}
    \end{subfigure}
    \vskip 1ex
    \begin{subfigure}[t]{\textwidth}
        \centering
        \includegraphics[width=0.95\textwidth]{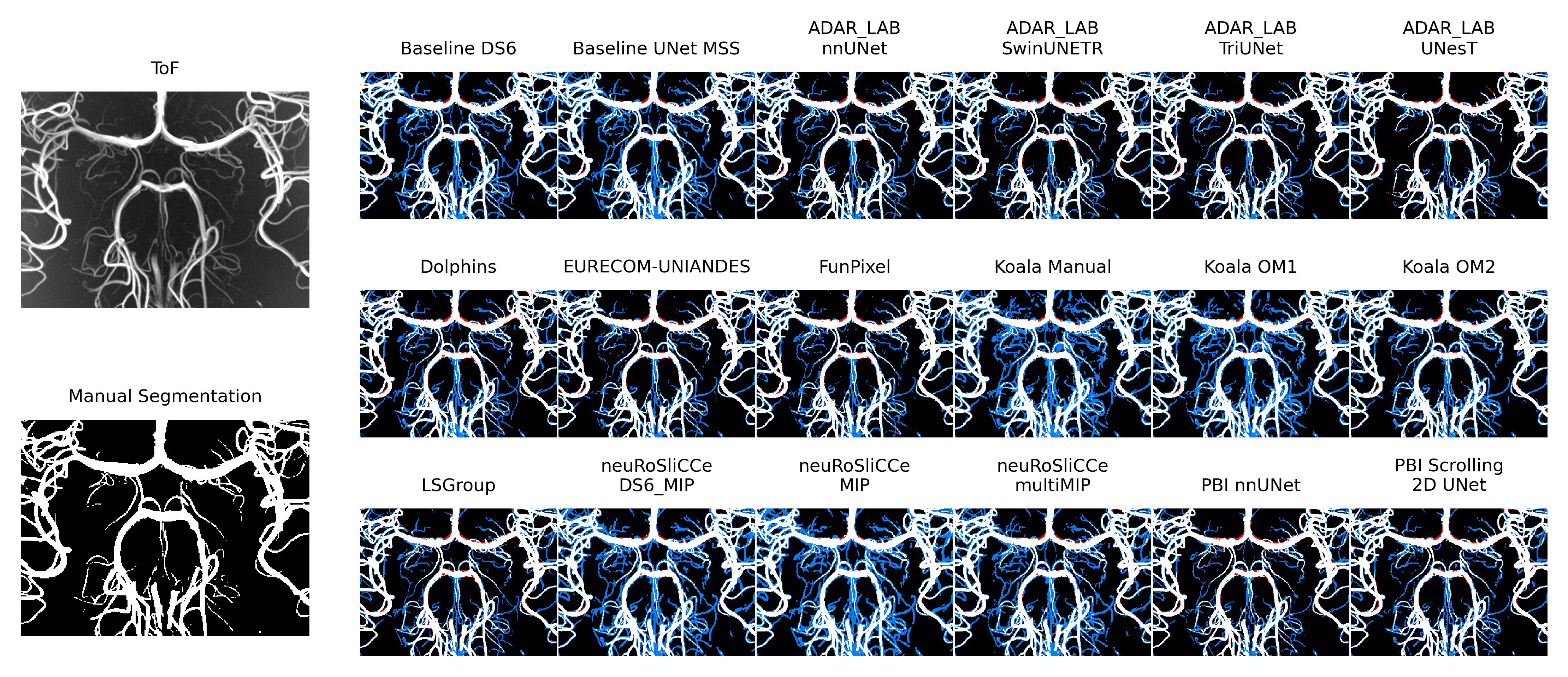}
        \caption{Cropped region around the Circle of Willis.}
        \label{fig:qual_open_crop}
    \end{subfigure}
    \caption{Qualitative comparison of the segmentations produced by the evaluated methods on a representative subject from the open dataset, visualised as maximum intensity projections (MIPs). Vessels in white correspond to correctly segmented voxels, red denotes under-segmentation relative to the reference annotation, and blue denotes over-segmentation.}
    \label{fig:qual_open}
\end{figure*}

The performance of the submitted deep learning methods was evaluated on the open dataset, comprising MRI volumes held out from the training set but sharing identical acquisition properties. The metrics assessed included the Dice coefficient (Dice), Jaccard index (Jaccard), volumetric similarity coefficient (VolSim), mutual information (MI), and the balanced average Hausdorff distance (bAVD, as termed by the \textit{EvaluateSegmentation} pipeline; also known as bAHD in the literature), each reported as median ± interquartile range (IQR) in Table \ref{tab:open_performance_median_iqr}. The per-subject Dice and bAVD score distributions are shown as box plots in Figures~\ref{fig:box_Forrest_Dice} and~\ref{fig:box_Forrest_bAVD}; the corresponding Jaccard, VolSim and MI box plots are provided in Appendix~\ref{app:extra_boxplots} (Figures~\ref{fig:box_Forrest_Jaccard}, \ref{fig:box_Forrest_VolSim} and \ref{fig:box_Forrest_MI}). A qualitative visualisation of the corresponding segmentations for a representative subject is shown in Figure~\ref{fig:qual_open}.

\input{Tables/Forrest/median_iqr_ranked}

The baselines, \baseDSSx[0] and \baseUNetMSS[0], yielded moderate performance with Dice scores of 0.808 ± 0.044 and 0.791 ± 0.039, respectively. While these baselines provided a solid foundation, they were surpassed by several of the submitted methods. For instance, \ADARnnUNet[0] and \ADARSwinUNETR[0] achieved Dice scores of 0.832 ± 0.070 and 0.832 ± 0.066, respectively, indicating improved segmentation performance over the baselines.

Notably, \pbiScroll[0] and \pbinnUNet[0] also outperformed the baselines, with Dice scores of 0.829 ± 0.058 and 0.825 ± 0.063. Their Jaccard values and volumetric similarities further corroborated their enhanced performance. Conversely, some methods did not surpass the baseline performance. The neuRoSliCCe series, comprising \neuromultiMIP[0], \neuroDSMIP[0], and \neuroMIP[0], yielded Dice scores ranging from 0.754 ± 0.020 to 0.783 ± 0.035, lower than those of the baselines. \koalaTwo[0] and \ADARUNesT[0] performed below the baselines as well.

\koalaOne[0] and \koalaMan[0] yielded the lowest scores of the evaluated methods. \koalaOne[0] achieved a Dice of 0.546 ± 0.064 and a Jaccard of 0.376 ± 0.061, considerably below the baselines and the other methods. The high bAVD of 8.728 ± 6.924 for \koalaOne[0] points to substantial boundary inaccuracies. Similarly, \koalaMan[0] reported a Dice of 0.653 ± 0.045 and a bAVD of 3.285 ± 1.517, below the performance of the baseline methods.

The \ADARTriUNet[0] method demonstrated the highest performance among all the evaluated techniques. Specifically, it achieved a Dice of 0.838 ± 0.066 and a Jaccard of 0.722 ± 0.096, outperforming the baselines and other methods on these overlap metrics. The VolSim for this method was also high at 0.959 ± 0.014, indicating strong agreement with the ground truth volumes. It further exhibited a low bAVD of 0.314 ± 0.224, reflecting accurate boundary delineation.

Similarly, the \lsgroup[0] method showed competitive performance, with a Dice of 0.837 ± 0.075 and a Jaccard of 0.720 ± 0.110. Its volumetric similarity was the highest among all methods at 0.968 ± 0.047, and it achieved the lowest bAVD of 0.309 ± 0.168. These results suggest that \lsgroup[0] is highly effective in both volumetric accuracy and boundary precision.

The mutual information metric remained relatively consistent across most methods, with values clustering around 0.060 ± 0.003. The baselines and the top-performing methods shared similar MI scores, suggesting that this metric was less discriminative among the evaluated techniques.

In light of all metrics and the limited discriminative power of mutual information, \ADARTriUNet[0] and \lsgroup[0] were the best performing algorithms across the board. 

\begin{figure*}[!htbp]
    \centering
    \includegraphics[width=0.77\textwidth]{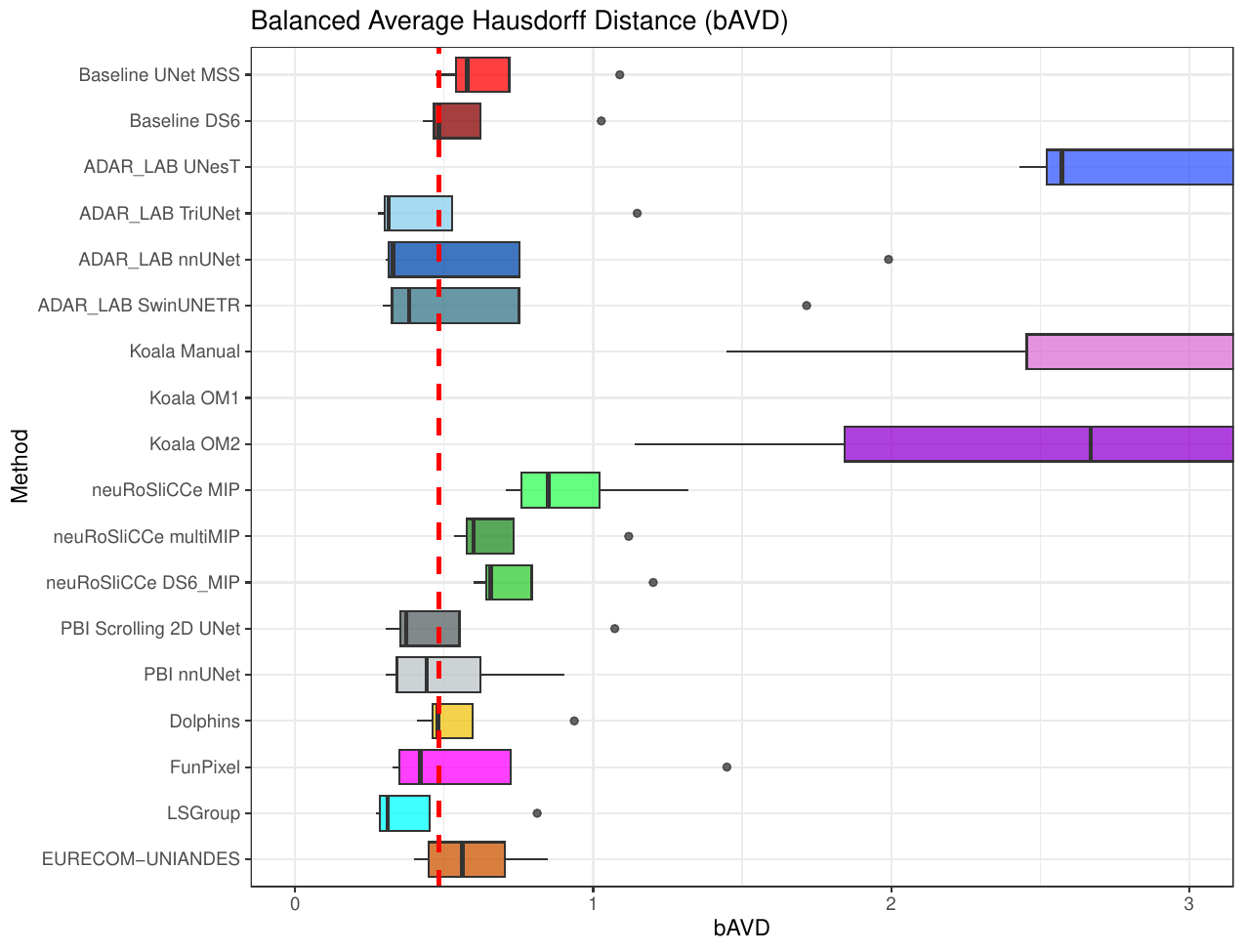}
    \caption{Balanced average Hausdorff distances on the test subset from the open dataset. The plot was confined to $bAVD \leq 3$, as three methods yielded extreme values, thereby rendering the remainder of the comparisons incomprehensible. The red dashed line denotes the median of the better-performing baseline method (i.e., DS6).}
    \label{fig:box_Forrest_bAVD}
\end{figure*}

\subsubsection{Secret Dataset}

\begin{figure*}[!htbp]
    \centering
    \includegraphics[width=0.77\textwidth]{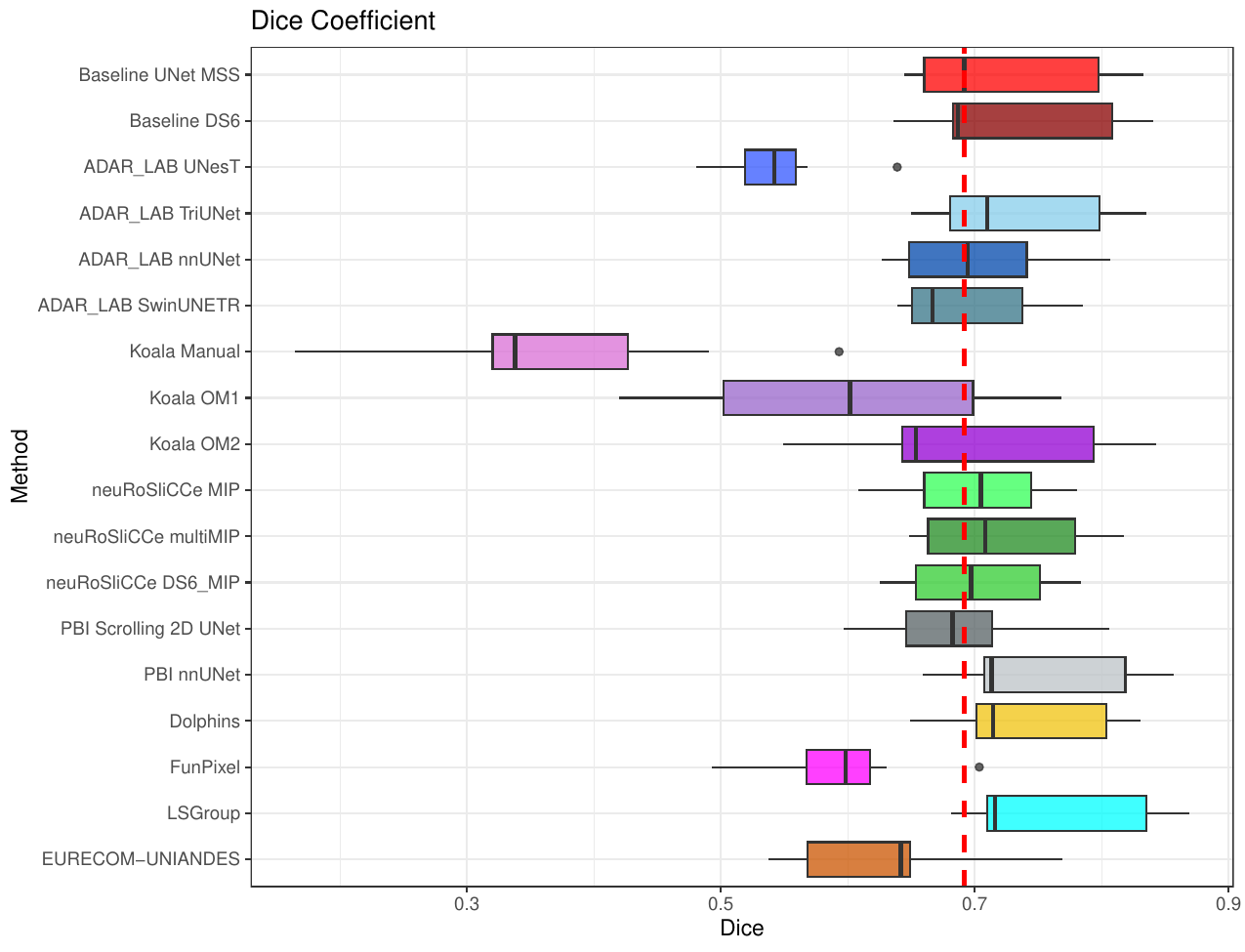}
    \caption{Dice scores on the secret dataset. The red dashed line denotes the median of the better-performing baseline method (i.e., UNet MSS).}
    \label{fig:box_HendrikROT_Dice}
\end{figure*}

\begin{figure*}[!htbp]
    \centering
    \begin{subfigure}[t]{\textwidth}
        \centering
        \includegraphics[width=0.95\textwidth]{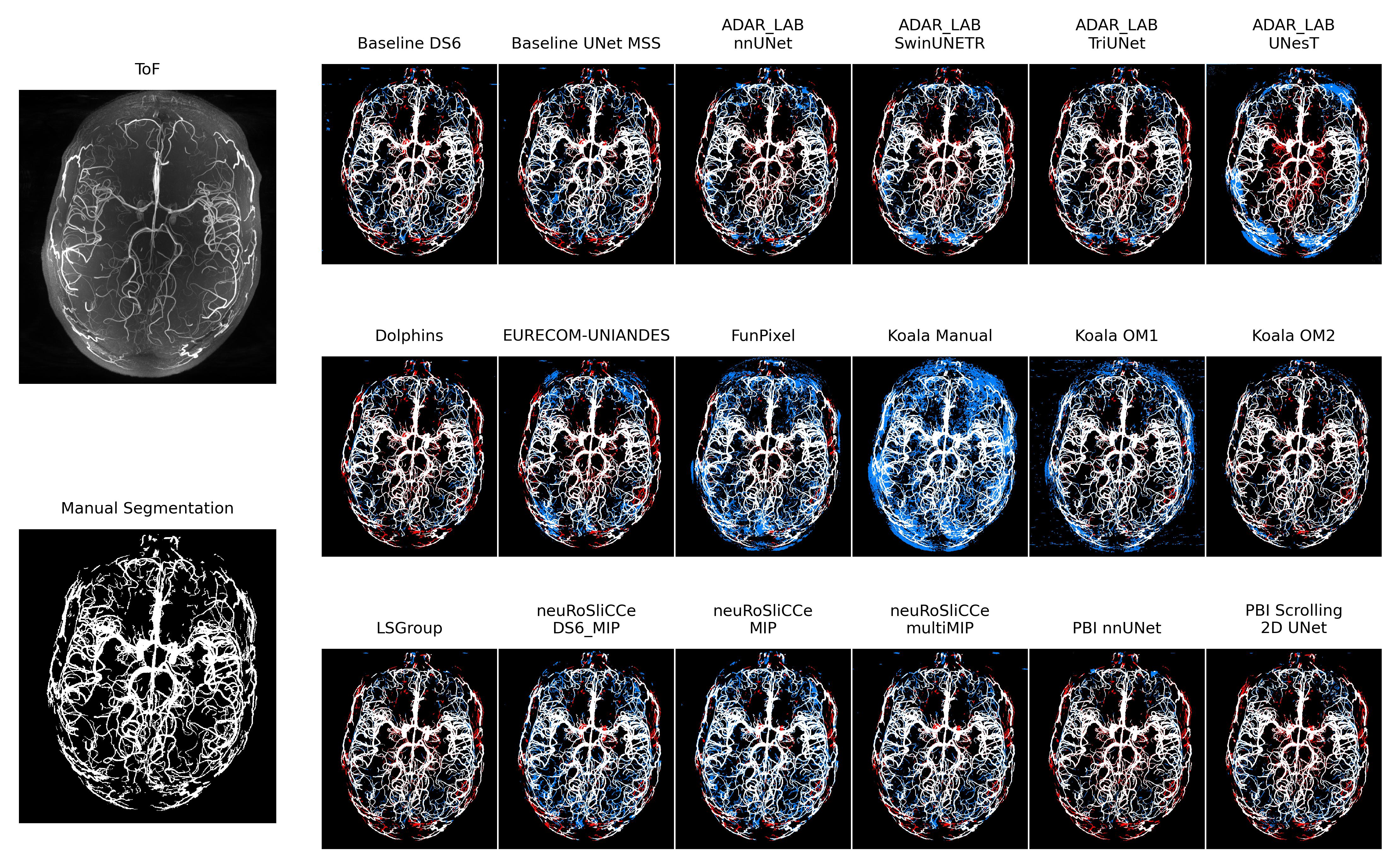}
        \caption{Full maximum intensity projection.}
        \label{fig:qual_secret_full}
    \end{subfigure}
    \vskip 1ex
    \begin{subfigure}[t]{\textwidth}
        \centering
        \includegraphics[width=0.95\textwidth]{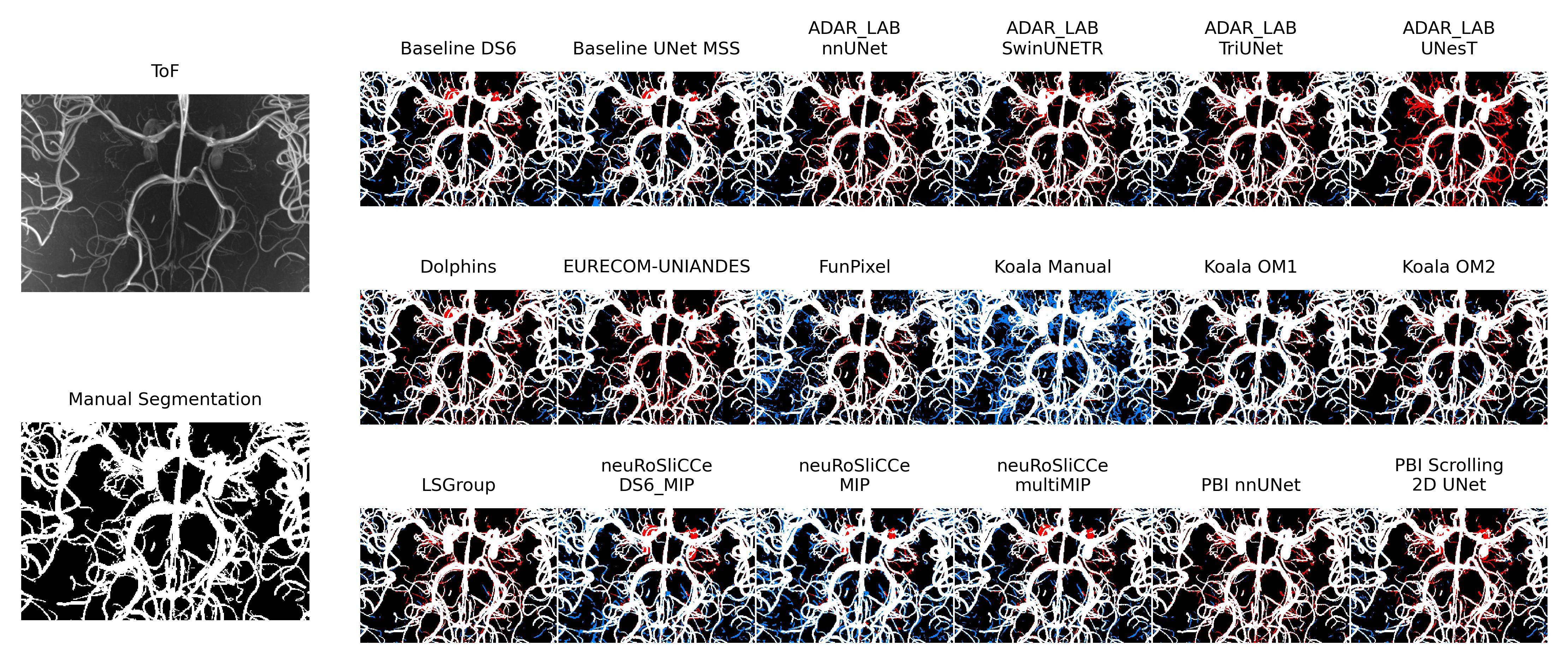}
        \caption{Cropped region around the Circle of Willis.}
        \label{fig:qual_secret_crop}
    \end{subfigure}
    \caption{Qualitative comparison of the segmentations produced by the evaluated methods on a representative subject from the secret dataset, visualised as maximum intensity projections (MIPs). Vessels in white correspond to correctly segmented voxels, red denotes under-segmentation relative to the reference annotation, and blue denotes over-segmentation.}
    \label{fig:qual_secret}
\end{figure*}

The next set of evaluations were performed on the secret dataset, which, whilst sharing the same resolution and field strength as the training data, originates from a different study and may possess distinct properties. The median ± IQR of the scores (Dice, Jaccard, VolSim, MI, and bAVD) are reported in Table \ref{tab:secret_performance_median_iqr}. The per-subject Dice and bAVD score distributions are shown as box plots in Figures~\ref{fig:box_HendrikROT_Dice} and~\ref{fig:box_HendrikROT_bAVD}; the corresponding Jaccard, VolSim and MI box plots are provided in Appendix~\ref{app:extra_boxplots} (Figures~\ref{fig:box_HendrikROT_Jaccard}, \ref{fig:box_HendrikROT_VolSim} and \ref{fig:box_HendrikROT_MI}). A qualitative visualisation of the segmentations on a representative subject from this dataset is shown in Figure~\ref{fig:qual_secret}.

\input{Tables/HendrikROT/median_iqr_ranked}

The baselines, \baseUNetMSS[0] and \baseDSSx[0], yielded Dice scores of 0.692 ± 0.137 and 0.687 ± 0.125, respectively. Although these baselines provided a solid reference point, they were outperformed by several of the submitted methods. However, certain methods did not exceed the baseline performance. For instance, \ADARSwinUNETR[0] achieved a Dice of 0.667 ± 0.086 and a Jaccard of 0.500 ± 0.102, both lower than those of the baselines. Similarly, \koalaTwo[0] reported a Dice of 0.654 ± 0.151 and a Jaccard of 0.485 ± 0.185, below the baseline metrics.

The \pbiScroll[0] method yielded a Dice of 0.683 ± 0.068 and a Jaccard of 0.518 ± 0.078, slightly below the baselines. Its VolSim was 0.867 ± 0.085, and it had a relatively high bAVD of 1.262 ± 3.010, indicating less precise boundary segmentation. \eurecom[0] and \funpixel[0] also did not outperform the baselines, with Dice scores of 0.642 ± 0.081 and 0.598 ± 0.050, respectively. Their bAVD scores were considerably higher, suggesting difficulty in boundary accuracy.

\koalaMan[0] exhibited the lowest performance metrics, with a Dice of 0.338 ± 0.107 and a Jaccard of 0.203 ± 0.083. Its volumetric similarity was 0.398 ± 0.145, and it reported a high bAVD of 17.325 ± 14.446, indicating substantial boundary inaccuracies.

The \lsgroup[0] method exhibited the highest overall performance on the secret dataset, achieving a Dice of 0.716 ± 0.125 and a Jaccard of 0.558 ± 0.168, surpassing both baseline methods and most other evaluated techniques. Its VolSim was 0.930 ± 0.096, indicating strong concordance with the ground truth volumes. Notably, \lsgroup[0] attained a bAVD of 0.730 ± 0.197, reflecting precise boundary delineation and marking the lowest bAVD among all methods on this dataset.

Close contenders included \dolphins[0] and \pbinnUNet[0], with Dice scores of 0.715 ± 0.103 and 0.713 ± 0.111, respectively. Their Jaccard values were 0.556 ± 0.133 and 0.554 ± 0.147, demonstrating competitive overlap metrics. The VolSim values for these methods were also high, 0.936 ± 0.127 for \dolphins[0] and 0.949 ± 0.162 for \pbinnUNet[0], suggesting robust volumetric accuracy. Their bAVD scores, 0.874 ± 0.165 for \dolphins[0] and 0.843 ± 0.164 for \pbinnUNet[0], were lower than those of the baselines, indicating improved boundary accuracy.

The \ADARTriUNet[0] method, which previously excelled on the open dataset, achieved a Dice of 0.710 ± 0.118 and a Jaccard of 0.550 ± 0.149 on the secret dataset. These results, whilst commendable and above both baselines, suggest a slight decrease in performance compared with its results on the open dataset. The VolSim for \ADARTriUNet[0] was 0.917 ± 0.093, and its bAVD was 0.947 ± 0.340, both indicating satisfactory but not superior performance relative to the top methods.

The neuRoSliCCe series, namely \neuromultiMIP[0], \neuroMIP[0], and \neuroDSMIP[0], demonstrated consistent performance, with Dice scores of 0.708 ± 0.116, 0.705 ± 0.084, and 0.697 ± 0.098, respectively. These methods outperformed the baselines in terms of Dice and Jaccard but did not surpass the top-performing methods. Their VolSim values ranged from 0.876 ± 0.057 to 0.901 ± 0.065, and their bAVD scores were comparable to the baselines, suggesting moderate boundary accuracy.

It is worth noting that, for VolSim, no submitted method exceeded the \baseDSSx[0] baseline on the secret dataset (0.950 ± 0.142); even the best-performing submissions (\pbinnUNet[0] at 0.949 ± 0.162 and \dolphins[0] at 0.936 ± 0.127) remained marginally below it. Since VolSim is a size-only agreement metric, this observation indicates that several methods segmented a total vessel volume slightly different from the reference even where their overlap and boundary metrics were strong.

Similar to the open dataset, MI showed less variation across methods, with values clustering around 0.026 ± 0.003. \lsgroup[0] had a slightly higher MI of 0.028 ± 0.004, potentially indicating better mutual dependence between the segmented and ground truth images. However, this metric did not considerably distinguish between the top-performing methods and the baselines.

In summary, on the secret dataset, several submitted methods, notably \lsgroup[0], \dolphins[0], and \pbinnUNet[0], consistently outperformed the baseline methods and the other submitted methods across multiple metrics. These methods demonstrated superior generalisation to data with different properties from the training set. The performance of \ADARTriUNet[0], whilst still above the baselines, was slightly diminished compared with its results on the open dataset, suggesting a degree of sensitivity to dataset variations.

The implications of these findings are pertinent for the field of medical image segmentation. The ability of methods like \lsgroup[0] and \dolphins[0] to maintain high performance on a dataset with different properties indicates strong generalisation, which is essential for clinical applicability. These results suggest that incorporating architectures and training strategies that promote adaptability can enhance the robustness of segmentation models.

The fact that some methods did not exceed the baseline performance on the secret dataset highlights the continued need for research into techniques that improve generalisation, although it is also worth recalling that the reference annotations themselves are imperfect at the scale of single-voxel vessels and may occasionally disagree with legitimately captured small branches. It underscores the importance of developing models that are not only optimised for specific datasets but are also resilient to variations inherent in medical imaging data.

\begin{figure*}[htb]
    \centering
    \includegraphics[width=0.77\textwidth]{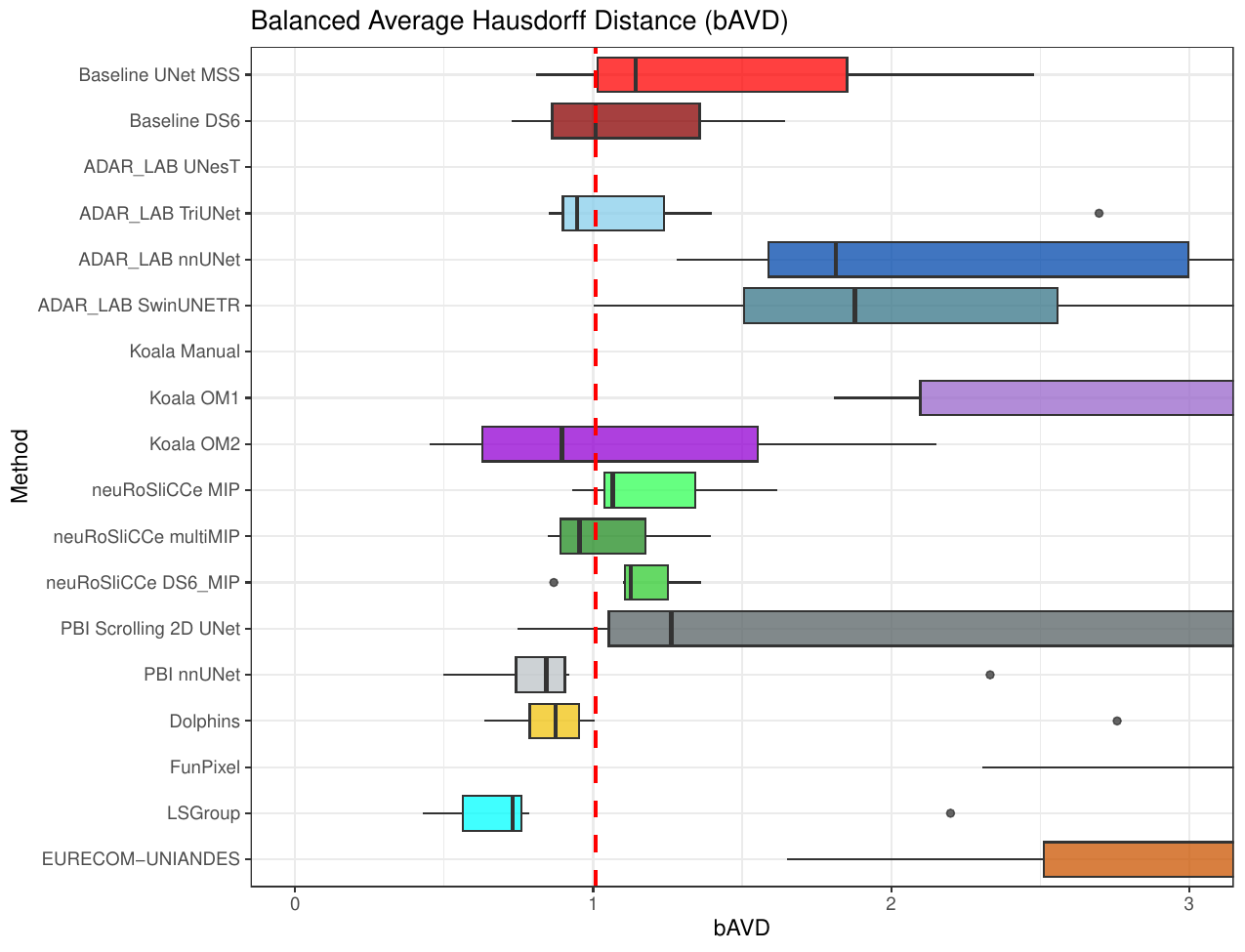}
    \caption{Balanced average Hausdorff distances on the secret dataset. The plot was confined to $bAVD \leq 3$, as six methods yielded extreme values, thereby rendering the remainder of the comparisons incomprehensible. The red dashed line denotes the median of the better-performing baseline method (i.e., DS6).}
    \label{fig:box_HendrikROT_bAVD}
\end{figure*}

\subsubsection{Overall Performance}
\begin{figure*}[!htbp]
    \centering
    \includegraphics[width=0.77\textwidth]{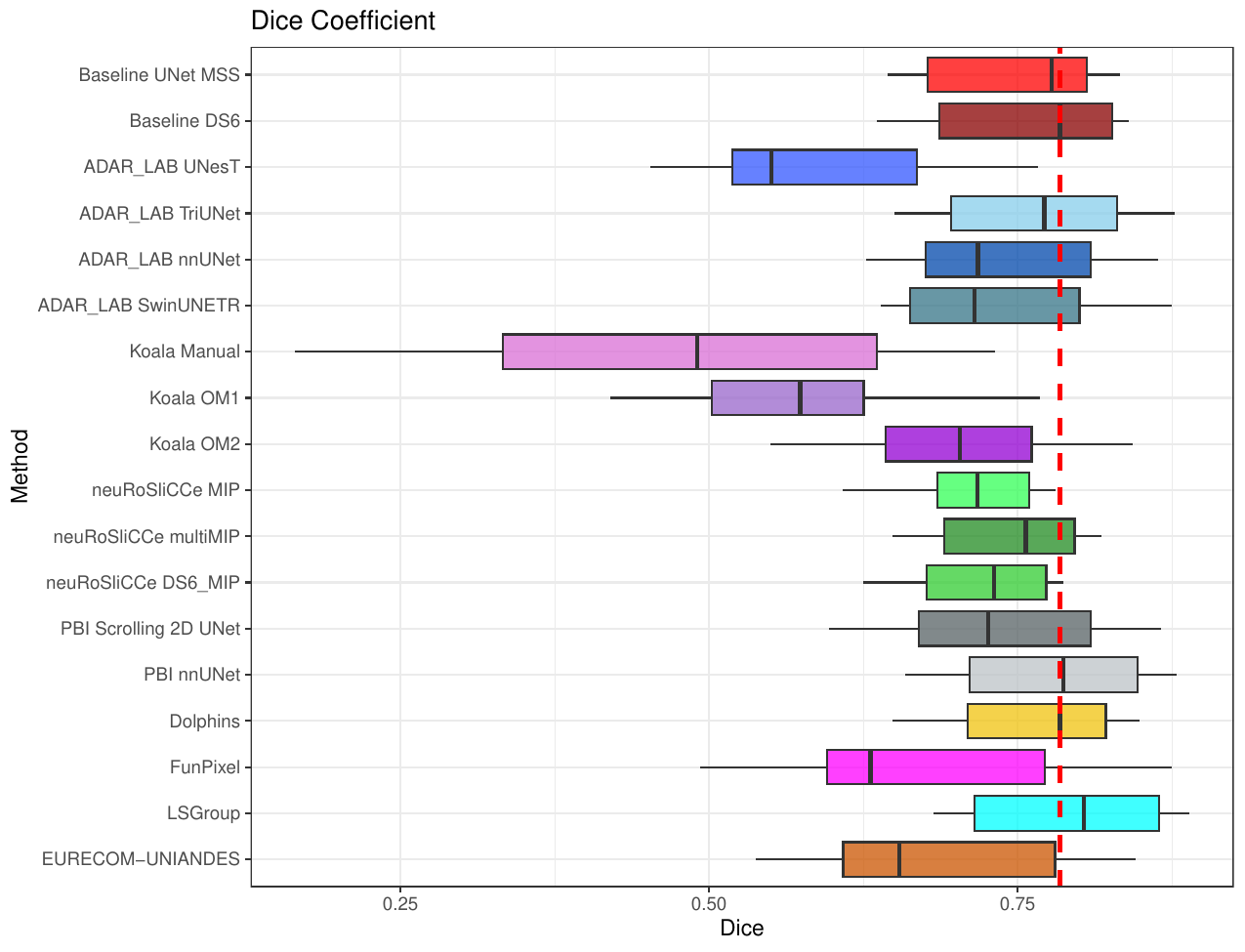}
    \caption{Dice scores on the combined datasets. The red dashed line denotes the median of the better-performing baseline method (i.e., DS6).}
    \label{fig:box_AllRes_Dice}
\end{figure*}

The final quantitative evaluation judged the overall performance of the methods across both datasets by aggregating the results from the open and secret datasets. The scores are reported in Table \ref{tab:overall_performance_median_iqr}. The per-subject Dice and bAVD score distributions are shown as box plots in Figures~\ref{fig:box_AllRes_Dice} and~\ref{fig:box_AllRes_bAVD}; the corresponding Jaccard, VolSim and MI box plots are provided in Appendix~\ref{app:extra_boxplots} (Figures~\ref{fig:box_AllRes_Jaccard}, \ref{fig:box_AllRes_VolSim} and \ref{fig:box_AllRes_MI}).

The baselines, \baseDSSx[0] and \baseUNetMSS[0], yielded Dice scores of 0.784 ± 0.140 and 0.778 ± 0.129, respectively. Several of the submitted methods did not surpass the baseline performance. For example, \pbiScroll[0] achieved a Dice of 0.726 ± 0.140 and a Jaccard of 0.570 ± 0.176, both lower than those of the baselines, and its bAVD was relatively high at 1.073 ± 1.536, indicating less precise boundary segmentation.

\koalaOne[0], \funpixel[0], and \ADARUNesT[0] yielded the lowest performance metrics. \koalaOne[0] reported a Dice of 0.574 ± 0.123 and a Jaccard of 0.403 ± 0.120, considerably below the baselines and the top methods. \funpixel[0] had a Dice of 0.631 ± 0.177 and a Jaccard of 0.461 ± 0.207, with a high bAVD of 4.184 ± 5.953. \ADARUNesT[0] achieved a Dice of 0.551 ± 0.149 and a Jaccard of 0.380 ± 0.152, with a bAVD of 5.576 ± 8.189, indicating substantial boundary inaccuracies.

The \lsgroup[0] method demonstrated the highest overall performance across the combined datasets, achieving a Dice of 0.804 ± 0.150 (approximately 2\% higher than that of \baseDSSx[0] and 3\% higher than \baseUNetMSS[0]) and a Jaccard of 0.672 ± 0.205, surpassing both baseline methods and all other evaluated techniques. Its VolSim was 0.942 ± 0.076, indicating strong agreement with the ground truth volumes. Additionally, \lsgroup[0] attained the lowest bAVD of 0.583 ± 0.380, reflecting precise boundary delineation.

Close competitors included \pbinnUNet[0] and \dolphins[0]. \pbinnUNet[0] achieved a Dice of 0.787 ± 0.136 and a Jaccard of 0.649 ± 0.183, while \dolphins[0] reported a Dice of 0.784 ± 0.112 and a Jaccard of 0.645 ± 0.147. Both methods outperformed the baselines, with \pbinnUNet[0] showing a slightly higher VolSim of 0.955 ± 0.090 compared with 0.943 ± 0.056 for \dolphins[0]. Their bAVD scores were also lower than those of the baselines, indicating better boundary accuracy.

The \ADARTriUNet[0] method, which previously excelled on the open dataset, achieved an overall Dice of 0.772 ± 0.135 and a Jaccard of 0.628 ± 0.176. Although these results surpass the baselines, they are slightly lower than those of the top-performing methods. The VolSim for \ADARTriUNet[0] was high at 0.949 ± 0.050, suggesting robust volumetric accuracy. However, its bAVD was 0.917 ± 0.529, which, whilst better than the baselines, was higher than that of \lsgroup[0], indicating less precise boundary segmentation.

\neuromultiMIP[0] and \neuroDSMIP[0] demonstrated moderate performance. \neuromultiMIP[0] achieved a Dice of 0.757 ± 0.106 and a Jaccard of 0.609 ± 0.134, outperforming the baselines in volumetric similarity with a score of 0.896 ± 0.087 but with a higher bAVD of 0.908 ± 0.311. \neuroDSMIP[0] had a Dice of 0.731 ± 0.097 and a Jaccard of 0.576 ± 0.119, with a bAVD of 1.114 ± 0.406, indicating less accurate boundary delineation compared with the top methods.

Notably, \ADARnnUNet[0] and \ADARSwinUNETR[0], which performed well on the open dataset, did not maintain superior performance in the overall evaluation. \ADARnnUNet[0] had a Dice of 0.718 ± 0.134 and a Jaccard of 0.560 ± 0.169, while \ADARSwinUNETR[0] had a Dice of 0.715 ± 0.137 and a Jaccard of 0.556 ± 0.171. Both methods had higher bAVD scores of 1.643 ± 1.505 and 1.508 ± 1.404, respectively, suggesting decreased boundary accuracy.

In summary, the \lsgroup[0] method consistently outperformed the baselines across multiple metrics in the overall evaluation. Its superior Dice, Jaccard, volumetric similarity, and lowest bAVD indicate its effectiveness in both overlap and boundary accuracy. \pbinnUNet[0] and \dolphins[0] also demonstrated strong performance, exceeding the baselines and achieving competitive metrics.

The baselines provided a solid performance benchmark but were surpassed by several submitted methods. However, certain methods, such as \pbiScroll[0], \ADARnnUNet[0], and \ADARSwinUNETR[0], did not consistently outperform the baselines in the overall evaluation. This suggests that, whilst some methods perform well on specific datasets, their generalisation across diverse data may be limited.

Methods like \koalaOne[0], \funpixel[0], and \ADARUNesT[0] did not exceed the baseline performance. This underscores the challenges in developing robust segmentation algorithms as well as ground truth segmentations. The lower Dice and Jaccard scores, along with higher bAVD values, could indicate difficulties in accurately capturing vessel structures and delineating boundaries, or imperfections in the manual segmentation of these large and complex datasets.

The implications of the findings of this challenge are significant for the advancement of medical image segmentation. The consistent superiority of the \lsgroup[0] method suggests that its architecture and training strategies effectively capture the complexities of vessel segmentation across diverse datasets. This robustness is important for routine application in clinical and neuroscientific studies, where models must perform reliably on data with varying properties.

The results highlight the importance of developing methods with strong generalisation capabilities. Whilst some methods may excel on familiar datasets, their performance may diminish when applied to data with different characteristics. Future research should therefore focus on enhancing the adaptability of segmentation models to ensure consistent performance across various imaging conditions.

In conclusion, the overall evaluation demonstrates that certain advanced methods can surpass the baseline performance and offer improved segmentation accuracy. \lsgroup[0], in particular, shows promise for clinical application owing to its superior performance across multiple metrics and datasets. However, the variability in performance among the different methods underscores the continuing need for the development of robust, generalisable segmentation algorithms in the field of medical image analysis.

\input{Tables/AllRes/median_iqr_ranked}

\begin{figure*}[!htbp]
    \centering
    \includegraphics[width=0.77\textwidth]{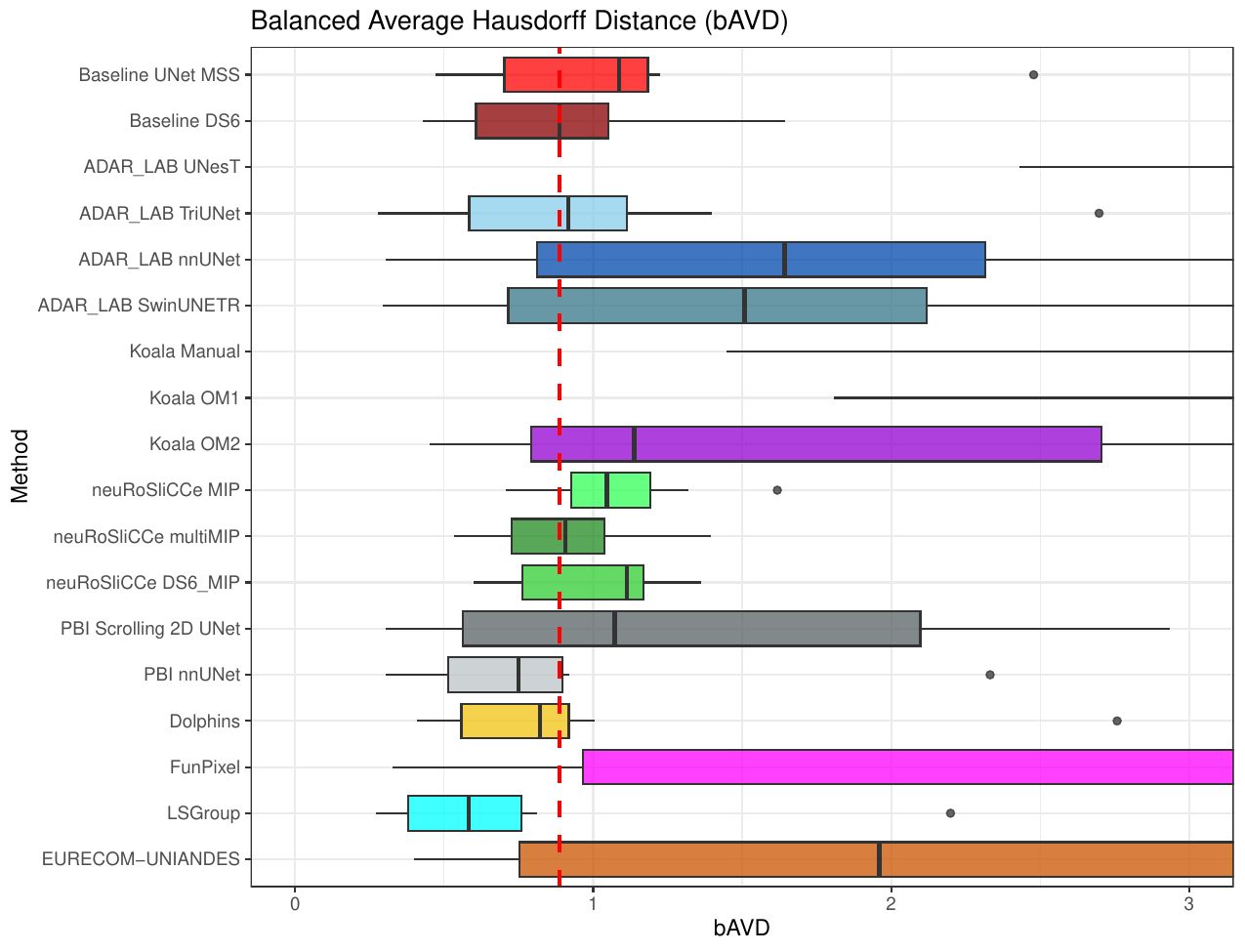}
    \caption{Balanced average Hausdorff distances on the combined datasets. The plot was confined to $bAVD \leq 3$, as eight methods yielded extreme values, thereby rendering the remainder of the comparisons incomprehensible. The red dashed line denotes the median of the better-performing baseline method (i.e., DS6).}
    \label{fig:box_AllRes_bAVD}
\end{figure*}

\paragraph{Conclusion of Quantitative Analysis}
In conclusion, the comprehensive evaluation of the submitted methods for vessel segmentation across both the open and the secret datasets has yielded significant insights into their performance and generalisation capabilities. The \lsgroup[0] method consistently demonstrated superior performance, achieving the highest metrics across both datasets. Specifically, it attained the highest Dice coefficient and Jaccard index, along with the lowest balanced average Hausdorff distance (bAVD), indicating excellent overlap and boundary accuracy. This suggests that \lsgroup[0] possesses robust generalisation capabilities, effectively capturing vessel structures even when confronted with data possessing different properties from the training set.

The baselines, \baseDSSx[0] and \baseUNetMSS[0], provided solid performance benchmarks, with respectable Dice coefficients and Jaccard indices. However, several submitted methods, including \lsgroup[0], \pbinnUNet[0], and \dolphins[0], consistently surpassed these baselines. \pbinnUNet[0] and \dolphins[0] exhibited strong performance, outperforming the baselines on key metrics such as the Dice coefficient, Jaccard index, and bAVD. This indicates that the incorporation of advanced architectures and training strategies can lead to meaningful improvements over standard baseline models.

Conversely, certain methods did not surpass the baseline performance on one or both datasets. Notably, \ADARUNesT[0], \koalaOne[0], \funpixel[0], and \koalaMan[0] remained below the baseline metrics, with lower Dice coefficients and higher bAVD values. These findings could stem from limitations in the models' segmentation capabilities and might indicate challenges in generalising to datasets with different characteristics. However, imperfections in the manual segmentations, i.e. missing small vessel segments, could also introduce a bias into the scores computed, particularly where a method has captured genuinely present small vessels that are not represented in the reference annotation.

The baselines, whilst surpassed by several methods, demonstrated relatively good performance, particularly considering their simplicity compared with more complex architectures and the number of years since their publication. Their consistent performance across both datasets underscores their reliability as reference models in vessel segmentation tasks. The ability of certain advanced methods to exceed the baseline performance nevertheless indicates that further enhancements in network design and training can yield substantial gains in segmentation accuracy.

Overall, these results emphasise the continuing need for segmentation methods that combine high accuracy with strong generalisation capabilities. The superior performance of \lsgroup[0] suggests that it effectively addresses the complexities inherent in vessel segmentation across varied datasets. This has meaningful implications for clinical applications, where models must reliably perform on data from different sources and with varying properties.

Future research should focus on enhancing the adaptability of segmentation models, ensuring consistent performance across diverse imaging conditions. The development of robust, generalisable algorithms is essential for advancing the utility of deep learning in medical image analysis and ultimately improving diagnostic processes and patient outcomes.

\subsubsection{Statistical significance across methods}
\label{sec:stat_results}
The medians and interquartile ranges reported above describe typical behaviour, but they do not by themselves establish whether the differences between methods are reliable. To address this, the statistical pipeline described in Section~\ref{sec:stat_testing} was applied to the combined dataset ($N = 11$ subjects), where the pooled sample size provides enough power for the rank-based procedures to be informative. Per-dataset tests were not conducted given the very small test subsets ($n = 4$ for the open and $n = 7$ for the secret dataset).

\paragraph{Friedman omnibus tests.} The Friedman omnibus test was highly significant for every metric (Table~\ref{tab:friedman_omnibus}), with all Holm-uncorrected $p$-values well below $10^{-7}$. The null hypothesis of equal method performance can therefore be rejected on all five metrics, and the post-hoc pairwise comparisons are justified. The identical $\chi^2$ and $p$-value obtained for Dice and Jaccard is a consequence of Jaccard being a strictly monotone transform of Dice (see Eq.~\ref{eq:jaccard}), so the per-subject rankings on the two metrics are identical and the rank-based statistic is unchanged. VolSim yielded the lowest (though still highly significant) $\chi^2$, indicating that the between-method separation on volumetric similarity, while real, is more modest than on the overlap or boundary metrics.

\input{Tables/Stats/friedman_omnibus}

\paragraph{Conover-Friedman post-hoc tests.} The Conover-Friedman pairwise comparisons with Holm correction identified a substantial number of statistically distinguishable method pairs: $76/153$ pairs for both Dice and Jaccard (identical, as expected), $66/153$ for bAVD, $59/153$ for MI and $20/153$ for VolSim. The much smaller number of significant pairs on VolSim echoes the weaker omnibus signal: several methods arrive at comparable total segmented volumes despite differing substantially in spatial overlap, which is consistent with the narrow spread of VolSim medians in Table~\ref{tab:overall_performance_median_iqr}. The full list of significantly different pairs for each metric is provided in the appendix (Table~\ref{tab:conover_summary}, Appendix~\ref{app:conover_full}), and the complete $18 \times 18$ Holm-adjusted $p$-value matrices per metric are released alongside the paper for querying any specific pair of interest.

\begin{figure*}[!htbp]
    \centering
    \includegraphics[width=0.95\textwidth]{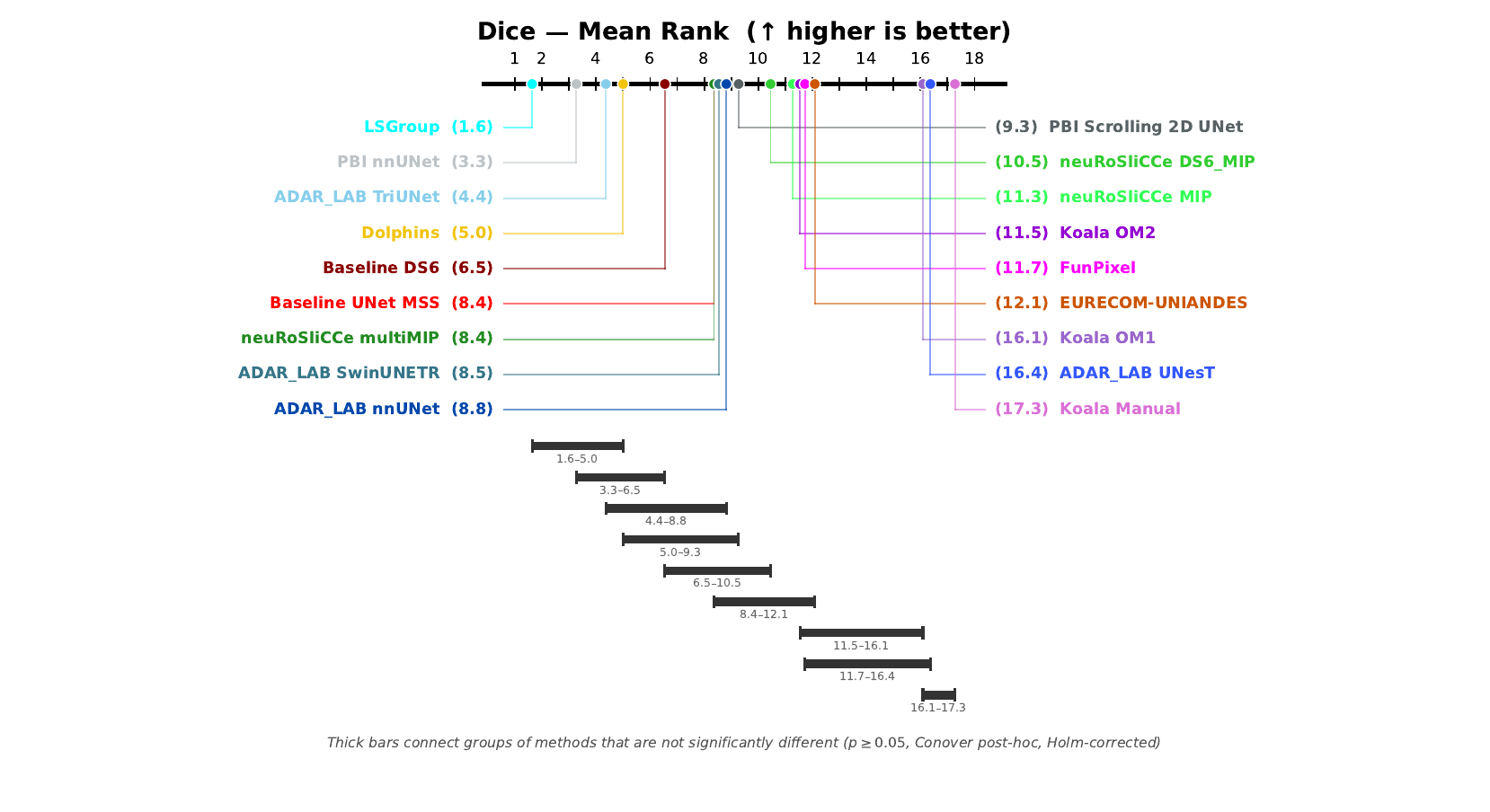}
    \caption{Critical Difference diagram for the Dice coefficient on the combined dataset ($N = 11$). Methods are placed on the axis at their mean rank (lower rank, further to the left, indicates better performance). Horizontal clique bars join groups of methods that are pairwise \emph{not} significantly different according to the Conover-Friedman test with Holm correction at $\alpha = 0.05$.}
    \label{fig:cd_dice}
\end{figure*}

\begin{figure*}[!htbp]
    \centering
    \includegraphics[width=0.95\textwidth]{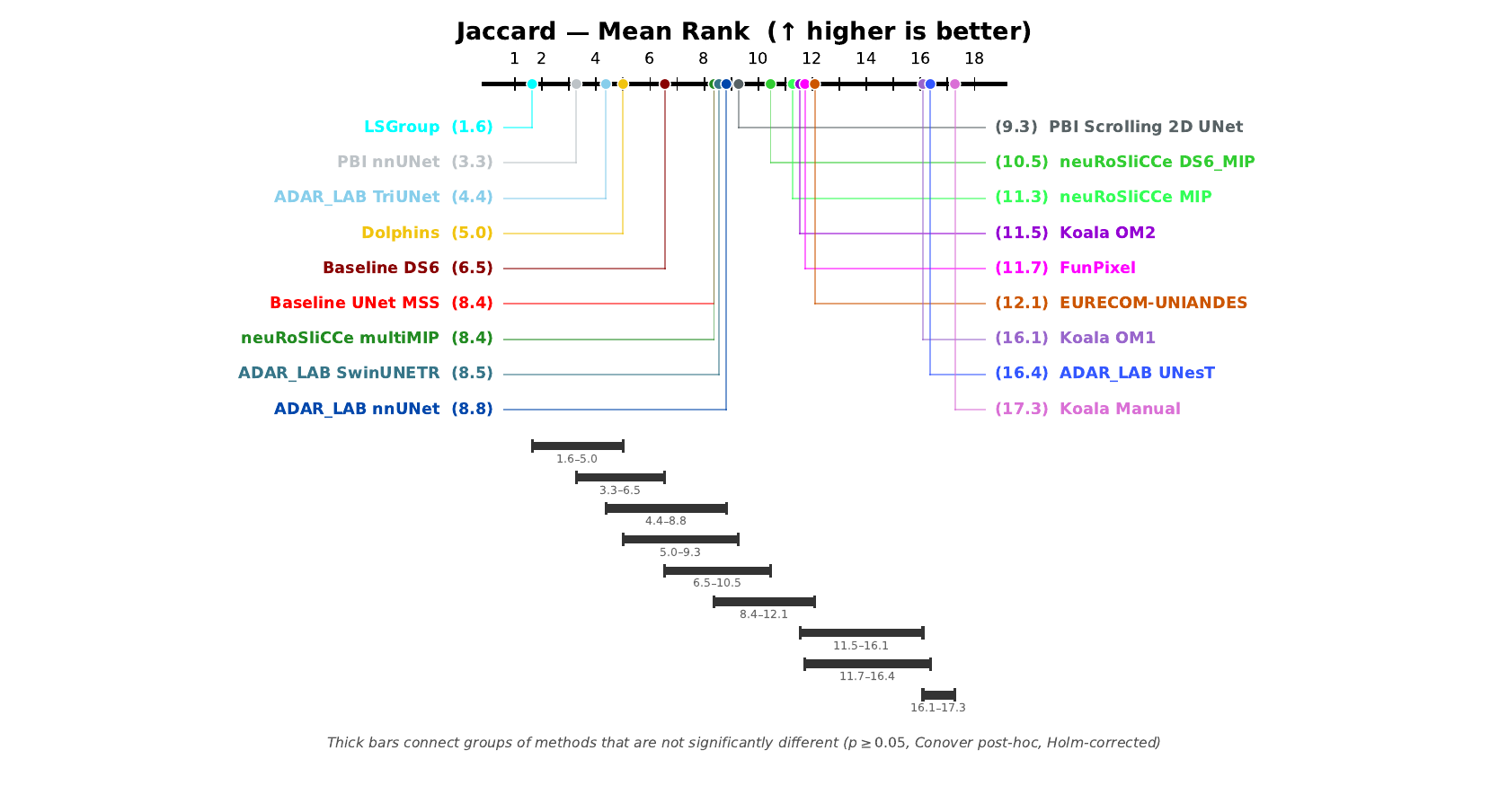}
    \caption{Critical Difference diagram for the Jaccard index on the combined dataset ($N = 11$). Clique bars show methods that are not significantly different after Conover-Friedman with Holm correction. As Jaccard is a monotone function of Dice, the per-subject ranking is identical to that in Figure~\ref{fig:cd_dice}.}
    \label{fig:cd_jaccard}
\end{figure*}

\begin{figure*}[!htbp]
    \centering
    \includegraphics[width=0.95\textwidth]{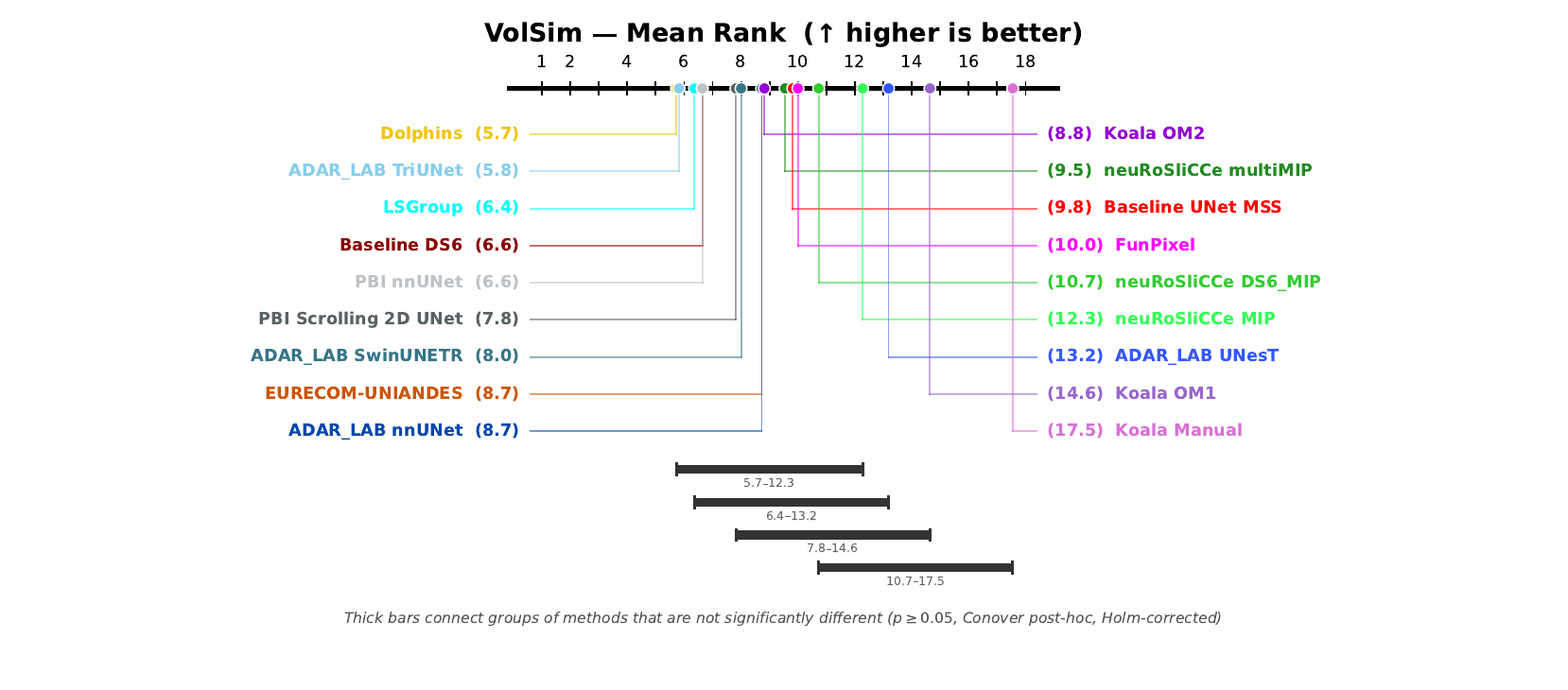}
    \caption{Critical Difference diagram for volumetric similarity on the combined dataset ($N = 11$). The comparatively long clique bars reflect the limited discriminative power of VolSim across methods observed in the Conover-Friedman analysis.}
    \label{fig:cd_volsim}
\end{figure*}

\begin{figure*}[!htbp]
    \centering
    \includegraphics[width=0.95\textwidth]{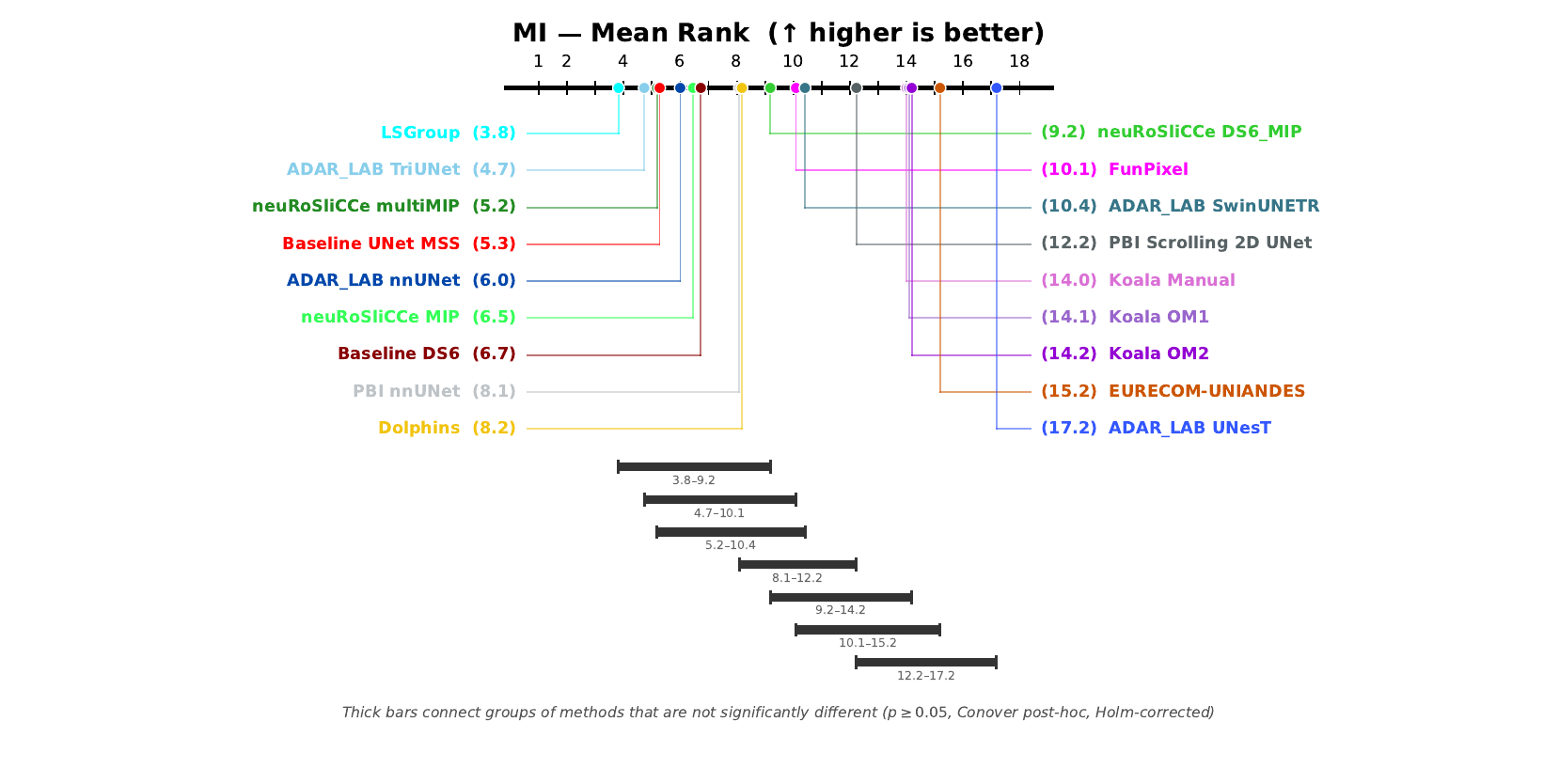}
    \caption{Critical Difference diagram for mutual information on the combined dataset ($N = 11$). While significant differences are present, the numerical spread of mean ranks is narrow, consistent with the narrow range of raw MI scores reported in Table~\ref{tab:overall_performance_median_iqr}.}
    \label{fig:cd_mi}
\end{figure*}

\begin{figure*}[!htbp]
    \centering
    \includegraphics[width=0.95\textwidth]{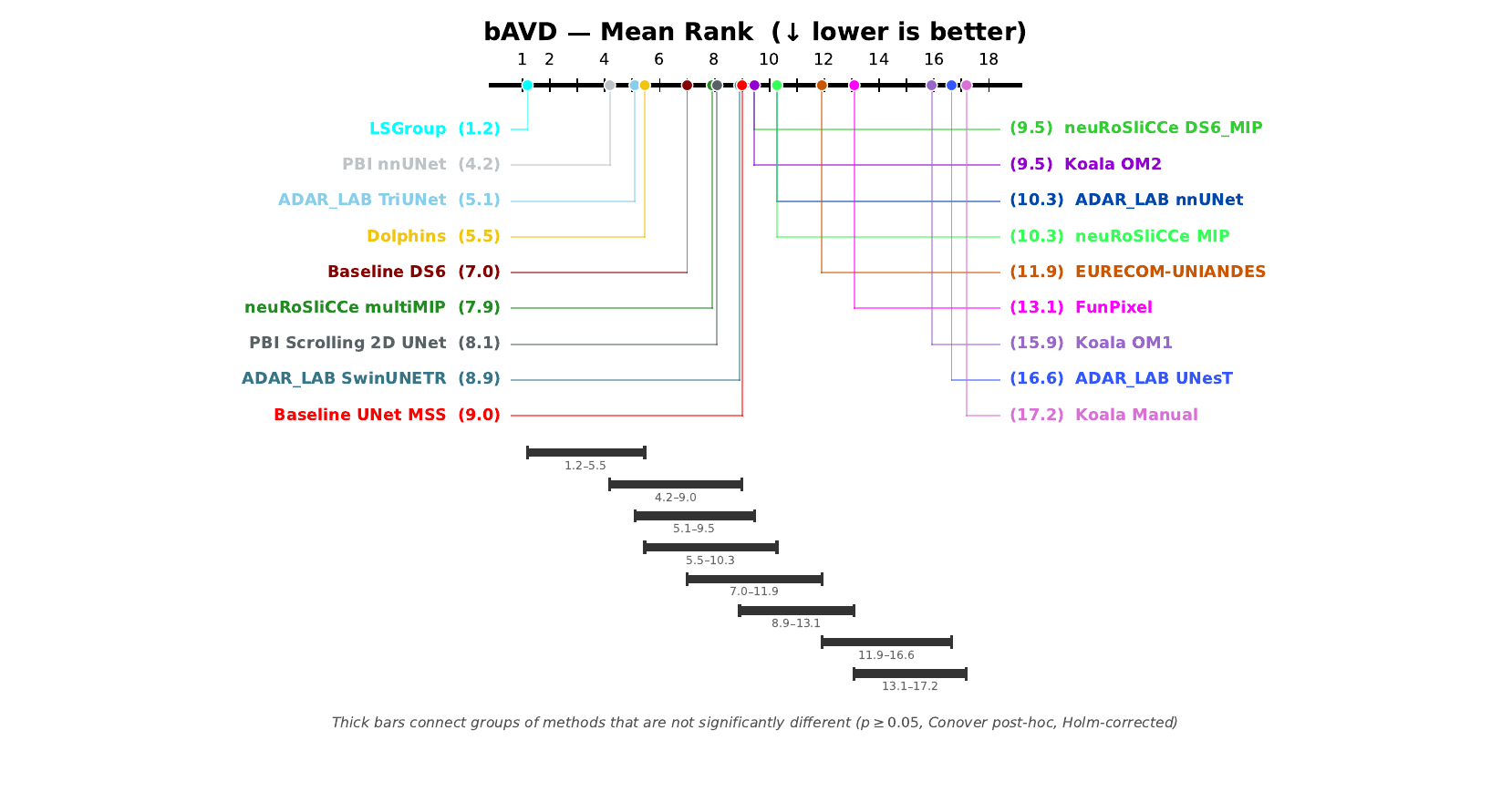}
    \caption{Critical Difference diagram for the balanced average Hausdorff distance on the combined dataset ($N = 11$). Lower mean rank corresponds to smaller bAVD, i.e.\ better boundary agreement.}
    \label{fig:cd_bavd}
\end{figure*}

\paragraph{Critical Difference diagrams.} To communicate both the overall ranking of methods and the groups that cannot be statistically separated, Critical Difference diagrams were produced in the style of Dem\v{s}ar~\cite{demsar2006statistical} and are shown in Figures~\ref{fig:cd_dice} to~\ref{fig:cd_bavd}. Across the overlap metrics (Dice, Jaccard), \lsgroup[0] occupies the best mean rank, followed by \pbinnUNet[0] and \dolphins[0]; all three are joined by a clique bar to \baseDSSx[0], indicating that at $N = 11$ none of these four methods can be declared significantly different on Dice or Jaccard after Holm correction. The lower end of the ranking is occupied by \ADARUNesT[0], \koalaMan[0] and \koalaOne[0], which are themselves grouped by a clique bar but are significantly worse than the top group. The bAVD diagram (lower is better) tells a broadly consistent story, with \lsgroup[0], \pbinnUNet[0], \dolphins[0] and the baselines jointly occupying the leading region and the same three low-performing methods trailing with significantly larger boundary errors. The VolSim diagram exhibits markedly longer clique bars than the other four, a direct graphical counterpart of its smaller number of significant pairs: most methods produce segmentations whose total volume is statistically indistinguishable from that of the baselines. The MI diagram shows moderate separation but confirms the interpretation raised earlier that MI differences, while statistically detectable, are numerically small and of limited discriminative value for ranking purposes.

\paragraph{Wilcoxon signed-rank tests against the baselines.} To isolate the comparisons of direct practical interest, each non-baseline method was compared against the two challenge baselines on every metric using the Wilcoxon signed-rank test with Holm correction within each metric (Tables~\ref{tab:wilcoxon_vs_Baseline_DS6} and~\ref{tab:wilcoxon_vs_Baseline_UNet_MSS}). The direction of any significant difference was read from the corresponding medians in Table~\ref{tab:overall_performance_median_iqr}.

Against \baseDSSx[0], only \lsgroup[0] and \pbinnUNet[0] achieved significantly higher Dice and Jaccard, with \lsgroup[0] also showing significantly lower bAVD; these are the only submitted methods for which there is statistical evidence of an improvement over this baseline on both overlap and boundary metrics. \dolphins[0] and \ADARTriUNet[0], although numerically above \baseDSSx[0] on Dice, did not reach Holm-corrected significance at $N = 11$; their apparent advantages should therefore be regarded as not yet established in a formal statistical sense. At the opposite end, \koalaMan[0], \ADARUNesT[0] and \koalaOne[0] were significantly worse than \baseDSSx[0] on Dice, Jaccard and bAVD, confirming that their relative under-performance is not attributable to sampling variability. \koalaTwo[0] and \neuroDSMIP[0] also showed significantly lower Dice than \baseDSSx[0], consistent with their below-baseline medians.

The same picture largely holds against \baseUNetMSS[0] (Table~\ref{tab:wilcoxon_vs_Baseline_UNet_MSS}): \lsgroup[0] and \pbinnUNet[0] once again achieve significantly higher Dice, Jaccard and lower bAVD, with \dolphins[0] additionally showing a significantly lower bAVD. The three lowest-performing methods (\koalaMan[0], \ADARUNesT[0], \koalaOne[0]) are again significantly worse on the overlap and boundary metrics. VolSim yielded few significant decisions, consistent with the CD diagram, and MI detected isolated significant differences (e.g.\ \koalaTwo[0], \eurecom[0] and \ADARUNesT[0] against both baselines) that reflect small absolute shifts in a narrow score range.

\input{Tables/Stats/wilcoxon_vs_Baseline_DS6}
\input{Tables/Stats/wilcoxon_vs_Baseline_UNet_MSS}

\paragraph{Summary.} On the whole, the statistical analysis singles out \lsgroup[0] and \pbinnUNet[0] as the only two submitted methods with formal evidence of improvement over the challenge baselines on multiple metrics; \dolphins[0] sits just behind, within the top clique on Dice and Jaccard and matching or exceeding the baselines on bAVD. At the opposite end, the evidence is clear that \koalaMan[0], \ADARUNesT[0] and \koalaOne[0] are worse than the baselines, in agreement with the median-based ranking. For the methods in the middle of the league table the tests advise caution: many of the apparent differences near the centre of the ranking cannot be resolved at $N = 11$ once multiplicity is properly controlled.

\subsection{Qualitative Results}

The qualitative analysis of the segmentation methods was conducted based on expert ratings ranging from 0 to 5, where 0 indicates poor performance and 5 denotes excellent performance. The experts evaluated each method on two key aspects: \textit{Q1}, small vessel segmentation performance, and \textit{Q2}, noise-free segmentation. The evaluations were performed separately on the open dataset and the secret dataset and the results provide insights into how each method performs in capturing fine vascular structures and producing clean segmentation outputs. During rating, names of the segmentation methods were pseudonymised to prevent bias. The intraclass correlation coefficient (ICC), specifically the ICC(2,k) model, was employed to evaluate inter-rater agreement. This model assesses absolute agreement in a two-way random-effects framework, accounting for both consistency and absolute agreement among multiple raters. The scores, together with the 95\% confidence intervals, are presented in Table \ref{tab:icc}. Overall, inter-rater agreement varied across datasets, ranging from poor to good, with the secret dataset demonstrating higher agreement than the open dataset. There are several factors to consider when interpreting the agreement: i) Like automatic small vessel segmentation itself, qualitative rating of small vessel segmentation is non-trivial. While instructions on the rating procedure were provided to the raters, individual preferences might have caused absolute agreement, i.e. same score for same algorithm, to be lower, while overall trends, i.e. higher scores for same algorithms, prevailed. ii) The agreement in the secret dataset was higher because overall more variance in small vessel segmentation performance was observed. This rendered difference visually more striking than in the open dataset and, therefore, made differentiating the segmentations qualitatively easier and better reproducible across raters. Between the two evaluation questions, \textit{Q2} yielded higher ICC values for both datasets, as it is inherently less subjective to evaluate whether a segmentation is noisy than to assess the segmentation quality of small vessels. The ratings assigned to each volume-method combination by the three raters were averaged to mitigate rater biases, thereby yielding a single score for each volume-method. Additionally, the standard deviation of the ratings is provided. Note that overall, standard deviation of the scores was lower for well-performing algorithms. These consolidated scores are presented in Table \ref{tab:qualitative_rating_MD}, organised across datasets and evaluation questions.

\begin{table}[!tbp]
\centering
\caption{Inter-rater agreement evaluated using ICC(2,k)}
\label{tab:icc}
\begin{tabular}{@{}c c c c@{}}
\toprule
\begin{tabular}[c]{@{}c@{}}Evaluation \\ Criteria\end{tabular} & Dataset & ICC  & 95\% CI          \\ \midrule
Q1                                                             & Open         & 0.36 & {[}0.12, 0.54{]} \\
Q2                                                             & Open         & 0.53 & {[}0.13, 0.73{]} \\
Q1                                                             & Secret       & 0.45 & {[}0.23, 0.60{]} \\
Q2                                                             & Secret       & 0.70 & {[}0.43, 0.83{]} \\ \bottomrule
\end{tabular}
\end{table}

\subsubsection{Open Dataset}
\paragraph{Q1: Small Vessel Segmentation Performance}

The baseline methods, \baseDSSx[0] and \baseUNetMSS[0], achieved median scores of 3.5 ± 0.33 and 3.5 ± 0.58 respectively for small vessel segmentation on the open dataset. These scores indicate a moderate to good ability of the baselines to segment small vessels accurately.

Among the evaluated methods, \neuromultiMIP[0] outperformed the baselines with the highest median score of 4.0 ± 0.25. This suggests that \neuromultiMIP[0] was particularly effective in capturing fine vascular structures on the open dataset. The low IQR of 0.25 indicates consistent performance across different samples.

Methods such as \neuroDSMIP[0], \dolphins[0], and \neuroMIP[0] achieved median scores comparable to the baselines, with scores of 3.5 ± 0.33, 3.5 ± 0.33 and 3.33 ± 0.75, respectively. \pbinnUNet[0] and \lsgroup[0] scored slightly below the baselines with median scores of 3.33 ± 0.0 and 3.17 ± 0.33 respectively, indicating performance close to that of the baselines but with minor variability.

Several methods performed below the baseline level. \ADARnnUNet[0], \eurecom[0], and \ADARSwinUNETR[0] achieved median scores of 3.0, reflecting moderate performance but not surpassing the baselines. Methods such as \pbiScroll[0] and \koalaTwo[0] scored lower, with median scores of 2.83 ± 0.33 and 2.83 ± 0.58 respectively, suggesting challenges in accurately segmenting small vessels.

The lowest scores were observed for \ADARUNesT[0] and \koalaOne[0], with median scores of 1.17 ± 0.33 and 2.5 ± 1.58 respectively. These methods scored considerably lower than the baselines for this criterion; however, given that the manual reference itself is an imperfect gold standard for such fine vascular structures, these low qualitative scores should be interpreted as a reflection of limited agreement with the reference rather than necessarily indicating a lack of anatomically plausible segmentation.

\paragraph{Q2: Noise-Free Segmentation}
In terms of noise-free segmentation on the open dataset, the baseline methods, \baseDSSx[0] and \baseUNetMSS[0], achieved median scores of 3.3 ± 0.50 and 3.0 ± 0.25 respectively, reflecting moderate performance in producing moderately ``clean'' segmentation outputs without spurious artefacts, with \baseDSSx[0] performing slightly better than \baseUNetMSS[0].

\pbinnUNet[0], \lsgroup[0], and \dolphins[0] outperformed the baselines with median scores of 3.67 ± 0.0, 3.67 ± 0.0, and 3.67 ± 0.25 respectively. These results indicate that these methods were particularly effective in producing clean segmentation outputs with minimal noise or artefacts.

Methods such as \neuromultiMIP[0] matched \baseDSSx[0] with a median score of 3.33 ± 0.50, indicating comparable performance in noise-free segmentation. \neuroDSMIP[0], \pbiScroll[0], \eurecom[0], and \ADARTriUNet[0] achieved median scores of 3.0, similar to \baseUNetMSS[0].

Several methods scored below the baselines in noise-free segmentation. \ADARnnUNet[0] and \ADARSwinUNETR[0] received median scores of 2.83 ± 0.58 and 2.67 ± 0.0 respectively, suggesting some issues with noise in their segmentation outputs. \koalaTwo[0] and \funpixel[0] scored lower with median scores of 2.33 ± 0.25, indicating challenges in maintaining output cleanliness.

The lowest performance was observed for \koalaOne[0], \koalaMan[0], and \ADARUNesT[0], with median scores ranging from 1.17 ± 1.08 to 1.67 ± 0.25. These methods scored considerably below the baselines on this criterion.

\subsubsection{Secret Dataset}
\paragraph{Q1: Small Vessel Segmentation Performance}

While segmenting small vessels on the secret dataset, both baseline methods, \baseDSSx[0] and \baseUNetMSS[0], achieved the highest median scores of 3.67 ± 0.33, indicating strong performance in small vessel segmentation on this dataset.

The method \neuromultiMIP[0] matched the baselines with a median score of 3.67 ± 0.5, demonstrating effective segmentation of small vessels and good generalisation to data with different properties.

Methods such as \pbinnUNet[0], \lsgroup[0], \dolphins[0], \neuroDSMIP[0], \pbiScroll[0], \koalaTwo[0], and \neuroMIP[0] scored slightly below the baselines with median scores of 3.33, indicating performance close to that of the baselines but with some variability.

Lower scores were observed for \ADARTriUNet[0], \ADARnnUNet[0], \eurecom[0], and \ADARSwinUNETR[0], each with median scores of 3.0, suggesting moderate performance but not surpassing the baselines.

Methods that received substantially lower scores include \koalaOne[0], \ADARUNesT[0], \funpixel[0], and \koalaMan[0], with median scores ranging from 2.67 ± 0.5 to 1.33 ± 1.67. As noted above, these qualitative scores reflect the limited agreement with the imperfect manual reference and should not be read as a categorical statement about the anatomical validity of the produced segmentations on the secret dataset.

\paragraph{Q2: Noise-Free Segmentation}

For noise-free segmentation on the secret dataset, the baseline methods, \baseDSSx[0] and \baseUNetMSS[0], achieved median scores of 3.33 ± 0.67 and 3.33 ± 0.5 respectively, indicating moderate performance with some variability.

\pbinnUNet[0] and \lsgroup[0] outperformed the baselines with median scores of 3.67 ± 0.67 and 3.67 ± 0.5 respectively. This suggests that these methods were particularly adept at producing clean segmentation outputs on the secret dataset.

Methods such as \dolphins[0] and \koalaTwo[0] matched the baselines with median scores of 3.33, reflecting comparable effectiveness in noise-free segmentation.

Several methods performed below the baselines. \pbiScroll[0] and \ADARTriUNet[0] achieved median scores of 3.0, while \neuromultiMIP[0], \neuroDSMIP[0], \ADARnnUNet[0], and \neuroMIP[0] received scores ranging from 2.67 ± 0.33 to 3.0 ± 0.33. These results indicate challenges in maintaining output cleanliness on the secret dataset.

The lowest scores were observed for \eurecom[0], \ADARSwinUNETR[0], \koalaOne[0], \ADARUNesT[0], \funpixel[0], and \koalaMan[0], with median scores between 2.0 ± 0.17 and 0.33 ± 0.67. These methods scored considerably below the baselines in producing noise-free segmentations on the secret dataset.

\subsubsection{Overall Performance Across Datasets}
\paragraph{Q1: Small Vessel Segmentation Performance}

Across both datasets, the baseline methods maintained strong performance in small vessel segmentation, particularly on the secret dataset where they achieved the highest median scores. \neuromultiMIP[0] consistently matched or exceeded the baseline performance, indicating robust capability in capturing fine vascular structures across different datasets.

Methods such as \pbinnUNet[0], \lsgroup[0], \dolphins[0], and \neuroDSMIP[0] demonstrated performance close to the baselines on both datasets, suggesting stable but not superior small vessel segmentation abilities.

Several methods, including \ADARTriUNet[0], \ADARnnUNet[0], \eurecom[0], and \ADARSwinUNETR[0], consistently scored below the baselines, suggesting limitations in matching the manual reference, particularly when dealing with data possessing different properties from the training set.

Methods such as \ADARUNesT[0], \funpixel[0], \koalaOne[0], and \koalaMan[0] received the lowest median scores for small vessel segmentation on both datasets. Given that the manual reference itself is an imperfect gold standard for these fine vascular structures, these scores are best understood as reflecting deviation from that reference rather than an intrinsic failure to produce anatomically meaningful segmentations.

\paragraph{Q2: Noise-Free Segmentation}
For noise-free segmentation, the baseline methods provided consistent and moderate performance across both datasets. On the open dataset, methods like \lsgroup[0], \dolphins[0], and \pbinnUNet[0] outperformed the baselines, achieving higher median scores and demonstrating superior ability in producing clean segmentation outputs.

On the secret dataset, \pbinnUNet[0] and \lsgroup[0] again outperformed the baselines, indicating strong generalisation in maintaining output cleanliness across datasets. Methods such as \dolphins[0] and \koalaTwo[0] matched the baselines, suggesting reliable performance.

Several methods showed decreased performance on the secret dataset, including the \emph{neuRoSliCCe} variants, \ADARnnUNet[0], and \ADARSwinUNETR[0], indicating difficulties in producing noise-free segmentations when confronted with unfamiliar data.

Methods like \eurecom[0], \koalaOne[0], \ADARUNesT[0], \funpixel[0], and \koalaMan[0] consistently received the lowest median scores for noise-free segmentation across both datasets.

\paragraph{Conclusion of Qualitative Analysis} The qualitative evaluation reveals that while some methods demonstrate strengths in specific areas, the baseline methods remain strong contenders, particularly in small vessel segmentation on the secret dataset. \neuromultiMIP[0] consistently matched or outperformed the baselines in small vessel segmentation, indicating its potential effectiveness in capturing fine vascular details across different datasets. Methods such as \pbinnUNet[0] and \lsgroup[0] exhibited strong performance in noise-free segmentation, outperforming the baselines on both datasets. Their ability to produce clean segmentation outputs with minimal noise suggests robustness and adaptability to varying data properties. However, no single method consistently outperformed the baselines across all criteria and datasets. Several methods struggled with generalisation, particularly on the secret dataset, highlighting the challenges of developing segmentation algorithms that maintain high performance across diverse imaging conditions, and the inherent difficulty of qualitatively rating small vessel segmentation against an imperfect manual reference.

\subsubsection{Additional Challenge: 150$\mu m$}
\paragraph{}

An additional qualitative evaluation was conducted on an MRA volume with a resolution of 150 $\mu m$ - double the resolution of the volumes in both the open and secret datasets. Due to computational resource limitations provided for the challenge, only a limited number of methods, including the baseline models, were capable of segmenting this high-resolution volume. Models that have high computational requirements (e.g. nnUNet-based or Transformer-based methods) did not manage to run within the given computational limitations, while simpler methods (e.g. UNet MSS-based methods) did manage to run. The ratings are presented in Table \ref{tab:qualitative_150mm_rating_MD}. The inter-rater agreement, calculated using the ICC(2,k) model, was considerably higher in this instance, with values of 0.88 (CI: [0.54, 0.98]) for \textit{Q1} and 0.80 (CI: [0.27, 0.96]) for \textit{Q2}. However, it is important to note that there was only one volume, and only a few methods were involved in this evaluation.

In terms of small vessel segmentation performance (\textit{Q1}) on the higher-resolution volume, \baseDSSx[0] achieved a median score of 3.00, indicating moderate effectiveness in capturing small vessels at the increased resolution. This suggests that \baseDSSx[0] maintains reasonable performance even when processing higher-resolution data. \baseUNetMSS[0] received a lower score of 1.67, reflecting limited capability in segmenting small vessels at this resolution. This performance is considerably below that of \baseDSSx[0], indicating that \baseUNetMSS[0] may not be well suited to higher-resolution data in terms of capturing fine vascular details.

\eurecom[0] outperformed both baselines with the highest score of 3.67 for \textit{Q1}. This result indicates that \eurecom[0] was particularly effective in accurately segmenting small vessels on the higher-resolution volume. Its superior performance suggests that the method is adept at handling increased data complexity and capturing fine vascular structures.

\neuromultiMIP[0] received a score of 2.00, demonstrating some effectiveness in small vessel segmentation but not surpassing \baseDSSx[0]. \neuroDSMIP[0] and \neuroMIP[0] obtained scores of 1.67 and 1.00, respectively, indicating limited performance in small vessel segmentation at the higher resolution, underperforming compared to \baseDSSx[0] and \eurecom[0].

\baseDSSx[0] achieved the highest score of 3.67 for noise-free segmentation (\textit{Q2}), indicating good performance in producing clean segmentation outputs without excessive noise or artefacts. This suggests that \baseDSSx[0] is robust in maintaining output quality even with increased data resolution. \baseUNetMSS[0] received a score of 3.00, reflecting moderate performance in noise-free segmentation. While not as high as \baseDSSx[0], it indicates that \baseUNetMSS[0] can produce reasonably clean outputs at higher resolutions.

\neuromultiMIP[0] scored 3.33, closely matching \baseDSSx[0]. This result suggests that \neuromultiMIP[0] is effective in generating noise-free segmentations on high-resolution data, demonstrating strength in maintaining output cleanliness.

\neuroDSMIP[0] and \neuroMIP[0] obtained scores of 2.67, indicating moderate performance in noise-free segmentation but below that of the baselines and \neuromultiMIP[0]. \eurecom[0] received a lower score of 2.33 for \textit{Q2}, suggesting that, despite its strong performance in small vessel segmentation, it introduced more noise or artefacts into the segmentation outputs than the baselines and \neuromultiMIP[0]. This trade-off indicates that the method prioritised capturing fine details over maintaining noise-free outputs.

\funpixel[0] received a median score of 0.00 for both \textit{Q1} and \textit{Q2}, suggesting challenges when applying the model to higher resolutions than provided in the training data.

\paragraph{Final Conclusions Across Questions and Datasets}

The qualitative analysis reveals that while some methods demonstrate strong performance in specific areas, no single method consistently outperforms the baselines across all criteria and datasets. The baseline methods themselves exhibit solid and sometimes superior performance, particularly \baseUNetMSS[0], which achieved the highest median score for small vessel segmentation on the secret dataset.

Methods such as \neuroDSMIP[0] and \neuroMIP[0] show promise in small vessel segmentation, occasionally surpassing the baselines. However, their performance varies between datasets, indicating that their effectiveness may be influenced by the characteristics of the data.

In terms of noise-free segmentation, many methods match the baseline performance, suggesting that producing clean segmentation outputs is a common strength among the evaluated techniques. Nevertheless, some methods struggle with this aspect on the secret dataset, which may be attributed to differences in data properties that affect their ability to generalise.

The \koalaMan[0] method, while performing well in small vessel segmentation, particularly on the secret dataset, shows substantial variability and lower scores in noise-free segmentation. This highlights the trade-offs that may occur when prioritising fine-detail capture over output cleanliness, and is also consistent with the observation that the manual reference itself carries intrinsic variability in very small vessels.

Overall, the baselines provide a robust benchmark, and while certain methods can match or slightly exceed their performance in specific areas, none consistently outperform them across all criteria and datasets. The results underscore the challenges of developing segmentation methods that are both highly accurate in capturing small vessels and capable of producing noise-free outputs, especially when applied to datasets with varying properties.

Future work should focus on enhancing the generalisation capabilities of segmentation methods, ensuring that they maintain high performance levels across diverse datasets. Emphasising robustness and adaptability in method development will be crucial for advancing the practical applicability of deep learning models in medical image analysis.

\input{Tables/qualitative_MD.tex}

%% file: Tables/Forrest/median_iqr_ranked.tex
\begin{table*}[!htbp]
\centering
\caption{Performance metrics (Median ± IQR) on the held-out test subset from the open dataset. Methods are ordered by Dice. The arrows ($\uparrow$/$\downarrow$) next to each metric indicate only whether higher or lower values are preferable, not the ordering within that column.}
\label{tab:open_performance_median_iqr}
\begin{tabular}{llllll}
  \toprule
Method & Dice $\uparrow$ & Jaccard $\uparrow$ & VolSim $\uparrow$ & MI $\uparrow$ & bAVD $\downarrow$\\ 
  \midrule
Baseline DS6 & 0.808 ± 0.044 & 0.678 ± 0.061 & 0.902 ± 0.045 & 0.062 ± 0.002 & 0.482 ± 0.155 \\ 
  Baseline UNet MSS & 0.791 ± 0.039 & 0.655 ± 0.053 & 0.862 ± 0.03 & 0.062 ± 0.002 & 0.578 ± 0.18 \\ 
  \midrule
  ADAR\_LAB TriUNet & 0.838 ± 0.066 & 0.722 ± 0.096 & 0.959 ± 0.014 & 0.062 ± 0.003 & 0.314 ± 0.224 \\ 
  LSGroup & 0.837 ± 0.075 & 0.72 ± 0.11 & 0.968 ± 0.047 & 0.06 ± 0.004 & 0.309 ± 0.168 \\ 
  ADAR\_LAB nnUNet & 0.832 ± 0.07 & 0.713 ± 0.099 & 0.926 ± 0.047 & 0.062 ± 0.004 & 0.328 ± 0.441 \\ 
  ADAR\_LAB SwinUNETR & 0.832 ± 0.066 & 0.713 ± 0.093 & 0.941 ± 0.027 & 0.061 ± 0.003 & 0.382 ± 0.426 \\ 
  PBI Scrolling 2D UNet & 0.829 ± 0.058 & 0.708 ± 0.083 & 0.954 ± 0.031 & 0.06 ± 0.007 & 0.373 ± 0.2 \\ 
  FunPixel & 0.82 ± 0.045 & 0.694 ± 0.064 & 0.974 ± 0.042 & 0.059 ± 0.008 & 0.42 ± 0.373 \\ 
  PBI nnUNet & 0.825 ± 0.063 & 0.702 ± 0.091 & 0.959 ± 0.02 & 0.058 ± 0.005 & 0.442 ± 0.281 \\ 
  Dolphins & 0.805 ± 0.056 & 0.674 ± 0.078 & 0.955 ± 0.026 & 0.057 ± 0.003 & 0.479 ± 0.135 \\ 
  EURECOM-UNIANDES & 0.803 ± 0.043 & 0.67 ± 0.059 & 0.948 ± 0.023 & 0.056 ± 0.006 & 0.561 ± 0.256 \\ 
  neuRoSliCCe multiMIP & 0.783 ± 0.035 & 0.643 ± 0.046 & 0.859 ± 0.037 & 0.06 ± 0.003 & 0.598 ± 0.158 \\ 
  neuRoSliCCe DS6\_MIP & 0.768 ± 0.029 & 0.624 ± 0.038 & 0.838 ± 0.03 & 0.06 ± 0.003 & 0.656 ± 0.153 \\ 
  neuRoSliCCe MIP & 0.754 ± 0.02 & 0.605 ± 0.025 & 0.805 ± 0.039 & 0.06 ± 0.003 & 0.85 ± 0.262 \\ 
  Koala OM2 & 0.71 ± 0.044 & 0.55 ± 0.051 & 0.932 ± 0.049 & 0.049 ± 0.008 & 2.669 ± 1.597 \\ 
  ADAR\_LAB UNesT & 0.707 ± 0.093 & 0.547 ± 0.099 & 0.866 ± 0.087 & 0.051 ± 0.005 & 2.573 ± 2.531 \\ 
  Koala Manual & 0.653 ± 0.045 & 0.485 ± 0.051 & 0.771 ± 0.106 & 0.055 ± 0.005 & 3.285 ± 1.517 \\ 
  Koala OM1 & 0.546 ± 0.064 & 0.376 ± 0.061 & 0.637 ± 0.144 & 0.046 ± 0.003 & 8.728 ± 6.924 \\ 
   \bottomrule
\end{tabular}%
\end{table*}

%% file: Tables/HendrikROT/median_iqr_ranked.tex
\begin{table*}[!htbp]
\centering
\caption{Performance metrics (Median ± IQR) on the secret dataset. Methods are ordered by Dice. The arrows ($\uparrow$/$\downarrow$) next to each metric indicate only whether higher or lower values are preferable, not the ordering within that column.}
\label{tab:secret_performance_median_iqr}

\begin{tabular}{llllll}
  \toprule
Method & Dice $\uparrow$ & Jaccard $\uparrow$ & VolSim $\uparrow$ & MI $\uparrow$ & bAVD $\downarrow$\\ 
  \midrule
Baseline DS6 & 0.687 ± 0.125 & 0.523 ± 0.16 & 0.95 ± 0.142 & 0.025 ± 0.003 & 1.008 ± 0.495 \\ 
  Baseline UNet MSS & 0.692 ± 0.137 & 0.529 ± 0.171 & 0.899 ± 0.074 & 0.026 ± 0.003 & 1.142 ± 0.838 \\ 
  \midrule
  LSGroup & 0.716 ± 0.125 & 0.558 ± 0.168 & 0.93 ± 0.096 & 0.028 ± 0.004 & 0.73 ± 0.197 \\ 
  PBI nnUNet & 0.713 ± 0.111 & 0.554 ± 0.147 & 0.949 ± 0.162 & 0.026 ± 0.003 & 0.843 ± 0.164 \\ 
  Dolphins & 0.715 ± 0.103 & 0.556 ± 0.133 & 0.936 ± 0.127 & 0.026 ± 0.003 & 0.874 ± 0.165 \\ 
  ADAR\_LAB TriUNet & 0.71 ± 0.118 & 0.55 ± 0.149 & 0.917 ± 0.093 & 0.026 ± 0.004 & 0.947 ± 0.34 \\ 
  neuRoSliCCe multiMIP & 0.708 ± 0.116 & 0.548 ± 0.142 & 0.901 ± 0.065 & 0.027 ± 0.003 & 0.954 ± 0.286 \\ 
  neuRoSliCCe MIP & 0.705 ± 0.084 & 0.544 ± 0.101 & 0.876 ± 0.057 & 0.029 ± 0.004 & 1.065 ± 0.303 \\ 
  neuRoSliCCe DS6\_MIP & 0.697 ± 0.098 & 0.535 ± 0.117 & 0.899 ± 0.044 & 0.027 ± 0.003 & 1.126 ± 0.143 \\ 
  ADAR\_LAB nnUNet & 0.695 ± 0.093 & 0.532 ± 0.11 & 0.873 ± 0.048 & 0.026 ± 0.003 & 1.814 ± 1.407 \\ 
  ADAR\_LAB SwinUNETR & 0.667 ± 0.086 & 0.5 ± 0.102 & 0.902 ± 0.043 & 0.025 ± 0.003 & 1.878 ± 1.054 \\ 
  Koala OM2 & 0.654 ± 0.151 & 0.485 ± 0.185 & 0.858 ± 0.189 & 0.024 ± 0.004 & 0.896 ± 0.925 \\ 
  PBI Scrolling 2D UNet & 0.683 ± 0.068 & 0.518 ± 0.078 & 0.867 ± 0.085 & 0.025 ± 0.003 & 1.262 ± 3.01 \\ 
  Koala OM1 & 0.602 ± 0.197 & 0.43 ± 0.204 & 0.731 ± 0.235 & 0.026 ± 0.004 & 4.601 ± 5.207 \\ 
  EURECOM-UNIANDES & 0.642 ± 0.081 & 0.473 ± 0.084 & 0.824 ± 0.165 & 0.023 ± 0.003 & 4.31 ± 2.779 \\ 
  FunPixel & 0.598 ± 0.05 & 0.427 ± 0.05 & 0.685 ± 0.126 & 0.026 ± 0.003 & 5.879 ± 3.658 \\ 
  ADAR\_LAB UNesT & 0.542 ± 0.04 & 0.372 ± 0.038 & 0.757 ± 0.128 & 0.021 ± 0.001 & 7.955 ± 9.856 \\ 
  Koala Manual & 0.338 ± 0.107 & 0.203 ± 0.083 & 0.398 ± 0.145 & 0.023 ± 0.005 & 17.325 ± 14.446 \\ 
   \bottomrule
\end{tabular}
\end{table*}

%% file: Tables/AllRes/median_iqr_ranked.tex
\begin{table*}[!htbp]
\centering
\caption{Performance metrics (Median ± IQR) on both the datasets combined. Methods are ordered by Dice. The arrows ($\uparrow$/$\downarrow$) next to each metric indicate only whether higher or lower values are preferable, not the ordering within that column.}
\label{tab:overall_performance_median_iqr}
\begin{tabular}{llllll}
  \toprule
Method & Dice $\uparrow$ & Jaccard $\uparrow$ & VolSim $\uparrow$ & MI $\uparrow$ & bAVD $\downarrow$\\ 
  \midrule
Baseline DS6 & 0.784 ± 0.14 & 0.645 ± 0.182 & 0.914 ± 0.103 & 0.03 ± 0.033 & 0.886 ± 0.443 \\ 
  Baseline UNet MSS & 0.778 ± 0.129 & 0.636 ± 0.163 & 0.88 ± 0.078 & 0.03 ± 0.033 & 1.087 ± 0.482 \\ 
  \midrule
  LSGroup & 0.804 ± 0.15 & 0.672 ± 0.205 & 0.942 ± 0.076 & 0.032 ± 0.027 & 0.583 ± 0.38 \\ 
  PBI nnUNet & 0.787 ± 0.136 & 0.649 ± 0.183 & 0.955 ± 0.09 & 0.03 ± 0.027 & 0.749 ± 0.384 \\ 
  Dolphins & 0.784 ± 0.112 & 0.645 ± 0.147 & 0.943 ± 0.056 & 0.029 ± 0.027 & 0.822 ± 0.36 \\ 
  ADAR\_LAB TriUNet & 0.772 ± 0.135 & 0.628 ± 0.176 & 0.949 ± 0.05 & 0.029 ± 0.031 & 0.917 ± 0.529 \\ 
  neuRoSliCCe multiMIP & 0.757 ± 0.106 & 0.609 ± 0.134 & 0.896 ± 0.087 & 0.029 ± 0.031 & 0.908 ± 0.311 \\ 
  PBI Scrolling 2D UNet & 0.726 ± 0.14 & 0.57 ± 0.176 & 0.924 ± 0.106 & 0.027 ± 0.028 & 1.073 ± 1.536 \\ 
  neuRoSliCCe DS6\_MIP & 0.731 ± 0.097 & 0.576 ± 0.119 & 0.884 ± 0.076 & 0.028 ± 0.031 & 1.114 ± 0.406 \\ 
  neuRoSliCCe MIP & 0.717 ± 0.074 & 0.559 ± 0.091 & 0.851 ± 0.093 & 0.029 ± 0.03 & 1.047 ± 0.265 \\ 
  ADAR\_LAB nnUNet & 0.718 ± 0.134 & 0.56 ± 0.169 & 0.879 ± 0.069 & 0.028 ± 0.031 & 1.643 ± 1.505 \\ 
  ADAR\_LAB SwinUNETR & 0.715 ± 0.137 & 0.556 ± 0.171 & 0.919 ± 0.036 & 0.028 ± 0.031 & 1.508 ± 1.404 \\ 
  Koala OM2 & 0.703 ± 0.118 & 0.542 ± 0.141 & 0.929 ± 0.162 & 0.026 ± 0.019 & 1.138 ± 1.912 \\ 
  EURECOM-UNIANDES & 0.654 ± 0.172 & 0.486 ± 0.202 & 0.936 ± 0.153 & 0.025 ± 0.028 & 1.96 ± 3.636 \\ 
  FunPixel & 0.631 ± 0.177 & 0.461 ± 0.207 & 0.843 ± 0.291 & 0.028 ± 0.027 & 4.184 ± 5.953 \\ 
  Koala OM1 & 0.574 ± 0.123 & 0.403 ± 0.12 & 0.679 ± 0.254 & 0.027 ± 0.017 & 5.569 ± 6.348 \\ 
  ADAR\_LAB UNesT & 0.551 ± 0.149 & 0.38 ± 0.152 & 0.776 ± 0.181 & 0.023 ± 0.022 & 5.576 ± 8.189 \\ 
  Koala Manual & 0.49 ± 0.303 & 0.325 ± 0.266 & 0.504 ± 0.31 & 0.027 ± 0.029 & 11.104 ± 14.385 \\ 
   \bottomrule
\end{tabular}
\end{table*}

%% file: Tables/Stats/friedman_omnibus.tex
\begin{table}[!htbp]
\centering
\caption{Friedman omnibus test results per metric (18 methods, 11 subjects).}
\label{tab:friedman_omnibus}
\begin{tabular}{lrrl}
  \toprule
  Metric & $\chi^2$ & df & $p$-value \\
  \midrule
  Dice $\uparrow$ & 127.70 & 17 & \textbf{5.19e-19} \\
  Jaccard $\uparrow$ & 127.70 & 17 & \textbf{5.19e-19} \\
  VolSim $\uparrow$ & 67.54 & 17 & \textbf{5.70e-08} \\
  MI $\uparrow$ & 111.08 & 17 & \textbf{7.56e-16} \\
  bAVD $\downarrow$ & 121.79 & 17 & \textbf{7.01e-18} \\
  \bottomrule
\end{tabular}
\end{table}

%% file: Tables/Stats/wilcoxon_vs_Baseline_DS6.tex
\begin{table*}[!htbp]
\centering
\caption{Wilcoxon signed-rank $p$-values vs.\ Baseline DS6 (Holm-corrected). Bold indicates $p < 0.05$.}
\label{tab:wilcoxon_vs_Baseline_DS6}
\begin{tabular}{llllll}
  \toprule
  Method & Dice $\uparrow$ & Jaccard $\uparrow$ & VolSim $\uparrow$ & MI $\uparrow$ & bAVD $\downarrow$ \\
  \midrule
  LSGroup & \textbf{0.016} & \textbf{0.016} & 1.000 & 1.000 & \textbf{0.016} \\
  PBI nnUNet & \textbf{0.023} & \textbf{0.023} & 1.000 & 1.000 & 0.082 \\
  Dolphins & 0.737 & 0.874 & 1.000 & 1.000 & 0.107 \\
  ADAR\_LAB TriUNet & 0.737 & 0.874 & 1.000 & 1.000 & 0.699 \\
  neuRoSliCCe multiMIP & 0.130 & 0.096 & 1.000 & 1.000 & 0.826 \\
  neuRoSliCCe DS6\_MIP & \textbf{0.049} & \textbf{0.049} & 1.000 & 1.000 & 0.404 \\
  PBI Scrolling 2D UNet & 0.737 & 0.874 & 1.000 & 0.117 & 0.826 \\
  ADAR\_LAB nnUNet & 0.737 & 0.874 & 1.000 & 1.000 & 0.123 \\
  neuRoSliCCe MIP & 0.062 & \textbf{0.049} & 1.000 & 1.000 & 0.123 \\
  ADAR\_LAB SwinUNETR & 0.737 & 0.874 & 1.000 & 0.378 & 0.699 \\
  Koala OM2 & \textbf{0.023} & \textbf{0.023} & 1.000 & \textbf{0.027} & 0.615 \\
  EURECOM-UNIANDES & 0.062 & 0.055 & 1.000 & \textbf{0.016} & 0.123 \\
  FunPixel & 0.193 & 0.193 & 1.000 & 0.322 & 0.107 \\
  Koala OM1 & \textbf{0.016} & \textbf{0.016} & 0.103 & 0.204 & \textbf{0.016} \\
  ADAR\_LAB UNesT & \textbf{0.016} & \textbf{0.016} & 0.191 & \textbf{0.016} & \textbf{0.016} \\
  Koala Manual & \textbf{0.016} & \textbf{0.016} & \textbf{0.016} & \textbf{0.027} & \textbf{0.016} \\
  \bottomrule
\end{tabular}
\end{table*}

%% file: Tables/Stats/wilcoxon_vs_Baseline_UNet_MSS.tex
\begin{table*}[!htbp]
\centering
\caption{Wilcoxon signed-rank $p$-values vs.\ Baseline UNet MSS (Holm-corrected). Bold indicates $p < 0.05$.}
\label{tab:wilcoxon_vs_Baseline_UNet_MSS}
\begin{tabular}{llllll}
  \toprule
  Method & Dice $\uparrow$ & Jaccard $\uparrow$ & VolSim $\uparrow$ & MI $\uparrow$ & bAVD $\downarrow$ \\
  \midrule
  LSGroup & \textbf{0.016} & \textbf{0.016} & 0.913 & 1.000 & \textbf{0.016} \\
  PBI nnUNet & \textbf{0.016} & \textbf{0.016} & 0.913 & 1.000 & \textbf{0.023} \\
  Dolphins & 0.098 & 0.123 & 0.317 & 1.000 & \textbf{0.023} \\
  ADAR\_LAB TriUNet & 0.054 & 0.054 & 0.260 & 1.000 & 0.258 \\
  neuRoSliCCe multiMIP & 1.000 & 1.000 & 1.000 & 1.000 & 1.000 \\
  neuRoSliCCe DS6\_MIP & 0.098 & 0.098 & 1.000 & 0.336 & 1.000 \\
  PBI Scrolling 2D UNet & 1.000 & 1.000 & 1.000 & 0.107 & 1.000 \\
  ADAR\_LAB nnUNet & 1.000 & 1.000 & 1.000 & 1.000 & 1.000 \\
  neuRoSliCCe MIP & 0.109 & 0.123 & 1.000 & 1.000 & 1.000 \\
  ADAR\_LAB SwinUNETR & 1.000 & 1.000 & 0.914 & 0.137 & 1.000 \\
  Koala OM2 & 0.109 & 0.123 & 1.000 & \textbf{0.027} & 1.000 \\
  EURECOM-UNIANDES & 0.322 & 0.322 & 1.000 & \textbf{0.016} & 0.167 \\
  FunPixel & 0.322 & 0.322 & 1.000 & 0.167 & 0.098 \\
  Koala OM1 & \textbf{0.016} & \textbf{0.016} & 0.146 & \textbf{0.035} & \textbf{0.016} \\
  ADAR\_LAB UNesT & \textbf{0.016} & \textbf{0.016} & 0.645 & \textbf{0.016} & \textbf{0.016} \\
  Koala Manual & \textbf{0.016} & \textbf{0.016} & \textbf{0.016} & \textbf{0.027} & \textbf{0.016} \\
  \bottomrule
\end{tabular}
\end{table*}

%% file: Tables/qualitative_MD.tex
\begin{table*}[!htbp]
    \centering
    \caption{Expert rating (Median ± IQR) between 0 (unacceptable) to 5 (excellent) on open and secret datasets. \textbf{Q1}: Small vessel segmentation performance and \textbf{Q2}: Noise-free segmentation}
    \label{tab:qualitative_rating_MD}
    \begin{tabular}{@{}lcccc@{}}
\toprule
\multicolumn{1}{c}{}  & \multicolumn{2}{c}{Open Dataset} & \multicolumn{2}{c}{Secret Dataset} \\ \midrule
Method                & Q1              & Q2             & Q1               & Q2              \\ \midrule
Baseline DS6          & 3.50 ± 0.33     & 3.33 ± 0.50    & 3.67 ± 0.33      & 3.33 ± 0.67     \\
Baseline UNet MSS     & 3.50 ± 0.58     & 3.00 ± 0.25    & 3.67 ± 0.33      & 3.33 ± 0.50     \\ \midrule
PBI nnUNet            & 3.33 ± 0.00     & 3.67 ± 0.00    & 3.33 ± 0.33      & 3.67 ± 0.67     \\
neuRoSliCCe multiMIP  & 4.00 ± 0.25     & 3.33 ± 0.50    & 3.67 ± 0.50      & 3.00 ± 0.33     \\
LSGroup               & 3.17 ± 0.33     & 3.67 ± 0.00    & 3.33 ± 0.50      & 3.67 ± 0.50     \\
Dolphins              & 3.50 ± 0.33     & 3.67 ± 0.25    & 3.33 ± 0.50      & 3.33 ± 0.83     \\
neuRoSliCCe DS6 MIP   & 3.50 ± 0.33     & 3.00 ± 0.00    & 3.33 ± 0.33      & 2.67 ± 0.33     \\
PBI Scrolling 2D UNet & 2.83 ± 0.33     & 3.00 ± 0.50    & 3.33 ± 0.50      & 3.00 ± 0.67     \\
Koala OM2             & 2.83 ± 0.58     & 2.33 ± 0.25    & 3.33 ± 0.50      & 3.33 ± 0.33     \\
neuRoSliCCe MIP       & 3.33 ± 0.75     & 2.50 ± 0.33    & 3.33 ± 0.67      & 2.67 ± 0.33     \\
ADAR LAB TriUNet      & 2.67 ± 0.25     & 3.00 ± 0.25    & 3.00 ± 0.67      & 3.00 ± 0.50     \\
ADAR LAB nnUNet       & 3.00 ± 0.25     & 2.83 ± 0.58    & 3.00 ± 0.50      & 2.67 ± 0.50     \\
EURECOM-UNIANDES      & 3.00 ± 0.50     & 3.00 ± 0.50    & 3.00 ± 0.33      & 2.00 ± 0.17     \\
ADAR LAB SwinUNETR    & 3.00 ± 0.00     & 2.67 ± 0.00    & 3.00 ± 0.17      & 2.00 ± 0.50     \\
Koala OM1             & 2.50 ± 1.58     & 1.17 ± 1.08    & 2.67 ± 0.50      & 1.33 ± 1.17     \\
FunPixel              & 2.67 ± 0.50     & 2.33 ± 0.25    & 1.33 ± 1.67      & 0.67 ± 0.83     \\
ADAR LAB UNesT        & 1.17 ± 0.33     & 1.67 ± 0.25    & 1.67 ± 0.67      & 1.67 ± 0.50     \\
Koala Manual          & 2.67 ± 1.25     & 1.67 ± 1.00    & 1.33 ± 1.67      & 0.33 ± 0.67     \\ \bottomrule
\end{tabular}
\end{table*}

\begin{table}[!htbp]
    \centering
\caption{Expert rating between 0 (unacceptable) to 5 (excellent) on an additional challenge volume with double the resolution. \textbf{Q1}: Small vessel segmentation performance and \textbf{Q2}: Noise-free segmentation}
\label{tab:qualitative_150mm_rating_MD}
\begin{tabular}{@{}lrr@{}}
\toprule
Method               & \multicolumn{1}{c}{Q1} & \multicolumn{1}{c}{Q2} \\ \midrule
Baseline DS6         & 3.00                   & 3.67                   \\
Baseline UNet MSS    & 1.67                   & 3.00                   \\ \midrule
EURECOM-UNIANDES     & 3.67                   & 2.33                   \\
neuRoSliCCe multiMIP & 2.00                   & 3.33                   \\
neuRoSliCCe DS6 MIP  & 1.67                   & 2.67                   \\
neuRoSliCCe MIP      & 1.00                   & 2.67                   \\
FunPixel             & 0.00                   & 0.00                   \\ \bottomrule
\end{tabular}
\end{table}

%% file: Sections/6_discussion.tex
The results of the SMILE-UHURA challenge offer a number of instructive findings regarding the segmentation of mesoscopic vessels from ultra-high-resolution 7T ToF-MRA. Rather than a single clear winner, the benchmarking reveals a more complex picture: the relative ranking of methods shifts depending on the dataset, the evaluation modality (quantitative versus qualitative), and the specific aspect of segmentation quality under scrutiny. In this section, we examine the principal themes that emerge from these results.

\subsection{Quantitative versus qualitative evaluation discrepancies}

One of the most striking observations from this challenge is the partial disagreement between the quantitative metrics and the expert ratings. Methods that dominated the quantitative rankings did not always fare as well in the eyes of the expert raters, and the converse was equally true.

Consider, for instance, the \neuromultiMIP[0] method. It achieved the highest qualitative score for small vessel segmentation on the open dataset (Q1 median of 4.0), outperforming every other method including \lsgroup[0] and \ADARTriUNet[0], which led the quantitative league table on that same dataset. Yet its Dice score of $0.783 \pm 0.035$ was below the baselines. A plausible explanation lies in what each evaluation modality is actually measuring. The Dice coefficient and Jaccard index are voxel-level overlap metrics that penalise both false positives and false negatives in equal measure. A method that detects a great many small vessels -- precisely the property valued in Q1 -- but does so with some degree of over-segmentation or slight positional imprecision will incur a Dice penalty that does not necessarily correspond to a loss of clinical utility when viewed as a Maximum Intensity Projection. The MIP-based qualitative assessment effectively collapses the 3D volume into a 2D projection, which can mask small boundary errors that are conspicuous in a voxel-wise computation. In this setting, a method that preserves vessel continuity and captures fine branches will appear superior to one that is more conservative but has higher overlap scores.

The converse is illustrated by \ADARTriUNet[0], which topped the quantitative rankings on the open dataset (Dice $0.838 \pm 0.066$) but received a modest Q1 score of only $2.67 \pm 0.25$ on that same dataset, falling below the baselines. This disparity suggests that TriUNet's high overlap with the ground truth arises in part from accurate delineation of larger and medium-calibre vessels, which dominate the voxel count and therefore the Dice score, whilst its performance on the smallest vessels -- those with apparent diameters of only 1--2 voxels, which are the specific focus of Q1 -- is less compelling. The qualitative raters, tasked explicitly with assessing small vessel depiction, were evidently unimpressed by what the Dice coefficient regarded as a strong result.

This quantitative-qualitative discrepancy is not merely a methodological curiosity. It carries practical implications for challenge design and, more broadly, for the evaluation of segmentation algorithms intended for clinical use. The Dice coefficient, for all its ubiquity, is a rather blunt instrument when the structures of interest occupy a very small fraction of the image volume. In the SMILE-UHURA dataset, small vessels constitute only a modest proportion of the total vessel mask; consequently, a method could achieve a high Dice score whilst neglecting a substantial fraction of the mesoscopic vasculature. Metrics that are more sensitive to topological correctness, vessel connectivity, or branch-level recall may be needed to complement overlap-based evaluation in future iterations of this challenge.

The inter-rater agreement scores (Table~6) also warrant comment. ICC values for Q1 ranged from 0.36 on the open dataset to 0.45 on the secret dataset, reflecting the inherent difficulty of assessing small vessel quality subjectively. Agreement was higher for Q2 (noise-free segmentation), with ICC reaching 0.70 on the secret dataset, consistent with the expectation that noise is easier to identify consistently than subtle differences in small vessel depiction. The somewhat low agreement on Q1 suggests that the qualitative rankings should be interpreted with appropriate caution, though the broad trends across raters were consistent.

\subsection{Architectural insights and generalisation}

The performance gap between the open and secret datasets provides a useful window into the generalisation characteristics of the competing methods. On the open dataset, the top methods achieved Dice scores in the range of 0.83--0.84, whilst on the secret dataset the best score was $0.716 \pm 0.125$ (\lsgroup[0]). This drop of roughly twelve percentage points is substantial and reflects the distinct properties of the secret dataset: different acquisition parameters, prospective motion correction, and sparse venous saturation that were absent from the training data.

What is instructive is which methods weathered this domain shift most gracefully. \lsgroup[0], \dolphins[0], and \pbinnUNet[0] maintained relatively stable rankings across datasets, whereas \ADARTriUNet[0] -- the top quantitative performer on the open dataset -- dropped noticeably, falling from a Dice of 0.838 to 0.710 on the secret data and yielding the overall lead to \lsgroup[0]. \ADARnnUNet[0] and \ADARSwinUNETR[0] similarly declined, with their Dice scores on the secret dataset falling below the baselines. The Wilcoxon signed-rank tests on the combined dataset (Section~\ref{sec:stat_results}) corroborate this ranking only in part: \lsgroup[0] and \pbinnUNet[0] are significantly better than both baselines on Dice and Jaccard after Holm correction, whereas \dolphins[0] and \ADARTriUNet[0], despite having higher median Dice than \baseUNetMSS[0], fall short of Holm-corrected significance at $N = 11$. Apparent ordering near the middle of the league table should therefore be read as a tendency rather than as an established ranking.

A pattern emerges here. Methods built upon nnUNet or employing self-configuring pipelines (\pbinnUNet[0], \lsgroup[0]'s MS-nnUNet) tended to generalise better, presumably because nnUNet's automated adaptation of preprocessing, patch sizes, and augmentation strategies confers a degree of resilience to variations in data properties. By contrast, methods that relied on fixed architectural configurations without such self-adaptation (e.g. the ADAR\_LAB transformer-based models, which used MONAI-default settings) were more vulnerable to domain shift. The \ADARSwinUNETR[0], for instance, yielded a competitive Dice of 0.832 on the open dataset but dropped to 0.667 on the secret dataset, falling below both baselines.

Transformer-based architectures showed an interesting split. \dolphins[0], which pairs a Swin Transformer encoder with a CNN-based decoder and cross-attention mechanism, generalised well (overall Dice 0.784, matching the \baseDSSx[0] baseline and exceeding \baseUNetMSS[0]). \ADARSwinUNETR[0], employing a broadly similar Swin Transformer encoder, did not generalise as well. The divergence likely reflects differences in training strategy, data augmentation, and perhaps the inductive biases introduced by \dolphins[0]' cross-attention windows, rather than a fundamental limitation of transformers per se. Nevertheless, the results do suggest that transformers trained on small datasets without careful regularisation are susceptible to overfitting, a concern that has been raised in the broader medical image segmentation literature.

These generalisation patterns are closely related to the architectural strategies that proved most effective overall. The two methods that most consistently appeared near the top of the quantitative rankings -- \lsgroup[0] and \ADARTriUNet[0] -- share a common principle: they aggregate information across multiple scales or multiple models. Yet their qualitative standing tells a rather different story. \lsgroup[0], which dominated quantitatively across both datasets (overall Dice $0.804 \pm 0.150$, lowest bAVD $0.583 \pm 0.380$), received a Q1 score of only $3.17 \pm 0.33$ on the open dataset -- below both baselines ($3.50$) and well below \neuromultiMIP[0] ($4.0$). On the secret dataset, \lsgroup[0]'s Q1 improved to $3.33$, though this still fell short of the baselines' $3.67$. Its noise-free segmentation scores (Q2) were stronger, reaching $3.67$ on both datasets, suggesting that \lsgroup[0]'s principal advantage lies in volumetric accuracy and boundary precision rather than in the depiction of the very finest vessels. This pattern is the mirror image of the \neuromultiMIP[0] case discussed above, and it reinforces the point that quantitative and qualitative evaluations are measuring partially different aspects of segmentation quality.

\lsgroup[0]'s multi-scale aggregation block applies dilated convolutions at rates of 2, 4, 6, and 8 to the final decoder features of an nnUNet, producing feature maps with progressively larger receptive fields that are then concatenated. This design is well suited to the particular challenge of mesoscopic vessel segmentation, where vessels of interest span a wide range of apparent diameters (from a single voxel to several voxels) and where the same network must simultaneously handle large arteries and tiny lenticulostriate branches. By fusing features at multiple scales, MS-nnUNet avoids the trade-off that a fixed-receptive-field architecture must make between sensitivity to small structures and contextual awareness of larger ones.

\ADARTriUNet[0] takes the idea further by ensembling three distinct architectures -- nnUNet, SwinUNETR, and UNesT -- through multi-layer 3D convolutions. Each constituent model brings different inductive biases: nnUNet excels at self-configured convolutional segmentation, SwinUNETR captures long-range dependencies through shifted windows, and UNesT employs hierarchical transformers. The ensemble thus spans a broad space of learned representations. On the open dataset, where all three constituent models are individually strong, this strategy paid dividends handsomely, yielding the highest Dice. On the secret dataset, however, the advantage was partly eroded, possibly because if the individual models overfit to the training domain, their errors become correlated, diminishing the benefit of ensembling.

The success of these approaches aligns with a broader observation in the medical imaging literature: for tasks involving structures at multiple spatial scales, architectures that explicitly model multi-scale information tend to outperform those that do not. The \eurecom[0] method (JoB-VS), which also employs a multi-scale lattice structure, performed respectably on the open dataset (Dice 0.803) but suffered considerable degradation on the secret dataset (0.642), suggesting that multi-scale processing alone is insufficient without adequate generalisation strategies.

An alternative to full 3D multi-scale processing is to approximate volumetric context through multi-orientation 2D inference. The \pbiScroll[0] method, rather than processing the volume with 3D convolutions, slides a 2D UNet along six anatomical directions (AP, PA, LR, RL, IS, SI), combining the resulting predictions through summation and normalisation. On the open dataset, this approach performed well, achieving a Dice of $0.829 \pm 0.058$ -- above the baselines and close to the top-performing methods. On the secret dataset, however, performance dropped to $0.683 \pm 0.068$, and the combined bAVD was $1.073 \pm 1.536$, suggesting poor boundary precision. By aggregating 2D predictions from multiple orientations, the method implicitly reconstructs 3D context, but the reconstruction is imperfect: each 2D slice is processed without awareness of its neighbours, and the combination relies on simple averaging rather than learned 3D features. On in-distribution data, where the training set provides ample examples of the expected intensity profiles and vessel appearances, this approximation is adequate. On out-of-distribution data, the per-slice predictions become less reliable, and their combination amplifies rather than averages away the errors. Full 3D methods, such as nnUNet, are better placed to exploit volumetric context for disambiguation under domain shift. That said, the \pbiScroll[0] carries a practical advantage: it can process arbitrarily large volumes without the memory constraints that hamper 3D methods, a trade-off that will become increasingly relevant as the field moves towards higher resolutions.

If multi-orientation 2D inference represents one pragmatic compromise, cross-domain transfer learning represents another attempt to compensate for limited training data. \ADARUNesT[0] is the only method in the challenge that was adapted from a pre-trained model developed for a different organ and a different imaging modality: it was originally trained for renal structure segmentation in 3D CT and was fine-tuned here for cerebral vessel segmentation in 7T ToF-MRA. The transfer was unsuccessful, with UNesT achieving an overall Dice of only $0.551 \pm 0.149$ and a bAVD of $5.576 \pm 8.189$, placing it among the worst-performing methods. The features learnt for renal CT segmentation -- where the task involves relatively large, well-defined anatomical structures with high tissue contrast -- have limited relevance to the segmentation of sub-voxel vessels in noisy, low-contrast MRA data. The inductive biases of the pre-trained hierarchical transformer, tuned to the spatial scales and intensity distributions of renal CT, are unlikely to transfer usefully to a domain where the structures of interest are orders of magnitude smaller relative to the field of view. Fine-tuning on only 14 volumes from a single site was evidently insufficient to overcome this domain gap. The result contrasts with the \dolphins[0] method, which also employs a Swin Transformer but was trained from scratch. \dolphins[0] achieved an overall Dice of $0.784 \pm 0.112$, far exceeding UNesT, suggesting that for this particular task, training a transformer from scratch on domain-specific data is preferable to fine-tuning a model pre-trained on a dissimilar domain. Transfer learning, for all its success in natural image processing and in some areas of medical imaging, is not a universal remedy, particularly when the source and target domains differ substantially in image content, resolution, and the spatial scale of relevant features.

Against this backdrop of varying architectural sophistication, the performance of the baselines merits attention. The \baseDSSx[0] and \baseUNetMSS[0] baselines, whilst conceptually simple relative to the more elaborate submissions, performed respectably across both datasets. \baseDSSx[0] achieved an overall Dice of $0.784 \pm 0.140$, which placed it above roughly half of the submitted methods. On the qualitative assessment, the baselines were amongst the strongest performers for small vessel segmentation (Q1) on the secret dataset, matching or exceeding several more complex methods. This resilience -- particularly of \baseDSSx[0], with its deformation-aware semi-supervised learning -- suggests that for this particular task, the bottleneck may lie less in architectural complexity than in training strategy and data handling. \baseDSSx[0]'s equivariance to elastic deformations appears to confer a form of implicit regularisation that benefits generalisation, especially when the training set is small (14 volumes). The lesson is one that bears repeating: increased model complexity does not guarantee improved performance, particularly in data-scarce regimes.

\subsection{Training objectives: loss functions and MIP-based losses}

Vessel segmentation in ultra-high-resolution MRA is a task defined by severe class imbalance: the vascular structures of interest occupy a small fraction of the image volume, and small mesoscopic vessels constitute a yet smaller fraction of that. The choice of loss function, therefore, has a non-trivial bearing on what a network learns to prioritise.

The methods in this challenge employed a range of loss formulations (see Table~1). The baselines and neuRoSliCCe variants used the Focal Tversky loss, which extends the Tversky index with a focal parameter that down-weights easy examples and concentrates the learning signal on hard, misclassified voxels. This is, at least in principle, well suited to detecting thin vessels that might otherwise be overwhelmed by the background class. The Koala methods used the standard Tversky loss with $\alpha = 0.3$ and $\beta = 0.7$, which weights false negatives more heavily than false positives, encouraging sensitivity at the expense of specificity. Most other methods, including nnUNet variants, TriUNet, and SwinUNETR, opted for combinations of Dice loss with cross-entropy (DiceCELoss), a formulation that balances volumetric overlap with voxel-wise classification accuracy.

It is difficult to isolate the effect of the loss function from other design choices (architecture, patch size, augmentation), but certain patterns are suggestive. The methods using Focal Tversky loss (baselines, neuRoSliCCe) tended to produce segmentations that were more conservative in volume -- their volumetric similarity scores on the open dataset were lower (in the range of 0.80--0.90) compared with those using Dice+CE combinations, which often achieved volumetric similarities above 0.93. This is consistent with the focal weighting encouraging the network to be selective rather than expansive in its predictions. Conversely, the Dice+CE-trained methods tended towards higher volumetric similarity and, in several cases, higher Dice scores, possibly because the cross-entropy term provides a stronger per-voxel gradient signal that helps maintain recall for smaller structures.

The \funpixel[0] method introduced an over-segmentation Dice loss (OSD) alongside 2D Dice losses computed on MIPs across three axes. On the open dataset, \funpixel[0] achieved the highest volumetric similarity of any method ($0.974 \pm 0.042$) and a respectable Dice of $0.820 \pm 0.045$, suggesting that the OSD term successfully encouraged the network to err on the side of inclusion. Yet this same tendency proved disastrous on the secret dataset, where \funpixel[0]'s Dice fell to $0.598 \pm 0.050$ and its bAVD ballooned to $5.879 \pm 3.658$, indicating that the over-segmentation bias did not transfer well to data with different noise and contrast characteristics.

A particularly instructive variant of this theme is the incorporation of projection-based losses into the training objective. The neuRoSliCCe methods and, to some extent, \funpixel[0] represent an interesting departure from conventional loss formulations by incorporating Maximum Intensity Projection comparisons directly into the training objective. The idea is to penalise voxel-level errors in the 3D volume alongside discrepancies visible in 2D projections, thereby encouraging the network to preserve the projected appearance of vessels -- which is, in essence, a proxy for vessel continuity.

This design choice helps explain the otherwise puzzling qualitative performance of \neuromultiMIP[0]. On the open dataset, its Q1 score of 4.0 (the highest of any method) was achieved despite a below-baseline Dice of 0.783. The MIP loss explicitly trains the network to produce segmentations whose projections resemble those of the ground truth, and since the qualitative evaluation was itself conducted on MIPs, \neuromultiMIP[0] was, in a sense, directly optimised for the qualitative assessment criterion. The three-axis MIP loss in multiMIP (as opposed to single-axis in \neuroMIP[0]) further improved this effect, yielding better Q1 scores than the single-axis variant.

There is an instructive lesson here about the relationship between training objectives and evaluation criteria. A method that optimises for MIP fidelity will tend to preserve the apparent connectivity and completeness of vessels in projection, even if its voxel-level overlap is somewhat imprecise. Whether this constitutes a genuine advantage depends on the intended downstream application. For clinical visualisation based on MIPs, the neuRoSliCCe approach may be preferable; for volumetric analyses requiring voxel-accurate segmentations, the Dice-optimised methods remain more appropriate.

\subsection{The role of preprocessing and post-processing}

Beyond the core network architecture and training objective, the choices made before and after the forward pass also contributed meaningfully to the observed variation across methods.

The Koala methods are notable for being the only submissions to incorporate N4ITK bias field correction and non-local means (NLM) denoising as explicit preprocessing steps, a choice motivated by their optimisation for ultra-high-field, high-resolution imaging. At 7T, the receive coil sensitivity profile produces a pronounced smooth, low-frequency intensity inhomogeneity; N4ITK removes this receive bias and, as a welcome side-effect, simplifies the subsequent z-score standardisation, which can then be performed across the whole volume without having to account for spatially varying intensity. NLM denoising addresses the elevated thermal noise that is characteristic of high-resolution MRA acquisitions at 7T.

Whether this preprocessing pipeline contributed to the quantitative performance of the Koala submissions is not straightforward to establish, and the available evidence points in opposite directions. On one hand, in ToF-MRA the image contrast arises from flow-related enhancement rather than tissue relaxation properties, and spatial intensity variation reflects not only coil sensitivity but also the saturation state of inflowing blood, which varies with vessel orientation, flow velocity, and distance from the slab boundary. In principle, a global intensity-correction algorithm that was designed for structural MRI therefore carries some risk of smoothing over intensity variations that are in fact informative of vascular signal. During the development of the baselines for this challenge we observed, in preliminary experiments on the same data, a small degradation in Dice when bias correction was applied prior to training, which led us not to include it in the baseline preprocessing. On the other hand, the participants have noted that aggressive bias-field correction applied to T2*-weighted images at 7T did not remove vessels in their experience, and Saunders et al.~\cite{saunders2021tofmra} reported that N4 bias correction was not detrimental to any of the six segmentation algorithms they evaluated on TOF-MRA at 1.5T, though a net benefit was observed only for a subset. Taken together, there is no conclusive evidence that N4ITK preprocessing contributed to the lower scores of the Koala methods, and it should not be inferred that its inclusion is universally inadvisable for 7T MRA segmentation.

A factor whose contribution is rather more plausible is the aggressive downsampling to a fixed $64^3$ patch size, which substantially reduces the effective resolution relative to the native $480 \times 640 \times 163$ volumes. \koalaTwo[0], the best-performing of the three Koala variants, achieved an overall Dice of only $0.703 \pm 0.118$, below the baselines, and the severe reduction in spatial fidelity at this patch size is likely to have discarded fine-grained information that is essential for resolving mesoscopic vessels at 300~\textmu m isotropic resolution.

It is also worth noting that \koalaOne[0] and \koalaTwo[0] were trained on the OMELETTE-generated labels supplied as part of the SMILE-UHURA dataset~\cite{OMELETTE} rather than on the manual annotations. These automated labels, produced through Frangi filtering and hysteresis thresholding, are inherently noisier and less exhaustive than the manual ground truth, and training on such silver-standard data necessarily imposes a ceiling on achievable accuracy. This design choice nonetheless reflects a fully automated pipeline that requires no manual intervention at any stage, and the comparatively encouraging performance of \koalaTwo[0] demonstrates the viability of eventually moving away from manual labelling altogether. The gap between \koalaTwo[0] and \koalaOne[0] (overall Dice 0.703 versus 0.574) further reflects the different Frangi sensitivity settings used to generate their respective labels, with the gamma~$= 0.02$ setting of OM2 evidently producing a more reliable training signal than the gamma~$= 0.01$ setting of OM1.

The interaction between preprocessing and label quality is most starkly illustrated by the \koalaMan[0] result -- perhaps the most puzzling finding in the entire challenge. This method used the same 3D UNet architecture and the same preprocessing pipeline (including N4ITK correction, non-local means denoising, and downsampling to $64^3$) as \koalaOne[0] and \koalaTwo[0], but was trained on the primary manual annotations -- the very labels against which all methods are evaluated. One might reasonably expect that training on the ``correct'' labels would yield the best performance within the Koala family. Instead, \koalaMan[0] produced the worst results of any submitted method, with an overall Dice of only $0.490 \pm 0.303$ and a bAVD of $11.104 \pm 14.385$.

This paradox admits several explanations, which are not mutually exclusive. First, the downsampling to $64^3$ patches -- a severe reduction from the original volume dimensions of $480 \times 640 \times 163$ -- means that the network operates at a fraction of the native resolution. At this reduced resolution, fine manual annotations of 1--2 voxel-diameter vessels are degraded to the point where they may contribute more noise than signal to the training process. The OMELETTE labels, by contrast, were generated at the original resolution using a multi-scale Frangi filter that tends to produce smoother, more spatially coherent annotations of larger vessels. When downsampled, these smoother labels may remain more self-consistent than the fine-grained manual masks, paradoxically providing a more learnable training signal at the reduced resolution. Second, the manual labels, being more complete in their depiction of small vessels, present a more challenging learning target. If the network's capacity is insufficient to learn the full complexity of the manual annotations at the reduced resolution, it may converge to a poor solution, whereas the simpler OMELETTE labels offer a more tractable objective. Third, it is possible that implementation-specific factors (e.g. the probability threshold of 0.1 used in post-processing, or the test-time adaptation procedure) interacted differently with the three label types, amplifying errors for the manual-label variant. Without access to intermediate training diagnostics, one cannot determine the relative contribution of each factor. What the result demonstrates rather clearly, however, is that the quality of training labels does not straightforwardly translate into segmentation performance when the training pipeline introduces aggressive spatial degradation.

One further nuance is worth mentioning. The Koala family is explicitly tuned to detect and enhance small, faint vasculature, and a portion of what the quantitative metrics record as over-segmentation in Figure~\ref{fig:qual_open} may in fact correspond to vessel-like structures that are visible in the original MIP but that were not captured by the manual reference. The reference annotations, however carefully produced, cannot be expected to include every marginal small vessel at this resolution, and some of the Dice penalty incurred by \koalaMan[0] will reflect this limitation of the reference rather than true segmentation failure. Even so, the qualitative assessment suggests that this cannot be the whole story: the raters, who viewed the MIPs directly and were explicitly asked to penalise noise and spurious foreground (Q2) as well as small vessel coverage (Q1), assigned \koalaMan[0] median Q2 scores of only $1.67$ on the open dataset and $0.33$ on the secret dataset, amongst the lowest of any method. These scores indicate substantial visible noise rather than an abundance of plausible but unlabelled vessels, so the contribution of reference incompleteness, whilst real, is unlikely to account for more than a small portion of the observed performance gap.

Turning to post-processing, several methods applied thresholding to predicted probability maps, but with quite different thresholds: the Koala methods used a probability threshold of 0.1, whereas \pbiScroll[0] used 0.3. These choices have direct consequences for the sensitivity--specificity trade-off. A threshold of 0.1 retains many uncertain voxels, increasing recall but also introducing noise; a threshold of 0.3 is more conservative. The Koala methods additionally removed small connected components of fewer than ten voxels, a step intended to suppress isolated false positives but which could also eliminate genuine small vessel fragments. \pbiScroll[0] combined predictions from six directional passes through summation and normalisation before thresholding, an implicit form of consensus filtering. \funpixel[0] employed five-fold majority voting, largest connected component analysis, and custom thresholding.

These post-processing choices are rarely the focus of challenge discussions, yet they can substantially affect the final metrics. A method with an otherwise good volumetric segmentation could be degraded by an overly aggressive connected-component filter that removes small vessels, whilst a method with excessive false positives could be rescued by a more stringent threshold. The absence of standardised post-processing across submissions makes it difficult to determine how much of the observed performance variation is attributable to the segmentation network itself and how much to decisions made after the forward pass. Future iterations of the challenge might consider either standardising post-processing or requiring participants to submit raw probability maps in addition to binary masks.

\subsection{Domain shift: an MR physics perspective}

The performance degradation observed on the secret dataset cannot be attributed to a single factor; it arises from a confluence of acquisition differences that collectively alter the image characteristics in ways that a model trained exclusively on the open dataset may not anticipate.

The open dataset was acquired without prospective motion correction or venous saturation. The secret dataset employed both: prospective motion correction to reduce blurring from head movement, and sparse venous saturation to suppress venous contamination whilst remaining within specific absorption rate limits. Prospective motion correction sharpens the images and preserves small vessel signal that would otherwise be smeared by motion, potentially revealing vessels that are absent or attenuated in the open dataset. A model that has learnt the open dataset's particular pattern of vessel appearances -- including whatever blurring and venous contamination is present -- may misinterpret the sharper, cleaner signals in the secret data. Venous saturation, by selectively suppressing venous signal, changes the overall distribution of bright voxels in the volume, which could affect methods that rely on intensity-based features for vessel detection.

Additionally, the two datasets originate from different studies and different subject cohorts (average age 26 versus 30 years), with different slab configurations (52 versus 60 slices per slab). Whilst these differences are modest, they contribute to the overall distributional shift between the training and test domains. The methods that generalised best -- \lsgroup[0], \dolphins[0], \pbinnUNet[0] -- are characterised either by self-configuring preprocessing (nnUNet) or by architectural features that encourage invariance to such variations (\dolphins[0]' cross-attention mechanism, \lsgroup[0]'s multi-scale receptive fields). Methods with more rigid configurations were correspondingly more fragile.

\subsection{Evaluation metrics and their discriminative power}

Amongst the five quantitative metrics employed in this challenge, mutual information (MI) was the least informative for distinguishing between methods. On the open dataset, values clustered tightly around $0.060 \pm 0.003$ for most methods, and on the secret dataset around $0.026 \pm 0.003$. The top-performing methods and the baselines yielded virtually identical MI scores, and even some of the weaker methods fell within the same narrow range.

This insensitivity likely stems from the nature of mutual information as a global, information-theoretic measure. It quantifies the statistical dependence between two binary images but does not account for the spatial distribution of agreements and disagreements. In a segmentation task dominated by background voxels (which all methods classify correctly with near-perfect accuracy), the mutual information is driven largely by this shared background, diluting the discriminative signal from the foreground. Metrics such as Dice, Jaccard, and bAVD, which focus specifically on the foreground class or on boundary agreement, are better suited to the particular challenges of vessel segmentation, where the foreground occupies a small fraction of the volume.

The Conover-Friedman post-hoc analysis quantifies this concretely: on the combined dataset, Dice and bAVD yielded 76 and 66 significantly different method pairs (out of 153) respectively, MI only 59, and volumetric similarity a mere 20. VolSim thus also emerges as relatively weak in its ability to separate methods, which is perhaps unsurprising given that it is, by design, insensitive to the spatial arrangement of voxels and responds only to the net cardinality of the segmentation. For ranking purposes, Dice, Jaccard and bAVD are therefore the most informative of the five, a finding that matches their prominence in the wider segmentation literature.

This finding has implications for future challenge design. Including mutual information in the evaluation battery adds little value for binary vessel segmentation at this scale; replacing or supplementing it with topology-aware metrics (such as the Betti number error or centreline Dice) would more directly capture the structural properties of the vascular tree.

\subsection{Scalability: the 150~\textmu m evaluation}

The supplementary qualitative evaluation on a 150~\textmu m resolution volume, whilst limited in scope (only a handful of methods could process it within the computational constraints), raises an important point about scalability. Transformer-based and nnUNet-based methods, which produced competitive results on the 300~\textmu m data, were unable to run on the higher-resolution volume owing to memory and computational limitations. By contrast, simpler \baseUNetMSS[0]-based architectures could be applied, though with mixed results.

\eurecom[0] achieved the highest Q1 score (3.67) on this volume, outperforming both baselines and the neuRoSliCCe variants. This is worth remarking upon, given that \eurecom[0] was not amongst the top quantitative performers on the 300~\textmu m datasets. Its multi-scale lattice structure (JoB-VS) may be better suited to the increased spatial detail available at 150~\textmu m, where the additional resolution reveals finer vascular branches that benefit from explicit multi-scale processing. This observation suggests that the optimal method may depend on the imaging resolution, and that evaluations conducted at a single resolution do not fully characterise a method's potential.

\subsection{Clinical relevance and implications for CSVD research}

It is reasonable to ask what Dice scores of 0.838 (open dataset) and 0.716 (secret dataset) mean in practical terms. These figures, taken in isolation, place the best-performing methods at a level that is competitive with, though not yet matching, expert inter-observer agreement typically reported for macroscopic vessel segmentation tasks. However, direct comparison is somewhat misleading, because mesoscopic vessel segmentation at 300~\textmu m is an intrinsically harder task than macroscopic segmentation: the vessels of interest have apparent diameters of only 1--2 voxels, the contrast between vessel lumen and background is poor, and partial volume effects are substantial. A Dice of 0.84 in this regime represents a rather different level of performance from a Dice of 0.84 for, say, aortic segmentation from CT angiography.

Whether this level of accuracy is sufficient for clinical or research use depends on the application. For population-level studies of vascular morphology -- where the goal is to compare aggregate measures such as total vessel length, branch density, or fractal dimension across cohorts -- a Dice in the range of 0.80 is likely adequate, provided the errors are not systematically biased towards particular vessel sizes or anatomical regions. The high volumetric similarity achieved by the top methods ($>0.94$) is encouraging in this regard, as it suggests that the total volume of segmented vasculature is close to the ground truth even when voxel-level agreement is imperfect.

For individual-level diagnostic applications -- such as identifying specific lenticulostriate artery loss in a patient suspected of harbouring small vessel disease -- the requirements are more exacting. The qualitative results suggest that even the best methods miss some of the smallest vessels, and the bAVD values indicate boundary imprecision at the single-voxel level. Whether a clinician could reliably distinguish pathological vessel loss from segmentation error is an open question that warrants investigation in a clinical validation study, which lies beyond the scope of this challenge but is an important direction for subsequent work.

The twelve-percentage-point drop in Dice from the open to the secret dataset is perhaps the more clinically relevant finding. In any real deployment scenario, segmentation algorithms will encounter data from scanners, sites, and acquisition protocols that differ from the training set. A method that achieves 0.84 on in-distribution data but falls to 0.72 on slightly different data is not yet reliable enough for unsupervised clinical deployment. The results suggest that domain adaptation, test-time augmentation, or training on multi-site data will be necessary steps before these methods can be integrated into clinical workflows with confidence.

These accuracy figures take on additional significance in the context of cerebral small vessel disease research. CSVD is amongst the most common neurological conditions in ageing populations, and the segmentation and quantification of mesoscopic vessels is a prerequisite for studying its pathophysiology in vivo. The SMILE-UHURA challenge represents the first public benchmark for this task at 7T resolution, and the results have several implications for the research community.

First, the challenge establishes that deep learning methods can segment mesoscopic vessels from 7T ToF-MRA at a level that meaningfully exceeds what is achievable by naive thresholding or classical filter-based approaches. The fact that even the baseline methods, trained on only 14 volumes, achieve Dice scores around 0.78--0.80 suggests that the annotated dataset is sufficient to support useful model training, and that further expansion of the training set is likely to yield continued improvement.

Second, the results point towards specific methodological priorities for building downstream CSVD analysis pipelines. The strong performance of multi-scale approaches (\lsgroup[0], \ADARTriUNet[0]) confirms that capturing the full range of vessel calibres within a single model is both feasible and beneficial. For pipelines that need to quantify, for instance, the density of perforating arteries in the basal ganglia -- a measurement of direct relevance to small vessel disease grading -- the \neuromultiMIP[0] approach, with its emphasis on small vessel continuity in projection, may be preferable to methods that optimise purely for voxel overlap. The choice of segmentation method should, in other words, be guided by the specific downstream analysis rather than by a single aggregate metric.

Third, the results raise an important question about the relationship between automated segmentation performance and the quality of the manual reference standard. Manual annotation of mesoscopic vessels at 300~\textmu m isotropic resolution is itself a demanding and inherently subjective task: the smallest vessels have apparent diameters of only one to two voxels, partial volume effects are substantial, and the boundary between vessel lumen and background is frequently ambiguous. The best-performing methods in this challenge achieve Dice scores of 0.838 on the open dataset, which falls within the range commonly reported for inter-annotator agreement in comparable vascular segmentation tasks. This observation suggests that the top deep learning methods may be approaching the ceiling of what a single annotator's reference can reliably measure. A portion of the residual discrepancy between automated predictions and the ground truth may therefore reflect legitimate differences in the placement of ambiguous voxels, rather than outright segmentation failures. This does not diminish the value of the annotations as a benchmarking resource -- they provide a consistent and carefully verified reference against which all methods are evaluated on equal terms -- but it does mean that the reported metrics should be understood as quantifying agreement with one particular set of annotation decisions, not as measuring deviation from an objectively defined ground truth. The practical consequence is that further gains in Dice beyond the present range will be difficult to interpret without multiple independent annotations, a point we return to in the limitations and outlook below.

\subsection{Public availability and reproducibility of the benchmarked methods}

A practical consideration that deserves separate attention is the extent to which the benchmarked methods are available for the wider community to use, adapt and build upon. The long-term value of a benchmarking challenge lies in the artefacts it leaves behind as much as in the rankings it produces: datasets that remain open for future comparison, and implementations that subsequent researchers can apply to their own data, fine-tune for related tasks, or integrate into larger analysis pipelines. On the dataset side this is now well covered, with both the open dataset and the OMELETTE-generated silver-standard labels freely available, and the secret dataset retained for fair evaluation of future submissions. On the method side the picture is rather more uneven.

Participants were encouraged to accompany their submissions with public code repositories or containerised images, but the degree of compliance varied. Some teams provided complete, well-documented public releases, for example the Koala submissions, which have since been extended into the openly distributed VesselBoost toolbox with Docker, OpenRecon, and Neurodesk~\cite{renton2024neurodesk} deployments. Others contributed working code but with limited documentation or external dependencies that make reuse effortful, and a further subset did not release any public implementation, which effectively restricts the reproducibility of their results to the challenge environment itself. This uneven availability is not unusual for segmentation challenges, but it does limit the practical value of the benchmark for researchers who wish to apply the highest-ranking methods directly to their own ToF-MRA data.

To partially address this gap, the organisers have mirrored several of the submitted methods on their own institutional repositories and Docker registries, with appropriate attribution and with the consent of the respective teams, so that the full benchmark configuration remains retrievable even if a team's original repository becomes unavailable in future. This mirroring is not yet comprehensive; we have prioritised the methods that already included permissive licences and runnable packaging, and we intend to extend the effort to as many of the remaining submissions as the original authors will permit. Links to the mirrored resources, where available, are listed on the challenge page alongside pointers to the original repositories. Future iterations of the benchmark will make reproducible packaging a firmer expectation at submission time, encouraged by the availability of platforms such as Neurodesk that substantially reduce the effort required to distribute such packages.

\subsection{Limitations}

Several limitations of this challenge should be acknowledged, some of which bear directly on the interpretation of the results presented above.

The test sample sizes are small: the held-out test subset of the open dataset comprises only four MRA volumes, and the secret dataset comprises seven. With so few test subjects, the median and IQR values reported in the quantitative tables are estimated from very limited samples, and a single outlier volume can shift a method's apparent ranking substantially. The wide interquartile ranges observed for several methods (e.g.\ \koalaMan[0]'s Dice IQR of 0.303 on the combined evaluation) are symptomatic of this issue. For this reason, formal non-parametric significance testing (Friedman, Conover-Friedman and Wilcoxon signed-rank, all with Holm correction) was only performed on the combined dataset of $N = 11$ subjects, where the pooled sample size offered enough power to draw meaningful conclusions about differences between methods. Even at $N = 11$ the detection threshold is non-trivial, and non-significance should not be read as equivalence; the tests identify the clear winners and clear losers, but many methods sitting near the middle of the league table cannot be statistically separated from one another or from the baselines at this sample size. The per-dataset rankings, based only on four or seven subjects, should accordingly be regarded as indicative rather than definitive, and readers should attend to the spread of scores as much as to the medians.

Both datasets were acquired from healthy young adults, with mean ages of 26 years (open dataset) and 30 years (secret dataset). This demographic composition limits the generalisability of the findings to the populations where mesoscopic vessel segmentation is most clinically needed: older adults with, or at risk of, cerebral small vessel disease. Ageing is associated with changes in vascular morphology (increased tortuosity, reduced branch density), alterations in flow dynamics, and the accumulation of white matter lesions and microbleeds, all of which may affect the appearance of vessels in ToF-MRA and the difficulty of segmenting them. Whether the methods benchmarked here would maintain their relative rankings on data from elderly or pathological cohorts remains an open question.

The ground truth annotations were produced through a three-step process: automated pre-segmentation by thresholding, extensive manual refinement by a single annotator, and verification by a senior neurologist. Whilst this procedure is thorough, it does not address the question of inter-annotator variability. The manual refinement stage, in particular, involves subjective decisions about whether marginal voxels belong to vessel or background -- decisions that a second annotator might make differently, especially for the smallest vessels at the limit of visibility. Without a second independent annotation, one cannot estimate the inter-rater variability of the ground truth itself, nor determine whether the segmentation errors of the submitted methods fall within or beyond the range of plausible human disagreement. This is a meaningful gap: if, for example, the inter-annotator Dice for this task were around 0.85, then a method achieving a Dice of 0.84 against one annotator's ground truth would be performing at essentially human-level accuracy. Without such a reference, the absolute performance figures are difficult to contextualise. As noted above, the best methods in this challenge achieve Dice values of 0.838 on the open dataset, placing them squarely in the range where this consideration becomes practically relevant rather than merely theoretical. The quantitative rankings accordingly carry an inherent bias: any voxel where the annotator's boundary decision is debatable will be counted as an error for every submitted method, even those that may have made an equally defensible segmentation choice.

The qualitative evaluation was conducted using Maximum Intensity Projections rather than slice-by-slice volumetric inspection. This was a pragmatic compromise -- for each ToF volume, 18 segmentations required evaluation, and slice-by-slice review of the full high-resolution data would have been impractical. However, the use of MIPs introduces specific biases into the qualitative assessment. MIPs effectively collapse the 3D volume along one axis, which means that segmentation errors occurring in the through-projection direction (e.g. a vessel segment incorrectly placed one or two voxels deeper or shallower than the true position) are invisible in the projected image. Conversely, MIPs can exaggerate the appearance of false positives, since a single bright spurious voxel anywhere along the projection path will appear as prominently as a true vessel. The qualitative rankings may therefore overweight certain types of error and underweight others, relative to what a full volumetric assessment would reveal. The additional zoomed view of the Circle of Willis partially mitigates this issue for the lenticulostriate arteries, but does not address it for the broader vasculature.

The inter-rater agreement for the qualitative evaluation was variable, with ICC(2,k) values as low as 0.36 (95\% CI: [0.12, 0.54]) for Q1 on the open dataset. This level of agreement is conventionally classified as ``poor'' and means that a substantial portion of the variance in Q1 scores is attributable to rater disagreement rather than to genuine differences between methods. The agreement improved on the secret dataset (ICC 0.45 for Q1, 0.70 for Q2), plausibly because the greater variance in segmentation quality on that dataset made differences between methods more visually apparent. Q2 (noise assessment) consistently yielded higher agreement than Q1 (small vessel assessment), which is unsurprising: judging whether a segmentation is noisy is a more binary visual task than assessing the adequacy of small vessel depiction, which requires the rater to compare the segmentation against their own mental model of the expected vascular anatomy. The low Q1 agreement means that the qualitative rankings, particularly on the open dataset, should be treated as approximate guides to expert preference rather than as reliable orderings.

The computational constraints of the challenge environment (a single Nvidia A6000 GPU with 48~GB of memory, 64~GB of RAM) necessarily excluded methods or configurations that might have performed well with greater resources, creating a selection bias towards methods that could fit within the hardware budget.

Finally, the challenge evaluated methods on ToF-MRA data from 7T scanners exclusively. The extent to which these results transfer to other MRA sequences (e.g. phase-contrast MRA, black-blood MRI) or to lower field strengths (3T, where mesoscopic vessels are less well resolved) remains untested.

%% file: Sections/7_conclusion.tex
The SMILE-UHURA challenge demonstrates that reliable segmentation of mesoscopic cerebral vessels from 7T ToF-MRA is achievable using current deep learning methods, with top Dice scores of 0.838 on in-distribution data and 0.716 on out-of-distribution data. The best overall performance was achieved by \lsgroup[0] (MS-nnUNet with multi-scale dilated convolutions), which combined strong quantitative metrics with good generalisation across datasets. Ensemble and multi-scale strategies proved consistently advantageous, whilst heavy preprocessing and aggressive downsampling were detrimental. The baselines, despite their relative simplicity, performed respectably and in several qualitative criteria outperformed more complex submissions, underscoring that training strategy and regularisation matter at least as much as architectural complexity in this data-scarce regime.

The discrepancy between quantitative metrics and expert assessments constitutes one of the central findings of this work. It reinforces the need for evaluation frameworks that go beyond voxel-level overlap, particularly for tasks where the structures of interest are small and where clinical utility depends on properties -- such as vessel continuity and branch completeness -- that overlap metrics capture only indirectly.

Several directions for future work emerge from the findings and limitations of this challenge. First, the evaluation framework would benefit from the incorporation of topology-aware metrics, centreline-based evaluation, or vessel-specific measures that more faithfully reflect the quality attributes valued by clinical experts. Replacing mutual information, which proved to have negligible discriminative power in this setting, with metrics such as the Betti number error or centreline Dice would be a concrete step in this direction.

Second, expanding the SMILE-UHURA dataset to include subjects from older age groups and from cohorts with confirmed cerebral small vessel disease is essential. Both the open and secret datasets comprise healthy young adults, and the vascular morphology in these subjects is likely to differ from that of elderly patients with established CSVD, where vessel tortuosity, white matter hyperintensity load, and microbleed prevalence alter the imaging characteristics. Validating the benchmarked methods on pathological cohorts is a prerequisite for translating these tools into clinical practice.

Third, providing multiple independent annotations would make it possible to establish inter-rater baselines for this task. At present, the ground truth was refined by a single annotator and verified by a neurologist; without a second independent annotation, the absolute performance figures are difficult to contextualise against the expected range of human disagreement. This limitation has become practically consequential now that the best methods achieve Dice values in the range typically associated with inter-annotator agreement, raising the possibility that a portion of the measured error reflects annotator subjectivity rather than genuine segmentation failure. Multiple annotations would allow the community to determine whether the top methods have already reached human-level performance on this task, and would provide a more robust reference for measuring future progress.

Fourth, the twelve-percentage-point drop in Dice between in-distribution and out-of-distribution data highlights the need for improved generalisation. Domain adaptation techniques, test-time augmentation, and training on multi-site data with diverse acquisition protocols are all avenues that merit systematic investigation. Extending the evaluation to other MRA sequences (e.g. phase-contrast MRA, black-blood MRI) and to lower field strengths (3T) would further characterise the robustness of these methods.

Fifth, as ultra-high-field systems become more prevalent and push towards even finer resolutions, the demand for memory-efficient segmentation architectures will grow. The 150~\textmu m evaluation already revealed that the most accurate methods at 300~\textmu m could not process the higher-resolution volume within the challenge's computational constraints. Developing methods that scale gracefully to larger volumes without sacrificing accuracy will be essential for realising the full potential of ultra-high-resolution imaging.

Sixth, accurate mesoscopic segmentation is only a first step; downstream morphometric pipelines that convert the segmentation into clinically meaningful vascular descriptors are needed to translate these methods into CSVD research practice. The LUMEN pipeline~\cite{li2025lumen}, which extracts lenticulostriate artery morphology from 7T ToF-MRA, is a good example of the kind of analysis that benefits directly from an improved upstream segmentation. Pairing the top-performing SMILE-UHURA methods with LUMEN-style morphometric analyses, and measuring their effect on vessel-level descriptors such as branch count, tortuosity and diameter distribution, is a natural next step towards clinical use.

The challenge remains open for new submissions, and the dataset and benchmarks presented here are intended to serve as a continuing resource for the development and fair comparison of mesoscopic vessel segmentation methods. The organisers hope that this work will encourage further collaboration between the medical imaging, computer science, and CSVD research communities towards segmentation tools that are ultimately fit for clinical purpose.

%% file: Sections/x_ack.tex
\section{Acknowledgements}
\noindent The organisers extend their gratitude to Paul Yushkevich and Maria A. Zululaga of the ISBI 2023 Challenge Committee for their invaluable guidance and support. The organisers also wish to thank Ignacio Larrabide, Camila García, and other members of the co-located challenge, \textbf{SHINY-ICARUS}: \textbf{S}egmentation over T\textbf{h}ree-D\textbf{i}mensional Rotational A\textbf{n}giograph\textbf{y} of the \textbf{I}nternal \textbf{C}arotid \textbf{A}rte\textbf{r}y with Ane\textbf{u}ry\textbf{s}m, for their support with the shared on-site organisation. Finally, the organisers express their sincere appreciation to Sage Bionetworks for their \textbf{SYPANSE} platform, which hosted the challenge dataset.%
H.M. was supported in by the German Research Foundation (DFG) under project numbers 425899996, 446268581,  and 501214112, as well as by the Deutsche Alzheimer Gesellschaft (DAlzG) e.V. (MD-DARS and BB-DARS project).
C.C.Y., T.L.H., Y.S.T., Y.Z.F., Y.C.Y., and J.D.H. received support in part from the Ministry of Science and Technology, Taiwan, under Grant No. MOST-111-2218-E-A49-022, as well as from the computing services of the National Center for High-Performance Computing (NCHC) and the Taiwan Computing Cloud (TWCC).%
K.X., S.L., F.L.R., and S.B. acknowledge Steffen Bollmann for his invaluable support with high-performance computing facilities, Docker containers, and Neurodesk.%
The research of R.E.J. leading to these results was funded by the French government, under the management of the Agence Nationale de la Recherche, as part of the "Investissements d’avenir" programme. This includes funding under reference ANR-19-P3IA-0001 (PRAIRIE 3IA Institute) and reference ANR-10-IAHU-0006 (Agence Nationale de la Recherche-10-IA Institut Hospitalo-Universitaire-6).%
The work of M.A.Z. was supported by the ANR JCJC project I-VESSEG (22-CE45-0015-01).

\section*{Conflict of interest and challenge integrity}
\noindent The authors declare no competing financial interests that could have inappropriately influenced this work. The funding sources listed above had no role in the design of the challenge, in the collection, analysis or interpretation of the data, in the writing of this manuscript, or in the decision to submit it for publication. Throughout the challenge, access to the held-out reference annotations of the \textit{Open Dataset} test subset and of the entire \textit{Secret Dataset} was strictly restricted to the organisers (S.~Chatterjee and H.~Mattern) and was never shared with any participating team, including teams from the organisers' own institutes; participants from organiser-affiliated groups (the baselines, the neuRoSliCCe submissions and the Koala submissions) were eligible to submit Docker containers under the same conditions as external teams but, as stated in Section~\ref{sec:governance}, were excluded from any prize ranking and are reported here only as reference comparators.

%% file: Sections/x_appendix.tex
\section{Additional quantitative box plots}
\label{app:extra_boxplots}

\begin{figure*}[!htbp]
    \centering
    \includegraphics[width=0.77\textwidth]{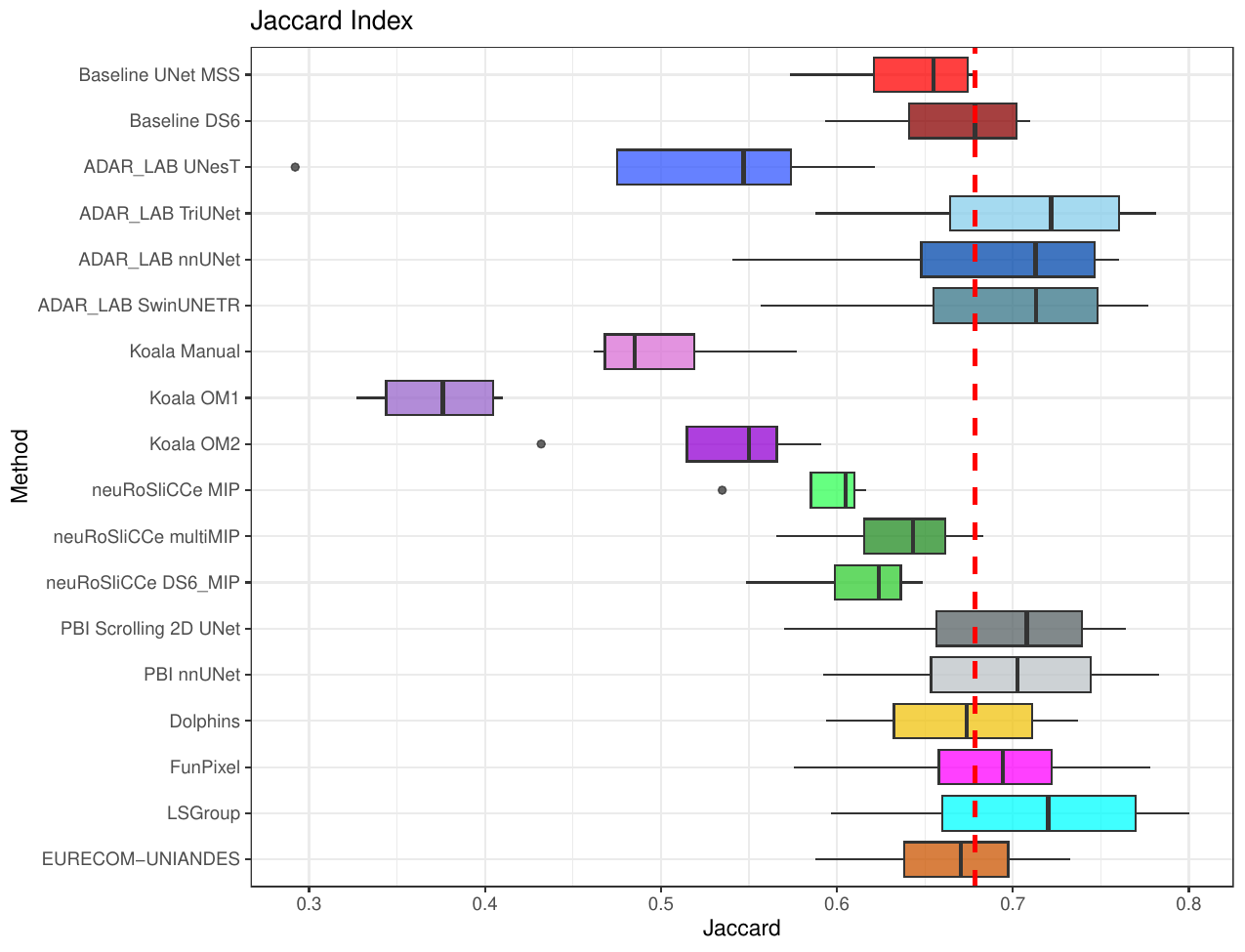}
    \caption{Jaccard index scores on the test subset from the open dataset. The red dashed line denotes the median of the better-performing baseline method (i.e., DS6).}
    \label{fig:box_Forrest_Jaccard}
\end{figure*}

\begin{figure*}[!htbp]
    \centering
    \includegraphics[width=0.77\textwidth]{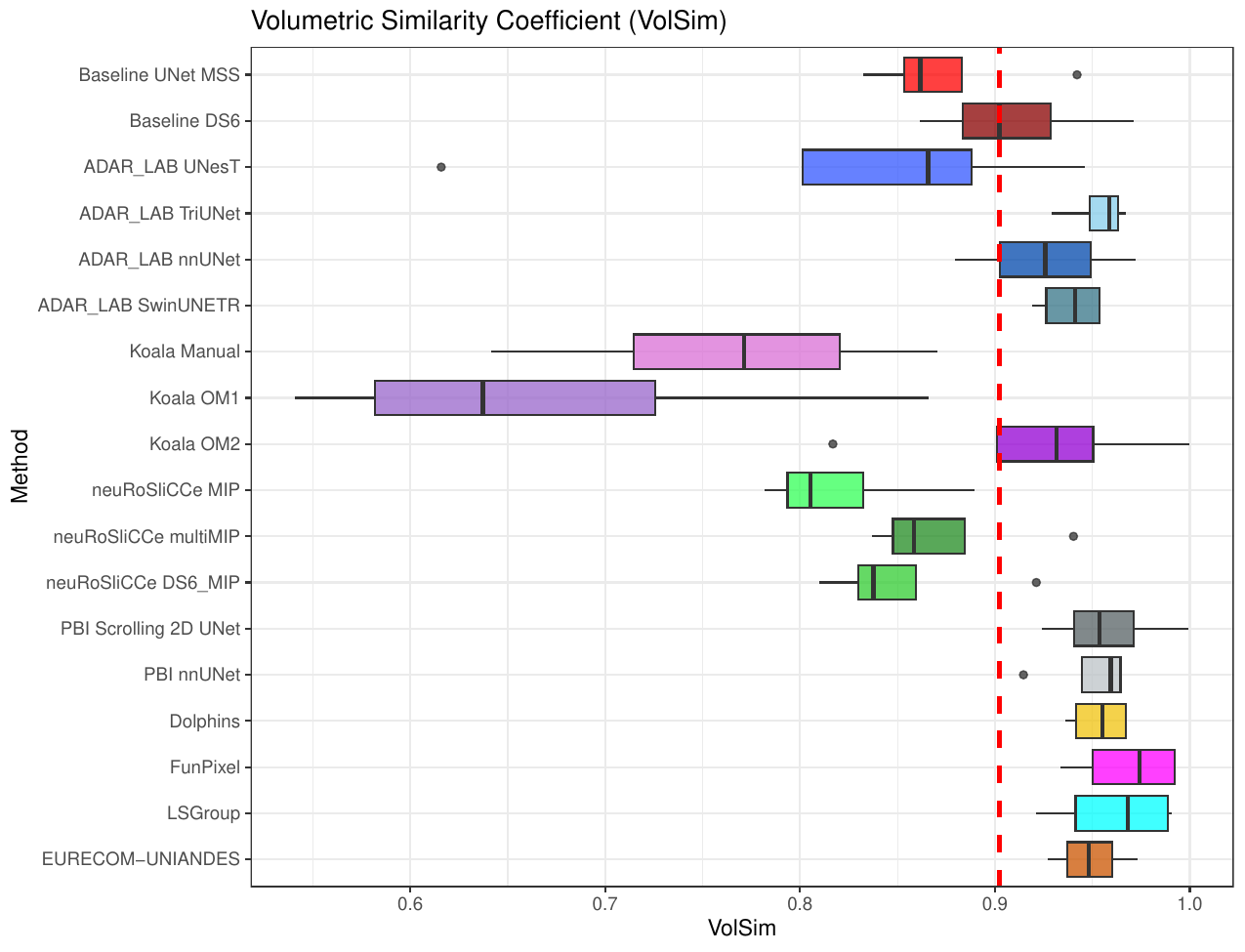}
    \caption{Volumetric similarity coefficients on the test subset from the open dataset. The red dashed line denotes the median of the better-performing baseline method (i.e., DS6).}
    \label{fig:box_Forrest_VolSim}
\end{figure*}

\begin{figure*}[!htbp]
    \centering
    \includegraphics[width=0.77\textwidth]{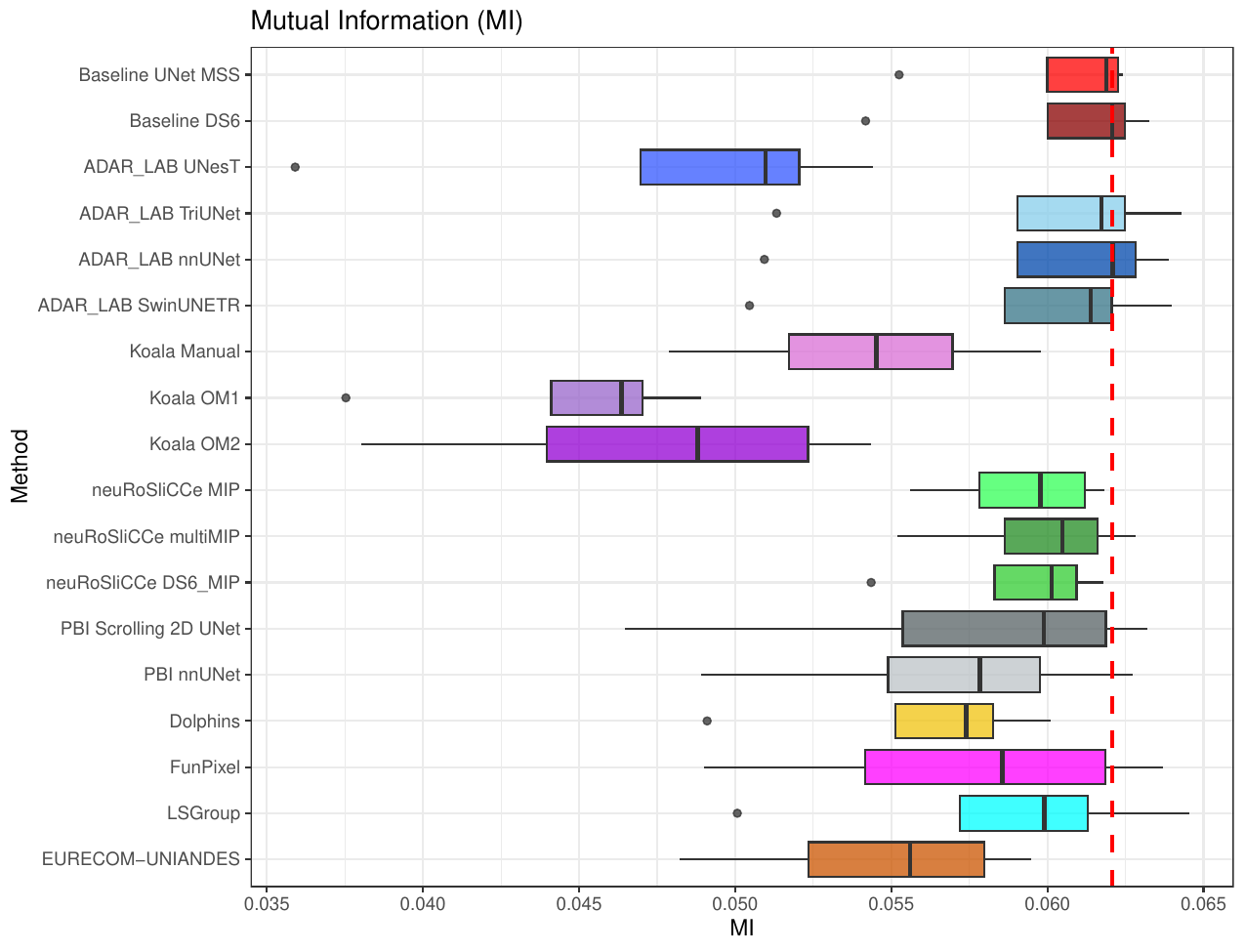}
    \caption{Mutual information scores on the test subset from the open dataset. The red dashed line denotes the median of the better-performing baseline method (i.e., DS6).}
    \label{fig:box_Forrest_MI}
\end{figure*}

\begin{figure*}[!htbp]
    \centering
    \includegraphics[width=0.77\textwidth]{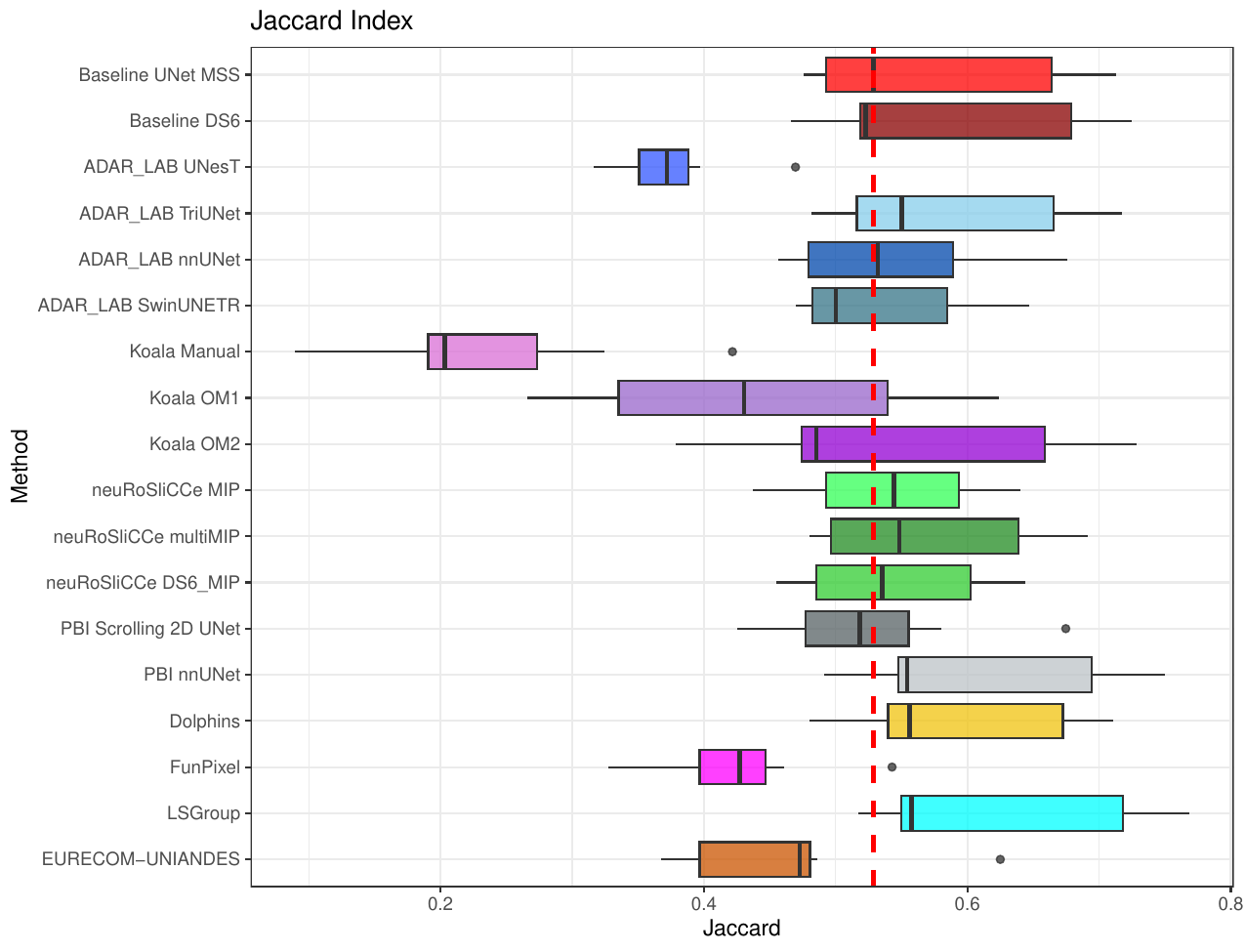}
    \caption{Jaccard index scores on the secret dataset. The red dashed line denotes the median of the better-performing baseline method (i.e., UNet MSS).}
    \label{fig:box_HendrikROT_Jaccard}
\end{figure*}

\begin{figure*}[!htbp]
    \centering
    \includegraphics[width=0.77\textwidth]{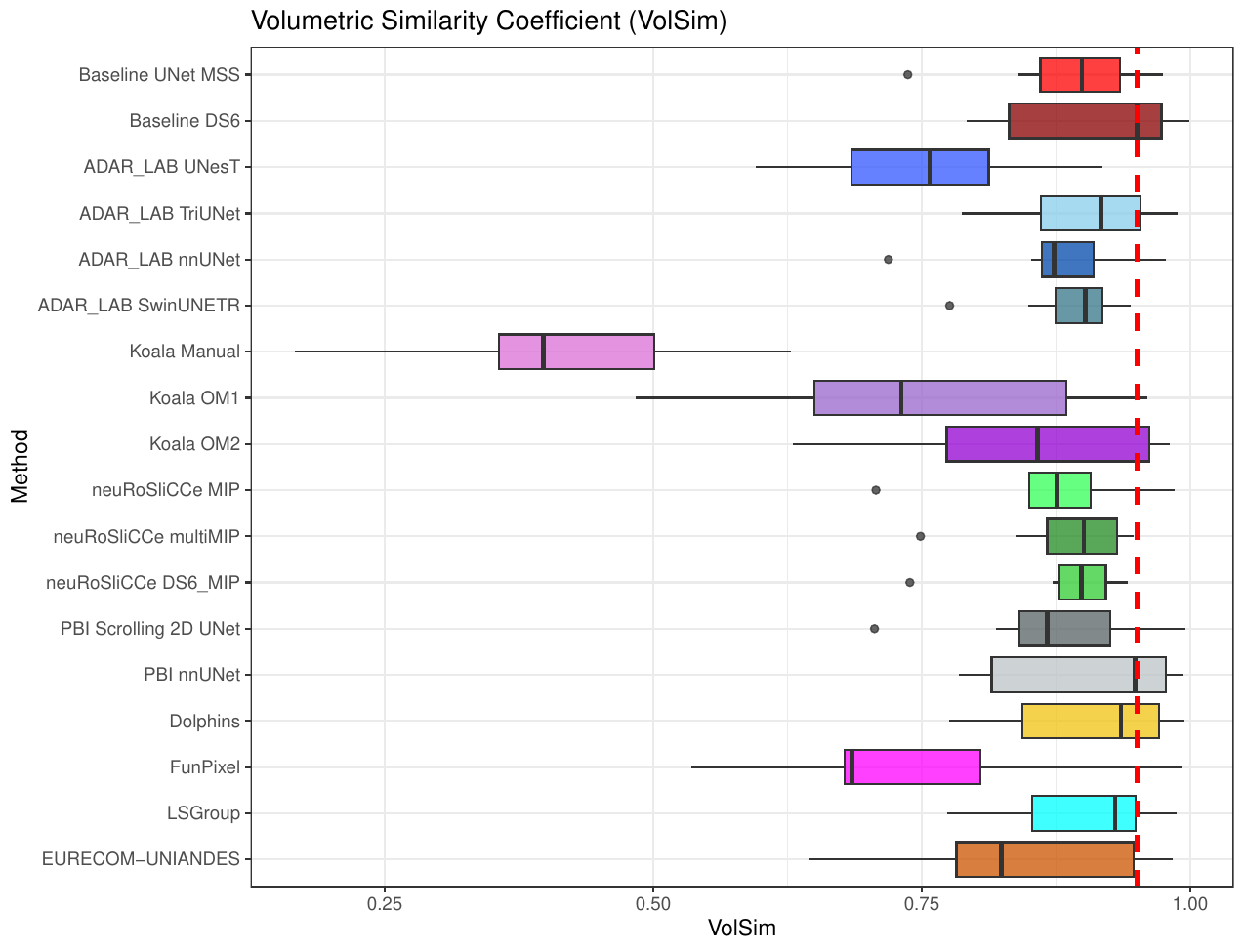}
    \caption{Volumetric similarity coefficients on the secret dataset. The red dashed line denotes the median of the better-performing baseline method (i.e., DS6).}
    \label{fig:box_HendrikROT_VolSim}
\end{figure*}

\begin{figure*}[!htbp]
    \centering
    \includegraphics[width=0.77\textwidth]{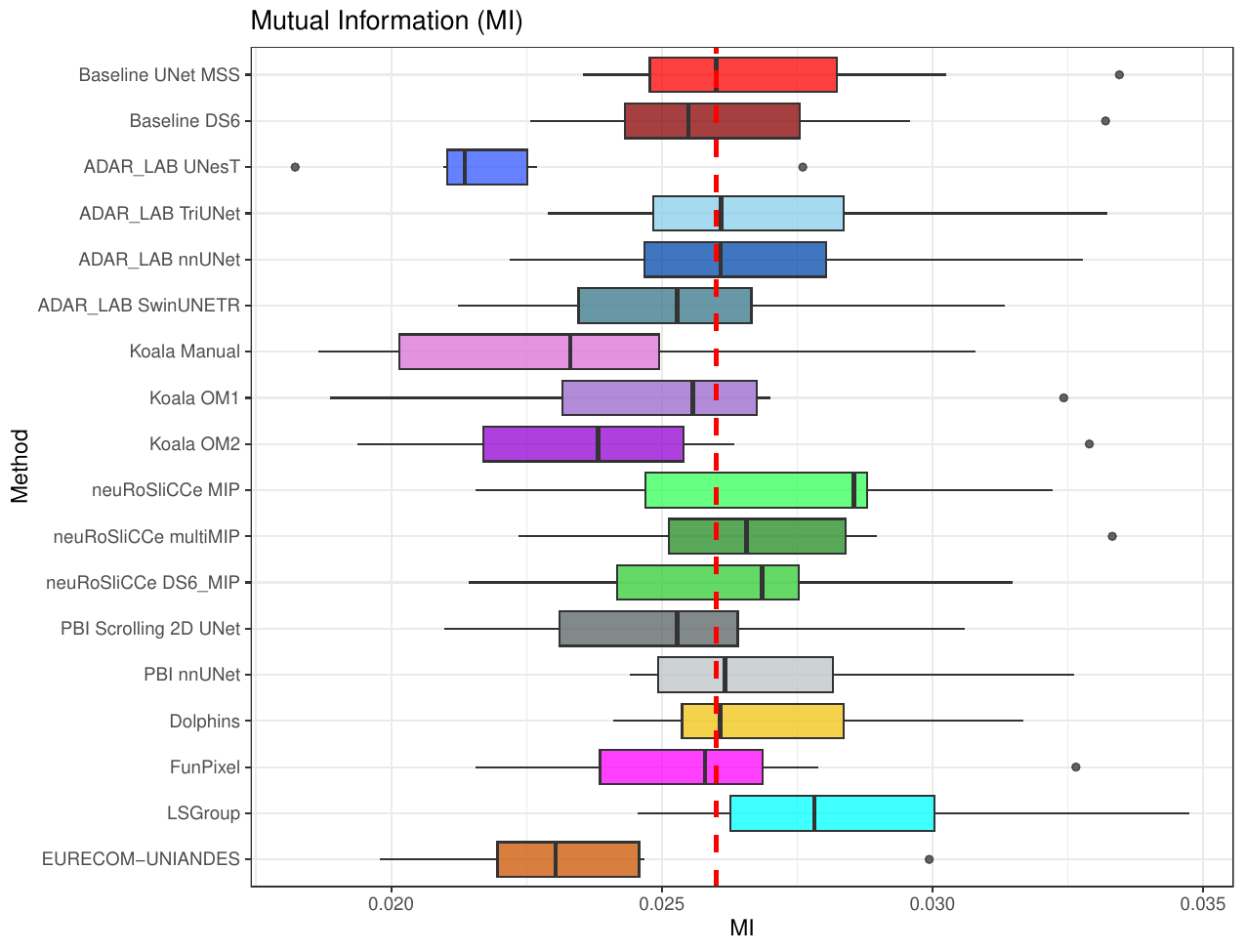}
    \caption{Mutual information scores on the secret dataset. The red dashed line denotes the median of the better-performing baseline method (i.e., UNet MSS).}
    \label{fig:box_HendrikROT_MI}
\end{figure*}

\begin{figure*}[!htbp]
    \centering
    \includegraphics[width=0.77\textwidth]{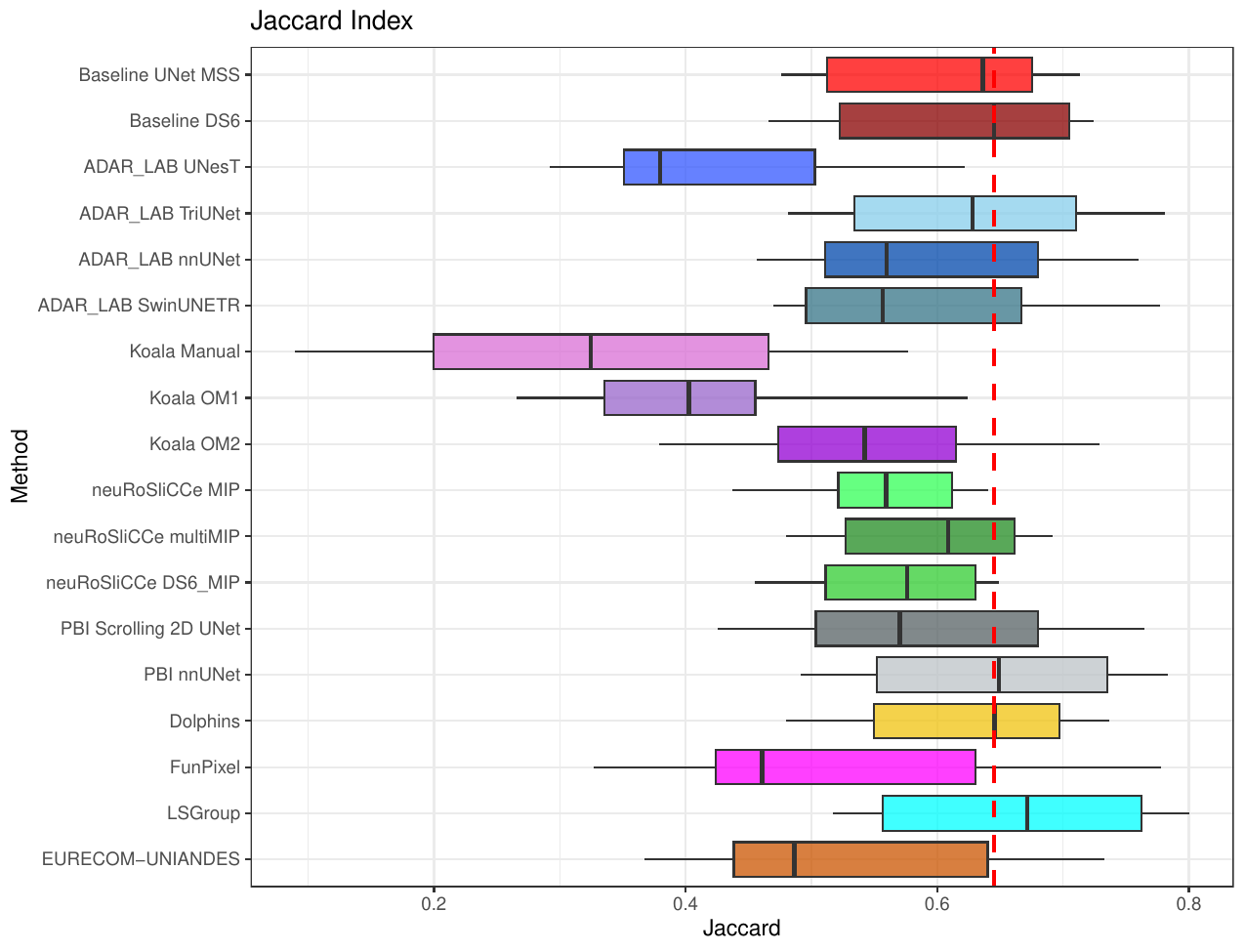}
    \caption{Jaccard index scores on the combined datasets. The red dashed line denotes the median of the better-performing baseline method (i.e., DS6).}
    \label{fig:box_AllRes_Jaccard}
\end{figure*}

\begin{figure*}[!htbp]
    \centering
    \includegraphics[width=0.77\textwidth]{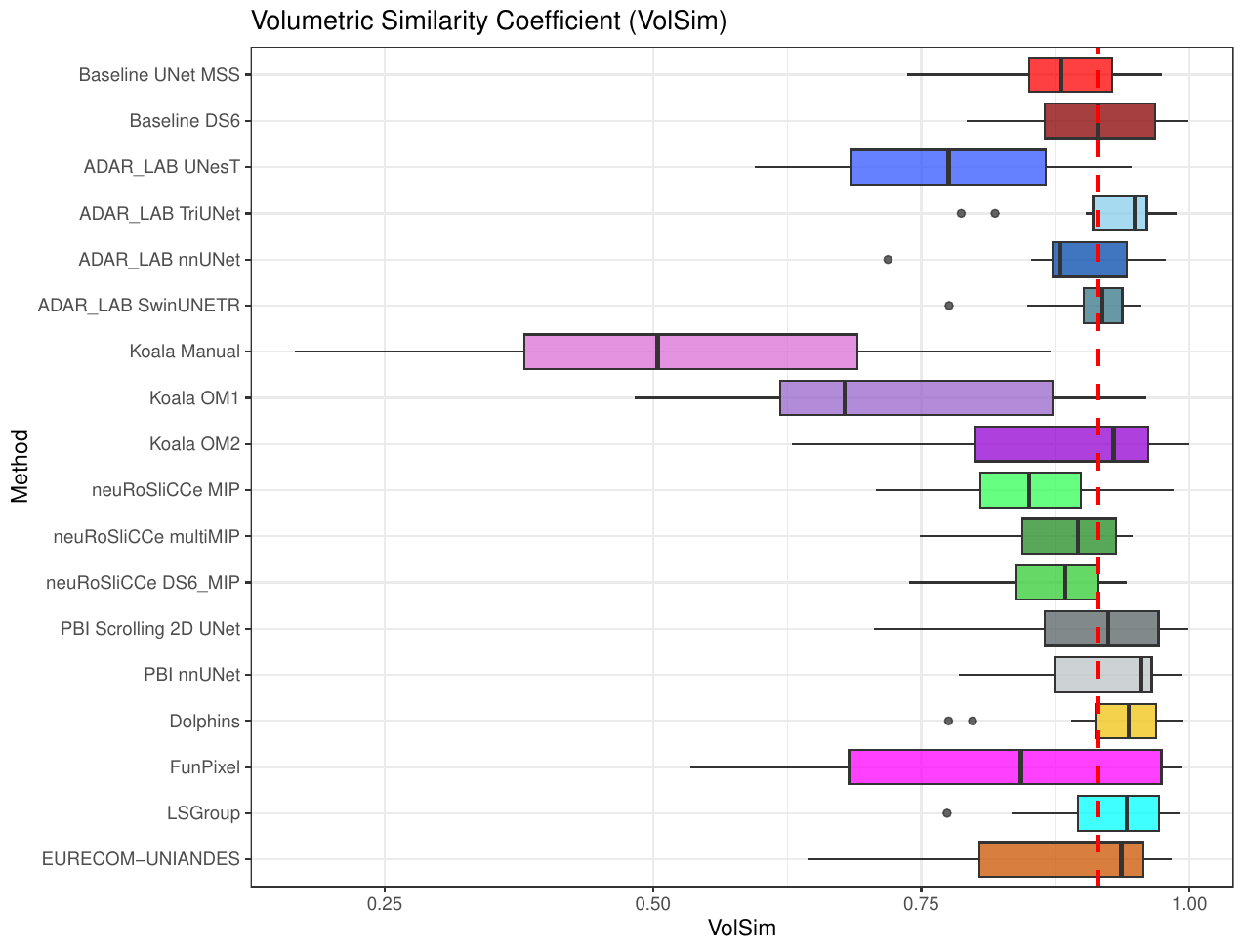}
    \caption{Volumetric similarity coefficients on the combined datasets. The red dashed line denotes the median of the better-performing baseline method (i.e., DS6).}
    \label{fig:box_AllRes_VolSim}
\end{figure*}

\begin{figure*}[!htbp]
    \centering
    \includegraphics[width=0.77\textwidth]{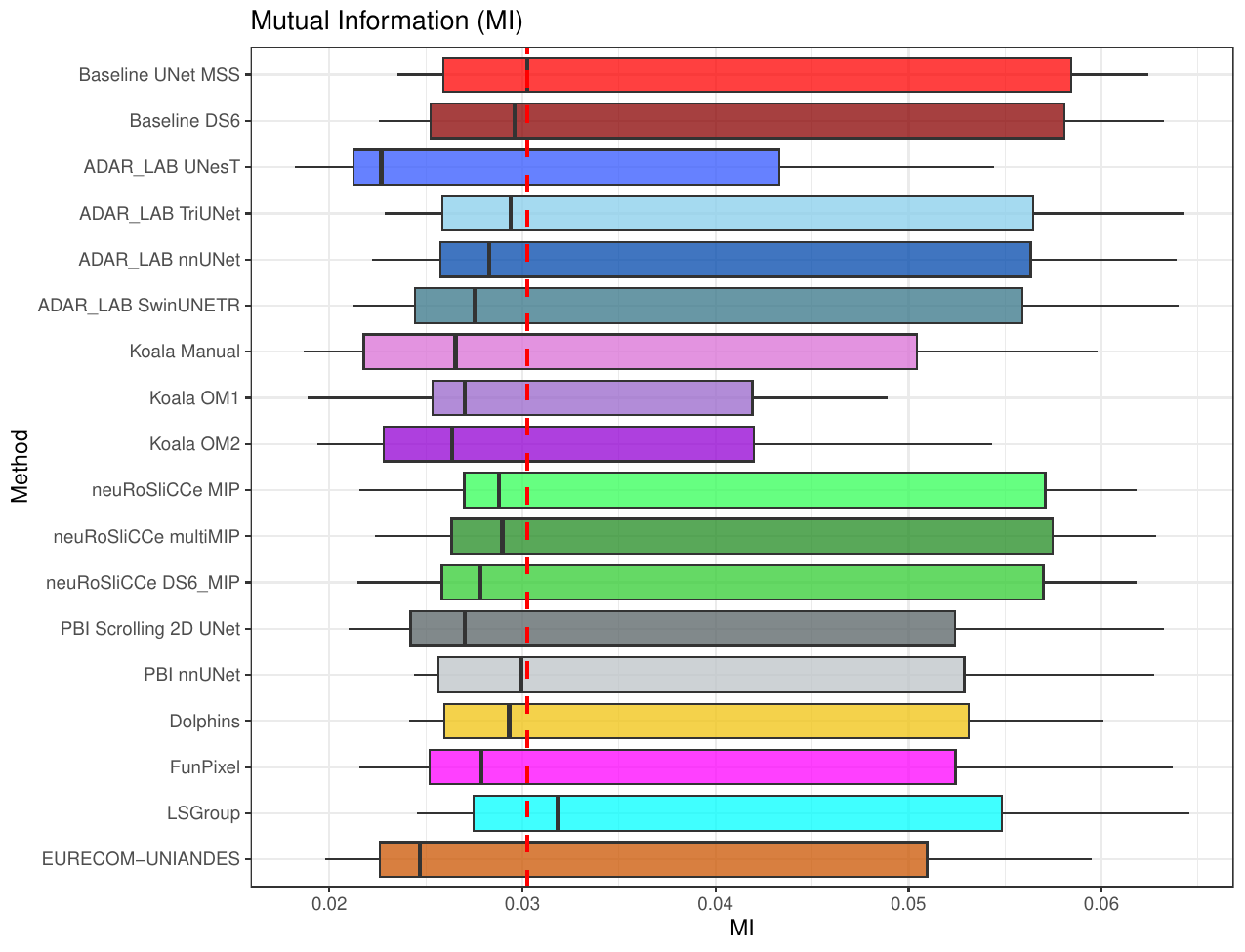}
    \caption{Mutual information scores on the combined datasets. The red dashed line denotes the median of the better-performing baseline method (i.e., UNet MSS).}
    \label{fig:box_AllRes_MI}
\end{figure*}

For completeness, the per-subject score distributions for the Jaccard index, volumetric similarity (VolSim) and mutual information (MI) are provided here as box plots for the open dataset (Figures~\ref{fig:box_Forrest_Jaccard}, \ref{fig:box_Forrest_VolSim} and \ref{fig:box_Forrest_MI}), the secret dataset (Figures~\ref{fig:box_HendrikROT_Jaccard}, \ref{fig:box_HendrikROT_VolSim} and \ref{fig:box_HendrikROT_MI}) and the combined dataset (Figures~\ref{fig:box_AllRes_Jaccard}, \ref{fig:box_AllRes_VolSim} and \ref{fig:box_AllRes_MI}). Because Jaccard is a strictly monotone function of Dice (Eq.~\ref{eq:jaccard}), the per-subject ordering of methods in the Jaccard box plots is identical to that of the corresponding Dice figures in the main text. The VolSim and MI box plots graphically corroborate the observations in Section~\ref{sec:stat_results} that these two metrics are comparatively less discriminative among the evaluated methods.

\subsection{Open dataset}

\subsection{Secret dataset}

\subsection{Combined dataset}

\section{Full pairwise statistical comparisons}
\label{app:conover_full}

The following table lists, for each metric, every pair of methods that was found to be significantly different after the Conover-Friedman post-hoc test with Holm correction at $\alpha = 0.05$ on the combined dataset ($N = 11$). Summary counts are discussed in Section~\ref{sec:stat_results} and visualised in the Critical Difference diagrams (Figures~\ref{fig:cd_dice} to~\ref{fig:cd_bavd}); the complete $18 \times 18$ Holm-adjusted $p$-value matrices per metric are released alongside the paper for querying any specific pair of interest.

\input{Tables/Stats/conover_summary}

\section{Author contributions}
\label{app:author_contributions}

\noindent Author contributions are reported following the Contributor Roles Taxonomy (CRediT). S.~Chatterjee and H.~Mattern jointly led the conception and organisation of the SMILE-UHURA challenge, including the curation and annotation of the \textit{Open} and \textit{Secret} datasets, the design of the evaluation pipeline, the implementation and training of the two baseline methods, the on-site logistics at ISBI 2023, and were the only co-authors with access to the held-out reference annotations throughout the challenge. S.~Chatterjee additionally implemented the statistical analysis and Critical Difference visualisation pipeline, prepared all figures and tables, and drafted the manuscript together with H.~Mattern; both authors revised the manuscript and coordinated input from the participating teams. O.~Speck, A.~N\"urnberger and S.~Oeltze-Jafra contributed to supervision, funding acquisition and to the editorial revision of the manuscript. All remaining co-authors, listed by their participating teams (ADAR\_Lab, Atlas, Joker, Koala, MIVR, neuRoSliCCe, NMSeg, PBI\_nnUNet, Pixel, RPI, Sanskar, SCAN, SuperPower, TUM-Brain, Vessel-Captain), contributed the methodology, software, training and validation of their respective submissions, the Docker packaging, the per-team write-up of their method description, and the review of the corresponding sections of this manuscript. A detailed per-author CRediT mapping is available from the corresponding authors on request.

%% file: Tables/Stats/conover_summary.tex
\onecolumn

{\setlength{\tabcolsep}{10pt}
\begin{longtable}{llll}
\caption{Conover post-hoc test (Holm-corrected) — significant pairs ($p < 0.05$).}
\label{tab:conover_summary}\\
  \toprule
  Metric & Method A & Method B & $p$-value \\
  \midrule
  \endfirsthead
  \multicolumn{4}{c}{\tablename\ \thetable{} (continued)}\\
  \toprule
  Metric & Method A & Method B & $p$-value \\
  \midrule
  \endhead
  \midrule
  \multicolumn{4}{r}{\footnotesize\itshape Continued on next page}\\
  \endfoot
  \endlastfoot
  Dice & ADAR\_LAB SwinUNETR & ADAR\_LAB UNesT & 3.58e-06 \\
  Dice & ADAR\_LAB SwinUNETR & Koala Manual & 1.17e-07 \\
  Dice & ADAR\_LAB SwinUNETR & Koala OM1 & 9.32e-06 \\
  Dice & ADAR\_LAB SwinUNETR & LSGroup & 8.12e-05 \\
  Dice & ADAR\_LAB SwinUNETR & PBI nnUNet & 0.0113 \\
  Dice & ADAR\_LAB TriUNet & ADAR\_LAB UNesT & 1.01e-13 \\
  Dice & ADAR\_LAB TriUNet & EURECOM-UNIANDES & 4.96e-06 \\
  Dice & ADAR\_LAB TriUNet & FunPixel & 1.76e-05 \\
  Dice & ADAR\_LAB TriUNet & Koala Manual & 1.53e-15 \\
  Dice & ADAR\_LAB TriUNet & Koala OM1 & 3.46e-13 \\
  Dice & ADAR\_LAB TriUNet & Koala OM2 & 3.30e-05 \\
  Dice & ADAR\_LAB TriUNet & PBI Scrolling 2D UNet & 0.0284 \\
  Dice & ADAR\_LAB TriUNet & neuRoSliCCe DS6\_MIP & 0.0011 \\
  Dice & ADAR\_LAB TriUNet & neuRoSliCCe MIP & 8.12e-05 \\
  Dice & ADAR\_LAB UNesT & ADAR\_LAB nnUNet & 9.32e-06 \\
  Dice & ADAR\_LAB UNesT & Baseline DS6 & 1.42e-09 \\
  Dice & ADAR\_LAB UNesT & Baseline UNet MSS & 1.87e-06 \\
  Dice & ADAR\_LAB UNesT & Dolphins & 1.77e-12 \\
  Dice & ADAR\_LAB UNesT & Koala OM2 & 0.0354 \\
  Dice & ADAR\_LAB UNesT & LSGroup & 2.76e-19 \\
  Dice & ADAR\_LAB UNesT & PBI Scrolling 2D UNet & 4.41e-05 \\
  Dice & ADAR\_LAB UNesT & PBI nnUNet & 6.55e-16 \\
  Dice & ADAR\_LAB UNesT & neuRoSliCCe DS6\_MIP & 0.0019 \\
  Dice & ADAR\_LAB UNesT & neuRoSliCCe MIP & 0.0182 \\
  Dice & ADAR\_LAB UNesT & neuRoSliCCe multiMIP & 1.87e-06 \\
  Dice & ADAR\_LAB nnUNet & Koala Manual & 3.37e-07 \\
  Dice & ADAR\_LAB nnUNet & Koala OM1 & 2.41e-05 \\
  Dice & ADAR\_LAB nnUNet & LSGroup & 3.30e-05 \\
  Dice & ADAR\_LAB nnUNet & PBI nnUNet & 0.0054 \\
  Dice & Baseline DS6 & EURECOM-UNIANDES & 0.0054 \\
  Dice & Baseline DS6 & FunPixel & 0.0145 \\
  Dice & Baseline DS6 & Koala Manual & 2.94e-11 \\
  Dice & Baseline DS6 & Koala OM1 & 4.39e-09 \\
  Dice & Baseline DS6 & Koala OM2 & 0.0225 \\
  Dice & Baseline DS6 & LSGroup & 0.0284 \\
  Dice & Baseline DS6 & neuRoSliCCe MIP & 0.0438 \\
  Dice & Baseline UNet MSS & Koala Manual & 5.83e-08 \\
  Dice & Baseline UNet MSS & Koala OM1 & 4.96e-06 \\
  Dice & Baseline UNet MSS & LSGroup & 0.0001 \\
  Dice & Baseline UNet MSS & PBI nnUNet & 0.0182 \\
  Dice & Dolphins & EURECOM-UNIANDES & 4.41e-05 \\
  Dice & Dolphins & FunPixel & 0.0001 \\
  Dice & Dolphins & Koala Manual & 2.89e-14 \\
  Dice & Dolphins & Koala OM1 & 5.95e-12 \\
  Dice & Dolphins & Koala OM2 & 0.0003 \\
  Dice & Dolphins & neuRoSliCCe DS6\_MIP & 0.0068 \\
  Dice & Dolphins & neuRoSliCCe MIP & 0.0006 \\
  Dice & EURECOM-UNIANDES & Koala Manual & 0.0145 \\
  Dice & EURECOM-UNIANDES & LSGroup & 9.59e-11 \\
  Dice & EURECOM-UNIANDES & PBI nnUNet & 8.26e-08 \\
  Dice & FunPixel & Koala Manual & 0.0054 \\
  Dice & FunPixel & LSGroup & 4.55e-10 \\
  Dice & FunPixel & PBI nnUNet & 3.37e-07 \\
  Dice & Koala Manual & Koala OM2 & 0.0032 \\
  Dice & Koala Manual & LSGroup & 3.42e-21 \\
  Dice & Koala Manual & PBI Scrolling 2D UNet & 1.87e-06 \\
  Dice & Koala Manual & PBI nnUNet & 8.90e-18 \\
  Dice & Koala Manual & neuRoSliCCe DS6\_MIP & 0.0001 \\
  Dice & Koala Manual & neuRoSliCCe MIP & 0.0014 \\
  Dice & Koala Manual & neuRoSliCCe multiMIP & 5.83e-08 \\
  Dice & Koala OM1 & LSGroup & 1.02e-18 \\
  Dice & Koala OM1 & PBI Scrolling 2D UNet & 0.0001 \\
  Dice & Koala OM1 & PBI nnUNet & 2.32e-15 \\
  Dice & Koala OM1 & neuRoSliCCe DS6\_MIP & 0.0042 \\
  Dice & Koala OM1 & neuRoSliCCe MIP & 0.0354 \\
  Dice & Koala OM1 & neuRoSliCCe multiMIP & 4.96e-06 \\
  Dice & Koala OM2 & LSGroup & 9.77e-10 \\
  Dice & Koala OM2 & PBI nnUNet & 6.70e-07 \\
  Dice & LSGroup & PBI Scrolling 2D UNet & 6.75e-06 \\
  Dice & LSGroup & neuRoSliCCe DS6\_MIP & 8.26e-08 \\
  Dice & LSGroup & neuRoSliCCe MIP & 3.03e-09 \\
  Dice & LSGroup & neuRoSliCCe multiMIP & 0.0001 \\
  Dice & PBI Scrolling 2D UNet & PBI nnUNet & 0.0014 \\
  Dice & PBI nnUNet & neuRoSliCCe DS6\_MIP & 3.30e-05 \\
  Dice & PBI nnUNet & neuRoSliCCe MIP & 1.87e-06 \\
  Dice & PBI nnUNet & neuRoSliCCe multiMIP & 0.0182 \\
  \midrule
  Jaccard & ADAR\_LAB SwinUNETR & ADAR\_LAB UNesT & 3.58e-06 \\
  Jaccard & ADAR\_LAB SwinUNETR & Koala Manual & 1.17e-07 \\
  Jaccard & ADAR\_LAB SwinUNETR & Koala OM1 & 9.32e-06 \\
  Jaccard & ADAR\_LAB SwinUNETR & LSGroup & 8.12e-05 \\
  Jaccard & ADAR\_LAB SwinUNETR & PBI nnUNet & 0.0113 \\
  Jaccard & ADAR\_LAB TriUNet & ADAR\_LAB UNesT & 1.01e-13 \\
  Jaccard & ADAR\_LAB TriUNet & EURECOM-UNIANDES & 4.96e-06 \\
  Jaccard & ADAR\_LAB TriUNet & FunPixel & 1.76e-05 \\
  Jaccard & ADAR\_LAB TriUNet & Koala Manual & 1.53e-15 \\
  Jaccard & ADAR\_LAB TriUNet & Koala OM1 & 3.46e-13 \\
  Jaccard & ADAR\_LAB TriUNet & Koala OM2 & 3.30e-05 \\
  Jaccard & ADAR\_LAB TriUNet & PBI Scrolling 2D UNet & 0.0284 \\
  Jaccard & ADAR\_LAB TriUNet & neuRoSliCCe DS6\_MIP & 0.0011 \\
  Jaccard & ADAR\_LAB TriUNet & neuRoSliCCe MIP & 8.12e-05 \\
  Jaccard & ADAR\_LAB UNesT & ADAR\_LAB nnUNet & 9.32e-06 \\
  Jaccard & ADAR\_LAB UNesT & Baseline DS6 & 1.42e-09 \\
  Jaccard & ADAR\_LAB UNesT & Baseline UNet MSS & 1.87e-06 \\
  Jaccard & ADAR\_LAB UNesT & Dolphins & 1.77e-12 \\
  Jaccard & ADAR\_LAB UNesT & Koala OM2 & 0.0354 \\
  Jaccard & ADAR\_LAB UNesT & LSGroup & 2.76e-19 \\
  Jaccard & ADAR\_LAB UNesT & PBI Scrolling 2D UNet & 4.41e-05 \\
  Jaccard & ADAR\_LAB UNesT & PBI nnUNet & 6.55e-16 \\
  Jaccard & ADAR\_LAB UNesT & neuRoSliCCe DS6\_MIP & 0.0019 \\
  Jaccard & ADAR\_LAB UNesT & neuRoSliCCe MIP & 0.0182 \\
  Jaccard & ADAR\_LAB UNesT & neuRoSliCCe multiMIP & 1.87e-06 \\
  Jaccard & ADAR\_LAB nnUNet & Koala Manual & 3.37e-07 \\
  Jaccard & ADAR\_LAB nnUNet & Koala OM1 & 2.41e-05 \\
  Jaccard & ADAR\_LAB nnUNet & LSGroup & 3.30e-05 \\
  Jaccard & ADAR\_LAB nnUNet & PBI nnUNet & 0.0054 \\
  Jaccard & Baseline DS6 & EURECOM-UNIANDES & 0.0054 \\
  Jaccard & Baseline DS6 & FunPixel & 0.0145 \\
  Jaccard & Baseline DS6 & Koala Manual & 2.94e-11 \\
  Jaccard & Baseline DS6 & Koala OM1 & 4.39e-09 \\
  Jaccard & Baseline DS6 & Koala OM2 & 0.0225 \\
  Jaccard & Baseline DS6 & LSGroup & 0.0284 \\
  Jaccard & Baseline DS6 & neuRoSliCCe MIP & 0.0438 \\
  Jaccard & Baseline UNet MSS & Koala Manual & 5.83e-08 \\
  Jaccard & Baseline UNet MSS & Koala OM1 & 4.96e-06 \\
  Jaccard & Baseline UNet MSS & LSGroup & 0.0001 \\
  Jaccard & Baseline UNet MSS & PBI nnUNet & 0.0182 \\
  Jaccard & Dolphins & EURECOM-UNIANDES & 4.41e-05 \\
  Jaccard & Dolphins & FunPixel & 0.0001 \\
  Jaccard & Dolphins & Koala Manual & 2.89e-14 \\
  Jaccard & Dolphins & Koala OM1 & 5.95e-12 \\
  Jaccard & Dolphins & Koala OM2 & 0.0003 \\
  Jaccard & Dolphins & neuRoSliCCe DS6\_MIP & 0.0068 \\
  Jaccard & Dolphins & neuRoSliCCe MIP & 0.0006 \\
  Jaccard & EURECOM-UNIANDES & Koala Manual & 0.0145 \\
  Jaccard & EURECOM-UNIANDES & LSGroup & 9.59e-11 \\
  Jaccard & EURECOM-UNIANDES & PBI nnUNet & 8.26e-08 \\
  Jaccard & FunPixel & Koala Manual & 0.0054 \\
  Jaccard & FunPixel & LSGroup & 4.55e-10 \\
  Jaccard & FunPixel & PBI nnUNet & 3.37e-07 \\
  Jaccard & Koala Manual & Koala OM2 & 0.0032 \\
  Jaccard & Koala Manual & LSGroup & 3.42e-21 \\
  Jaccard & Koala Manual & PBI Scrolling 2D UNet & 1.87e-06 \\
  Jaccard & Koala Manual & PBI nnUNet & 8.90e-18 \\
  Jaccard & Koala Manual & neuRoSliCCe DS6\_MIP & 0.0001 \\
  Jaccard & Koala Manual & neuRoSliCCe MIP & 0.0014 \\
  Jaccard & Koala Manual & neuRoSliCCe multiMIP & 5.83e-08 \\
  Jaccard & Koala OM1 & LSGroup & 1.02e-18 \\
  Jaccard & Koala OM1 & PBI Scrolling 2D UNet & 0.0001 \\
  Jaccard & Koala OM1 & PBI nnUNet & 2.32e-15 \\
  Jaccard & Koala OM1 & neuRoSliCCe DS6\_MIP & 0.0042 \\
  Jaccard & Koala OM1 & neuRoSliCCe MIP & 0.0354 \\
  Jaccard & Koala OM1 & neuRoSliCCe multiMIP & 4.96e-06 \\
  Jaccard & Koala OM2 & LSGroup & 9.77e-10 \\
  Jaccard & Koala OM2 & PBI nnUNet & 6.70e-07 \\
  Jaccard & LSGroup & PBI Scrolling 2D UNet & 6.75e-06 \\
  Jaccard & LSGroup & neuRoSliCCe DS6\_MIP & 8.26e-08 \\
  Jaccard & LSGroup & neuRoSliCCe MIP & 3.03e-09 \\
  Jaccard & LSGroup & neuRoSliCCe multiMIP & 0.0001 \\
  Jaccard & PBI Scrolling 2D UNet & PBI nnUNet & 0.0014 \\
  Jaccard & PBI nnUNet & neuRoSliCCe DS6\_MIP & 3.30e-05 \\
  Jaccard & PBI nnUNet & neuRoSliCCe MIP & 1.87e-06 \\
  Jaccard & PBI nnUNet & neuRoSliCCe multiMIP & 0.0182 \\
  \midrule
  VolSim & ADAR\_LAB SwinUNETR & Koala Manual & 0.0002 \\
  VolSim & ADAR\_LAB TriUNet & ADAR\_LAB UNesT & 0.0216 \\
  VolSim & ADAR\_LAB TriUNet & Koala Manual & 8.34e-07 \\
  VolSim & ADAR\_LAB TriUNet & Koala OM1 & 0.0011 \\
  VolSim & ADAR\_LAB UNesT & Dolphins & 0.0182 \\
  VolSim & ADAR\_LAB nnUNet & Koala Manual & 0.0011 \\
  VolSim & Baseline DS6 & Koala Manual & 7.16e-06 \\
  VolSim & Baseline DS6 & Koala OM1 & 0.0062 \\
  VolSim & Baseline UNet MSS & Koala Manual & 0.0107 \\
  VolSim & Dolphins & Koala Manual & 6.57e-07 \\
  VolSim & Dolphins & Koala OM1 & 0.0009 \\
  VolSim & EURECOM-UNIANDES & Koala Manual & 0.0011 \\
  VolSim & FunPixel & Koala Manual & 0.0153 \\
  VolSim & Koala Manual & Koala OM2 & 0.0013 \\
  VolSim & Koala Manual & LSGroup & 3.54e-06 \\
  VolSim & Koala Manual & PBI Scrolling 2D UNet & 0.0001 \\
  VolSim & Koala Manual & PBI nnUNet & 7.16e-06 \\
  VolSim & Koala Manual & neuRoSliCCe multiMIP & 0.0062 \\
  VolSim & Koala OM1 & LSGroup & 0.0035 \\
  VolSim & Koala OM1 & PBI nnUNet & 0.0062 \\
  \midrule
  MI & ADAR\_LAB SwinUNETR & ADAR\_LAB TriUNet & 0.0246 \\
  MI & ADAR\_LAB SwinUNETR & ADAR\_LAB UNesT & 0.0017 \\
  MI & ADAR\_LAB SwinUNETR & LSGroup & 0.0027 \\
  MI & ADAR\_LAB TriUNet & ADAR\_LAB UNesT & 9.23e-12 \\
  MI & ADAR\_LAB TriUNet & EURECOM-UNIANDES & 1.67e-08 \\
  MI & ADAR\_LAB TriUNet & Koala Manual & 9.71e-07 \\
  MI & ADAR\_LAB TriUNet & Koala OM1 & 7.20e-07 \\
  MI & ADAR\_LAB TriUNet & Koala OM2 & 5.32e-07 \\
  MI & ADAR\_LAB TriUNet & PBI Scrolling 2D UNet & 0.0002 \\
  MI & ADAR\_LAB UNesT & ADAR\_LAB nnUNet & 1.17e-09 \\
  MI & ADAR\_LAB UNesT & Baseline DS6 & 1.67e-08 \\
  MI & ADAR\_LAB UNesT & Baseline UNet MSS & 7.56e-11 \\
  MI & ADAR\_LAB UNesT & Dolphins & 2.36e-06 \\
  MI & ADAR\_LAB UNesT & FunPixel & 0.0007 \\
  MI & ADAR\_LAB UNesT & LSGroup & 2.51e-13 \\
  MI & ADAR\_LAB UNesT & PBI nnUNet & 1.76e-06 \\
  MI & ADAR\_LAB UNesT & neuRoSliCCe DS6\_MIP & 5.25e-05 \\
  MI & ADAR\_LAB UNesT & neuRoSliCCe MIP & 6.24e-09 \\
  MI & ADAR\_LAB UNesT & neuRoSliCCe multiMIP & 5.36e-11 \\
  MI & ADAR\_LAB nnUNet & EURECOM-UNIANDES & 1.31e-06 \\
  MI & ADAR\_LAB nnUNet & Koala Manual & 5.25e-05 \\
  MI & ADAR\_LAB nnUNet & Koala OM1 & 4.00e-05 \\
  MI & ADAR\_LAB nnUNet & Koala OM2 & 3.04e-05 \\
  MI & ADAR\_LAB nnUNet & PBI Scrolling 2D UNet & 0.0069 \\
  MI & Baseline DS6 & EURECOM-UNIANDES & 1.31e-05 \\
  MI & Baseline DS6 & Koala Manual & 0.0004 \\
  MI & Baseline DS6 & Koala OM1 & 0.0003 \\
  MI & Baseline DS6 & Koala OM2 & 0.0003 \\
  MI & Baseline DS6 & PBI Scrolling 2D UNet & 0.0375 \\
  MI & Baseline UNet MSS & EURECOM-UNIANDES & 1.11e-07 \\
  MI & Baseline UNet MSS & Koala Manual & 5.53e-06 \\
  MI & Baseline UNet MSS & Koala OM1 & 4.17e-06 \\
  MI & Baseline UNet MSS & Koala OM2 & 3.14e-06 \\
  MI & Baseline UNet MSS & PBI Scrolling 2D UNet & 0.0010 \\
  MI & Dolphins & EURECOM-UNIANDES & 0.0009 \\
  MI & Dolphins & Koala Manual & 0.0180 \\
  MI & Dolphins & Koala OM1 & 0.0147 \\
  MI & Dolphins & Koala OM2 & 0.0120 \\
  MI & EURECOM-UNIANDES & LSGroup & 5.97e-10 \\
  MI & EURECOM-UNIANDES & PBI nnUNet & 0.0007 \\
  MI & EURECOM-UNIANDES & neuRoSliCCe DS6\_MIP & 0.0120 \\
  MI & EURECOM-UNIANDES & neuRoSliCCe MIP & 5.53e-06 \\
  MI & EURECOM-UNIANDES & neuRoSliCCe multiMIP & 8.15e-08 \\
  MI & FunPixel & LSGroup & 0.0062 \\
  MI & Koala Manual & LSGroup & 4.31e-08 \\
  MI & Koala Manual & PBI nnUNet & 0.0147 \\
  MI & Koala Manual & neuRoSliCCe MIP & 0.0002 \\
  MI & Koala Manual & neuRoSliCCe multiMIP & 4.17e-06 \\
  MI & Koala OM1 & LSGroup & 3.14e-08 \\
  MI & Koala OM1 & PBI nnUNet & 0.0120 \\
  MI & Koala OM1 & neuRoSliCCe MIP & 0.0002 \\
  MI & Koala OM1 & neuRoSliCCe multiMIP & 3.14e-06 \\
  MI & Koala OM2 & LSGroup & 2.28e-08 \\
  MI & Koala OM2 & PBI nnUNet & 0.0097 \\
  MI & Koala OM2 & neuRoSliCCe MIP & 0.0001 \\
  MI & Koala OM2 & neuRoSliCCe multiMIP & 2.36e-06 \\
  MI & LSGroup & PBI Scrolling 2D UNet & 1.50e-05 \\
  MI & PBI Scrolling 2D UNet & neuRoSliCCe MIP & 0.0199 \\
  MI & PBI Scrolling 2D UNet & neuRoSliCCe multiMIP & 0.0008 \\
  \midrule
  bAVD & ADAR\_LAB SwinUNETR & ADAR\_LAB UNesT & 1.82e-05 \\
  bAVD & ADAR\_LAB SwinUNETR & Koala Manual & 2.90e-06 \\
  bAVD & ADAR\_LAB SwinUNETR & Koala OM1 & 0.0002 \\
  bAVD & ADAR\_LAB SwinUNETR & LSGroup & 1.82e-05 \\
  bAVD & ADAR\_LAB TriUNet & ADAR\_LAB UNesT & 8.64e-12 \\
  bAVD & ADAR\_LAB TriUNet & ADAR\_LAB nnUNet & 0.0289 \\
  bAVD & ADAR\_LAB TriUNet & EURECOM-UNIANDES & 0.0003 \\
  bAVD & ADAR\_LAB TriUNet & FunPixel & 7.35e-06 \\
  bAVD & ADAR\_LAB TriUNet & Koala Manual & 8.68e-13 \\
  bAVD & ADAR\_LAB TriUNet & Koala OM1 & 1.77e-10 \\
  bAVD & ADAR\_LAB TriUNet & neuRoSliCCe MIP & 0.0289 \\
  bAVD & ADAR\_LAB UNesT & ADAR\_LAB nnUNet & 0.0012 \\
  bAVD & ADAR\_LAB UNesT & Baseline DS6 & 1.94e-08 \\
  bAVD & ADAR\_LAB UNesT & Baseline UNet MSS & 2.38e-05 \\
  bAVD & ADAR\_LAB UNesT & Dolphins & 3.95e-11 \\
  bAVD & ADAR\_LAB UNesT & Koala OM2 & 0.0001 \\
  bAVD & ADAR\_LAB UNesT & LSGroup & 2.62e-19 \\
  bAVD & ADAR\_LAB UNesT & PBI Scrolling 2D UNet & 1.10e-06 \\
  bAVD & ADAR\_LAB UNesT & PBI nnUNet & 1.81e-13 \\
  bAVD & ADAR\_LAB UNesT & neuRoSliCCe DS6\_MIP & 0.0001 \\
  bAVD & ADAR\_LAB UNesT & neuRoSliCCe MIP & 0.0012 \\
  bAVD & ADAR\_LAB UNesT & neuRoSliCCe multiMIP & 5.71e-07 \\
  bAVD & ADAR\_LAB nnUNet & Koala Manual & 0.0002 \\
  bAVD & ADAR\_LAB nnUNet & Koala OM1 & 0.0091 \\
  bAVD & ADAR\_LAB nnUNet & LSGroup & 1.53e-07 \\
  bAVD & ADAR\_LAB nnUNet & PBI nnUNet & 0.0026 \\
  bAVD & Baseline DS6 & FunPixel & 0.0026 \\
  bAVD & Baseline DS6 & Koala Manual & 2.28e-09 \\
  bAVD & Baseline DS6 & Koala OM1 & 2.96e-07 \\
  bAVD & Baseline DS6 & LSGroup & 0.0056 \\
  bAVD & Baseline UNet MSS & Koala Manual & 3.91e-06 \\
  bAVD & Baseline UNet MSS & Koala OM1 & 0.0002 \\
  bAVD & Baseline UNet MSS & LSGroup & 1.36e-05 \\
  bAVD & Dolphins & EURECOM-UNIANDES & 0.0009 \\
  bAVD & Dolphins & FunPixel & 2.38e-05 \\
  bAVD & Dolphins & Koala Manual & 4.06e-12 \\
  bAVD & Dolphins & Koala OM1 & 7.67e-10 \\
  bAVD & EURECOM-UNIANDES & Koala Manual & 0.0236 \\
  bAVD & EURECOM-UNIANDES & LSGroup & 2.55e-10 \\
  bAVD & EURECOM-UNIANDES & PBI nnUNet & 1.82e-05 \\
  bAVD & FunPixel & LSGroup & 1.88e-12 \\
  bAVD & FunPixel & PBI Scrolling 2D UNet & 0.0444 \\
  bAVD & FunPixel & PBI nnUNet & 2.96e-07 \\
  bAVD & FunPixel & neuRoSliCCe multiMIP & 0.0289 \\
  bAVD & Koala Manual & Koala OM2 & 1.82e-05 \\
  bAVD & Koala Manual & LSGroup & 2.14e-20 \\
  bAVD & Koala Manual & PBI Scrolling 2D UNet & 1.53e-07 \\
  bAVD & Koala Manual & PBI nnUNet & 1.67e-14 \\
  bAVD & Koala Manual & neuRoSliCCe DS6\_MIP & 1.82e-05 \\
  bAVD & Koala Manual & neuRoSliCCe MIP & 0.0002 \\
  bAVD & Koala Manual & neuRoSliCCe multiMIP & 7.77e-08 \\
  bAVD & Koala OM1 & Koala OM2 & 0.0009 \\
  bAVD & Koala OM1 & LSGroup & 7.25e-18 \\
  bAVD & Koala OM1 & PBI Scrolling 2D UNet & 1.36e-05 \\
  bAVD & Koala OM1 & PBI nnUNet & 4.06e-12 \\
  bAVD & Koala OM1 & neuRoSliCCe DS6\_MIP & 0.0009 \\
  bAVD & Koala OM1 & neuRoSliCCe MIP & 0.0091 \\
  bAVD & Koala OM1 & neuRoSliCCe multiMIP & 7.35e-06 \\
  bAVD & Koala OM2 & LSGroup & 2.90e-06 \\
  bAVD & Koala OM2 & PBI nnUNet & 0.0236 \\
  bAVD & LSGroup & PBI Scrolling 2D UNet & 0.0002 \\
  bAVD & LSGroup & neuRoSliCCe DS6\_MIP & 2.90e-06 \\
  bAVD & LSGroup & neuRoSliCCe MIP & 1.53e-07 \\
  bAVD & LSGroup & neuRoSliCCe multiMIP & 0.0004 \\
  bAVD & PBI nnUNet & neuRoSliCCe DS6\_MIP & 0.0236 \\
  bAVD & PBI nnUNet & neuRoSliCCe MIP & 0.0026 \\
  \bottomrule
\end{longtable}
}

\twocolumn